\documentclass[prl,aps,superscriptaddress,twocolumn]{revtex4-2}

\usepackage{subfigure}
\usepackage{xcolor}
\usepackage{graphicx}
\usepackage{dcolumn}
\usepackage{bm}
\usepackage{amssymb}
\usepackage{comment}
\usepackage{physics}

\usepackage{booktabs}
\usepackage{amsthm}

\usepackage[colorlinks=true]{hyperref}
\hypersetup{linkcolor=blue,citecolor=blue,urlcolor=blue}

\newcommand{\be}{\begin{equation}}
\newcommand{\ee}{\end{equation}}

\newcommand{\bea}{\begin{eqnarray}}
\newcommand{\eea}{\end{eqnarray}}
\newcommand{\ba}{\begin{align}}
\newcommand{\ea}{\end{align}}

\newcommand{\beq}{\begin{equation}}
\newcommand{\eeq}{\end{equation}}

\begin{document}

\title{Majorana zero modes in half-quantum vortices of pair density wave superconductors}

\author{Xinyu Sun}
\affiliation{Chen-Ning Yang Institute for Advanced Study (YIAS), Tsinghua University, Beijing 100084, China}

\author{Hong Yao}
\email{yaohong@tsinghua.edu.cn}
\affiliation{Chen-Ning Yang Institute for Advanced Study (YIAS), Tsinghua University, Beijing 100084, China}

\date{\today}


\begin{abstract}
Pair-density-wave (PDW) superconductors admit half-quantum vortices that can bind Majorana zero modes, but whether such defects are energetically stabilized in microscopic models remains unclear.
We address this question in a spinless honeycomb-lattice model with a PDW phase.
Microscopic Hartree-Fock calculations determine the superfluid stiffness $\rho$ and PDW relative-phase stiffness $\kappa$, and show that $\kappa/\rho$ approaches unity near the continuous PDW--Dirac-semimetal transition, strongly reducing the long-wavelength cost of vortex fractionalization.
Combining this with microscopically extracted Ginzburg--Landau couplings, we show that vortex-core energetics favor fractionalization, producing a strongly enhanced splitting scale and a \textit{field-driven full-vortex--to--half-vortex lattice transition}.
We determine the resulting two-flavor half-vortex lattice structure and construct the associated Majorana lattice.
Majorana hybridization produces geometry- and flux-dependent bands with Dirac nodes and zero-energy Fermi lines.
Together, these results establish a microscopic route from PDW superconductivity to field-induced half-vortex and Majorana lattices.
\end{abstract}

\maketitle

\textit{Introduction}.---
Pair-density-wave (PDW) superconductivity, in which Cooper pairs condense at finite momentum, provides an enriched form of superconducting order that breaks both charge and translational symmetries~\cite{fulde1964superconductivity,larkin1965nonuniform,Berg2007Dynamical,Li2007Two,Berg2009Striped,Berg2009Theory,Berg2010Pair,Jaefari2012Pair,Soto2014Pair,Lee2014Amperean,Fradkin2015Colloquium,hamidian2016detection,ruan2018visualization,Edkins2019Magnetic,Agterberg2020Physics,Han2020Strong,Li2021Evolution,chen2021roton,Liu2021Discovery,huang2022pair,Han2022Pair,Zhang2022Pair,liu2023pair,zhao2023smectic,Wu2023Pair,Wuym2023Pair,Wang2025Pair}.
Its translational phase degree of freedom gives rise to unusual collective behavior~\cite{Wang2015Coexistence,Jian2020Mass,wu2025time,Nagashima2025Optically,wang2026anomalous}, fractional defects~\cite{Agterberg2011Conventional,Lesser2026Emblems}, and vestigial phases~\cite{Agterberg2015Emergent,zhou2022chern,Jonatan2023Nematic,wu2024d,Huecker2026Vestigial} such as charge-$4e$ superconductivity~\cite{berg2009charge,Jiang2017Charge,Jian2021Charge,Ge2024Charge,Zou2026Emergence}.
Of particular interest are half-quantum vortices, in which a half superconducting winding is bound to a PDW dislocation~\cite{agterberg2008dislocations,Radzihovsky2009Quantum,Mross2015Spin,Rosales2024Electronic}. 
Such defects are natural objects in multicomponent superconductors~\cite{Salomaa1985Half,Babaev2002Vortices,Chung2007Stability,Jang2011Observation} and can form vortex lattices even when full vortices are energetically competitive~\cite{Chung2009Fractional,Chung2010Entropy}.

Topological phenomena associated with PDW make fractional defects especially interesting~\cite{Cho2012Superconductivity,Cho2014Topological,Chan2017Non,Santos2019Pair}. 
In a topological superconductor, a vortex can bind a Majorana zero mode~\cite{Read2000Paired,Ivanov2001Non,Stone2006Fusion,Tewari2007Index,Tewari2008Testable,Fu2008Superconducting,Teo2010Topological,Wang2012The,Xu2014Artificial,Xu2015Experimental,Sun2016Majorana,Xu2016Topological}, while the two PDW components give rise to two half-vortex flavors with opposite PDW dislocations. 
An array of such defects can therefore realize a Majorana lattice with an internal structure absent in conventional vortex Majorana platforms~\cite{Biswas2013Majorana,Murray2015Majorana,Liu2015Electronic,Yoshida2016Generic}. 
A key question is whether half vortices can be energetically stabilized in a microscopic PDW. 
This question is particularly relevant near the continuous transition between a Dirac semimetal and a PDW state, where emergent supersymmetry has been predicted~\cite{Jian2015Emergent,Jian2017Emergence}, and the interplay between the superconducting and PDW translational sectors becomes important.

\begin{figure}[t]
    \centering
    \includegraphics[width=0.85\columnwidth]{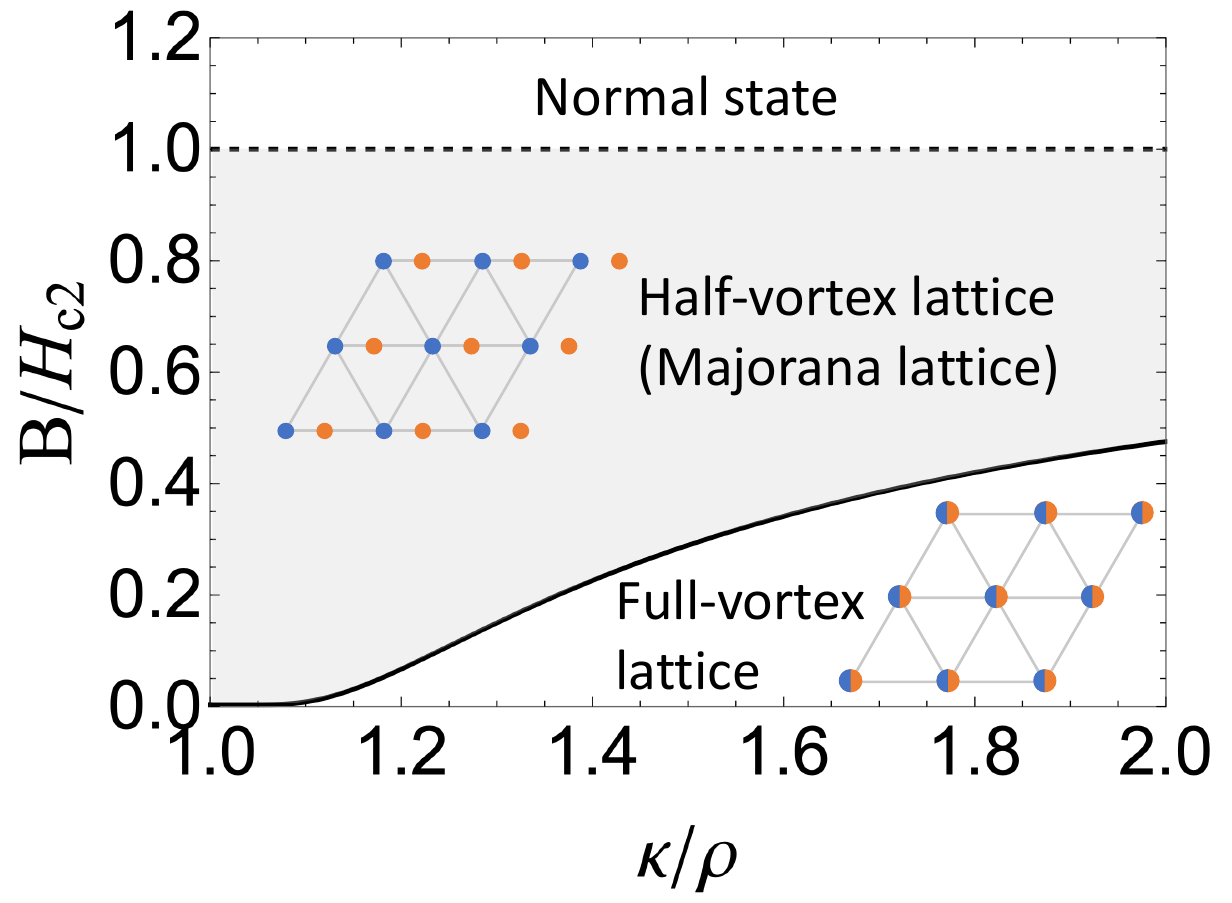}
    \caption{\textbf{Schematic phase diagram of the vortex lattice.}
    The solid black curve denotes the estimated critical field $B_c/H_{c2}$ separating the full-vortex and half-vortex lattices as a function of the stiffness ratio $\kappa/\rho$. 
    For $B<B_c$, the two half-vortex flavors remain bound into full vortices, whereas for $B>B_c$ they separate and form a rhombus-like half-vortex lattice, as shown schematically.
    The blue and orange markers denote the two half-vortex flavors. 
    The dashed line marks the upper critical field $H_{c2}$.
    The lower critical field $H_{c1}$ is not shown.}
\label{fig:phase_B}
\end{figure}

In this work, we investigate vortex fractionalization microscopically in the spinless honeycomb $t$--$V_1$--$V_2$ model, building on recent studies of PDW phases~\cite{Jian2015Emergent,Jiang2024Pair}.
Using self-consistent mean-field calculations, we determine the stiffness ratio $\kappa/\rho$, which approaches unity near the continuous PDW--Dirac-semimetal transition, strongly weakening the long-distance cost of vortex fractionalization.
We then extract the Ginzburg--Landau couplings directly from the microscopic state and find that the core energy favors separating the two component vortices.
Together, these effects produce a strongly enhanced splitting scale and the field-driven full-vortex--to--half-vortex lattice transition shown in Fig.~\ref{fig:phase_B}.
We further determine the resulting two-flavor half-vortex lattice, finding that the sign of $\kappa-\rho$ selects the relative displacement direction, while core energetics control the radial equilibrium separation and can stabilize the symmetric midpoint.
Finally, using the vortex-bound Majorana modes, we construct an effective Majorana lattice model, whose hybridization produces geometry- and flux-dependent Dirac nodes and zero-energy Fermi lines.
Overall, our results establish a microscopic route from PDW superconductivity to field-induced half-vortex and Majorana lattices with tunable low-energy band structure.

\textit{Microscopic model and PDW phase}.---
We study spinless fermions at half filling on the honeycomb lattice,
\begin{equation}
    H=-t\sum_{\langle ij\rangle}(c_i^\dagger c_j+\mathrm{H.c.})
    +V_1\sum_{\langle ij\rangle}n_i n_j
    +V_2\sum_{\langle\!\langle ij\rangle\!\rangle}n_i n_j ,
\label{eq:model}
\end{equation}
where $V_1$ and $V_2$ are the nearest- and next-nearest-neighbor interactions.
The noninteracting system is a Dirac semimetal with symmetry-related Dirac cones at $\pm\bm K$.
Interactions can destabilize the semimetal toward competing charge, quantum anomalous Hall (QAH), and superconducting orders~\cite{Raghu2008Topological,Jian2015Emergent,Jiang2024Pair}. 
Using self-consistent Hartree--Fock--Bogoliubov (HFB) calculations, we find an extended PDW phase for attractive $V_1$ and intermediate repulsive $V_2$.
In the following, we focus on this PDW phase, particularly near the PDW--DSM boundary, where the pairing is dominated by the two symmetry-related components at $\pm\bm K$.
For a nearest-neighbor bond orientation $\alpha$, we write the PDW order parameter as
\begin{equation}
    \Delta_{\alpha}(\bm R)=\Delta_{\alpha,+}e^{i\bm K\cdot\bm R}+\Delta_{\alpha,-}e^{-i\bm K\cdot\bm R},
\label{eq:pdw_order}
\end{equation}
where $\Delta_{\alpha,\pm}$ denote the two PDW components.
We next characterize the long-wavelength phase response of this two-component PDW.

\textit{PDW stiffnesses and critical behavior}.---
The two PDW components contain a common superconducting phase $\phi_{\rm sc}$ and a relative phase $\phi_K$ that translates the PDW pattern, with $\Delta_{\alpha,\pm}=|\Delta_{\alpha,\pm}|e^{i(\phi_{\rm sc}\pm\phi_K)}$~\cite{Radzihovsky2011Fluctuations}.
Their long-wavelength phase energy is
\begin{equation}
    S_{\rm ph}=\frac{1}{2}\int {\rm d}^2r\left[\rho\left(\bm\nabla\phi_{\rm sc}-\frac{2e}{\hbar c}\bm A\right)^2+\kappa(\bm\nabla\phi_K)^2\right],
\label{eq:phase_action}
\end{equation}
which defines the superfluid stiffness $\rho$ and the PDW relative-phase stiffness $\kappa$.
The vector potential couples only to the phase $\phi_{\rm sc}$, so an applied magnetic field induces a net superconducting-phase winding, while $\phi_K$ remains a neutral relative-phase mode. 
The two stiffnesses therefore control the relative long-wavelength cost of integer and half-vortex configurations, motivating their independent determination across the PDW phase.

\begin{figure}[t]
    \centering
    \includegraphics[width=\columnwidth]{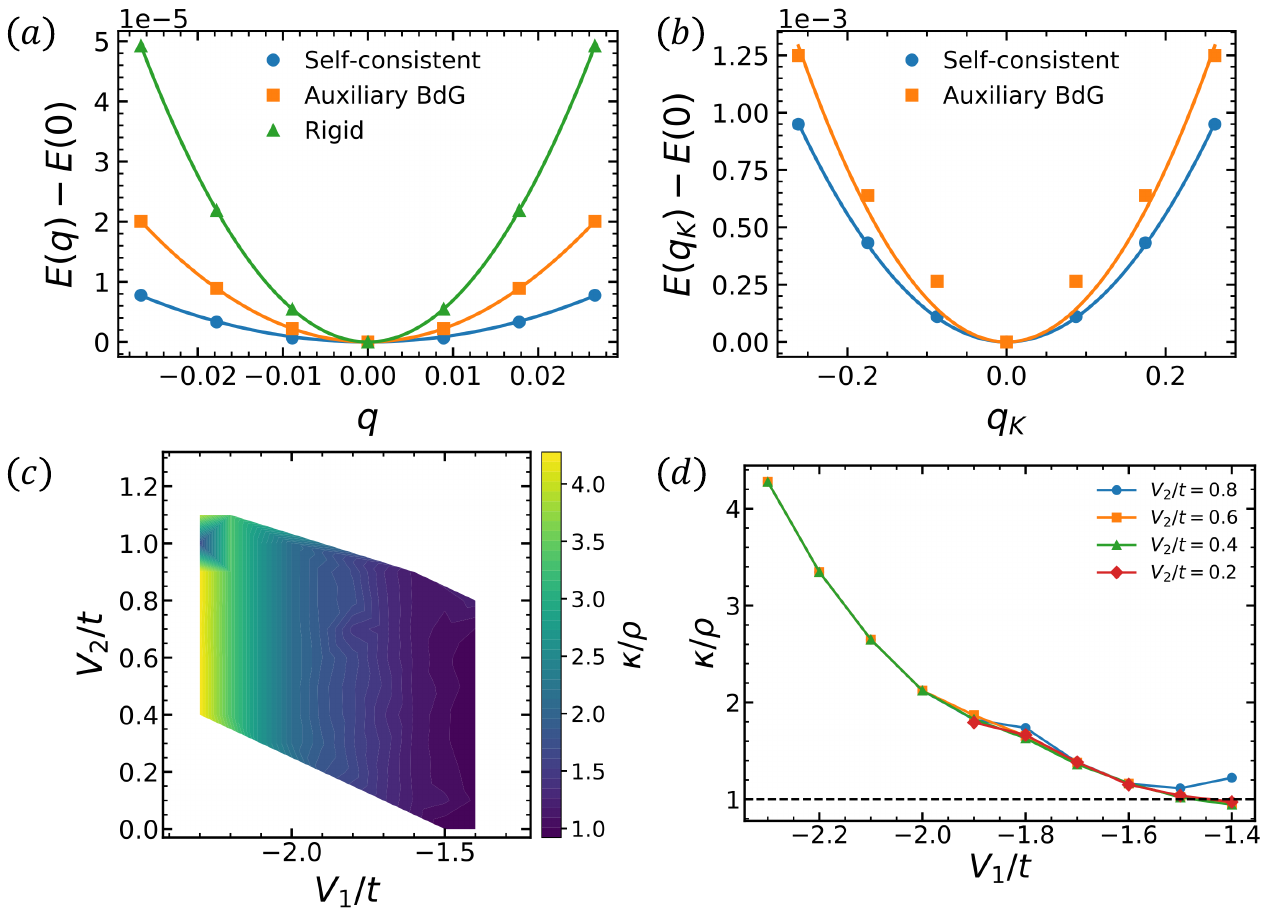}
    \caption{\textbf{PDW stiffnesses and their evolution across the PDW phase.}
    (a,b) Representative energy-curvature calculations at $(V_1,V_2)=(-1.7,0.6)t$. 
    (a) Energy cost of a global superconducting phase twist, obtained from fully self-consistent, auxiliary-BdG, and rigid-state calculations. 
    (b) Energy cost of a PDW relative-phase deformation, obtained from fully self-consistent and auxiliary-BdG calculations.
    Solid curves denote quadratic fits used to extract $\rho$ and $\kappa$.
    (c) Thermodynamic-limit fully self-consistent ratio $\kappa/\rho$ across the PDW region.
    (d) Representative cuts at fixed $V_2/t=0.2,0.4,0.6,0.8$, showing that $\kappa/\rho$ decreases toward unity upon approaching the PDW--DSM boundary. 
    The dashed line marks $\kappa/\rho=1$.}
\label{fig:stiffness}
\end{figure}

We extract both stiffnesses from the quadratic energy cost of uniform phase gradients~\cite{Waardh2018Suppression}.
A superconducting phase gradient is implemented through the corresponding Peierls twist, while a relative-phase gradient shifts the two PDW wave vectors from $\pm\mathbf K$ to $\pm(\mathbf K+\mathbf q)$.
For either channel $a=\mathrm{sc},K$, we fit the energy to $e_a(\bm q)-e(0)=\Upsilon_a q^2/2+O(q^4)$, with $\Upsilon_{\rm sc}=\rho$ and $\Upsilon_K=\kappa$.
We compare fully self-consistent (SCF), auxiliary-BdG, and, for the superconducting twist, rigid-state responses, corresponding to progressively less relaxation of the mean-field state~\footnote{In the fully self-consistent calculation, the HFB fields are reoptimized at each imposed deformation.
In the auxiliary-BdG scheme, the deformed BdG Hamiltonian is constructed from the undeformed mean-field saddle and diagonalized once, while in the rigid calculation the undeformed Gaussian state itself is kept fixed.
The rigid construction applies only to the superconducting phase twist.
See the SM~\cite{SM} for details.}.
Representative results in Figs.~\ref{fig:stiffness}(a) and (b) exhibit clear quadratic behavior, with relaxation systematically reducing the curvature.
We therefore use the fully self-consistent results and extrapolate them to the thermodynamic limit.
Both stiffnesses soften toward the continuous PDW--DSM transition, with $\kappa$ decreasing more rapidly than $\rho$.
Consequently, the ratio $\kappa/\rho$, shown in Figs.~\ref{fig:stiffness}(c) and (d), decreases from values well above unity inside the PDW phase toward $\kappa/\rho\simeq1$ near the DSM boundary.
See SM~\cite{SM} for details.

This behavior follows naturally from the two-component Ginzburg--Landau theory. 
Because the PDW components at $\pm\bm K$ are related by symmetry, their leading gradient terms are identical. 
Retaining the leading symmetry-allowed quartic-gradient corrections, we write
\begin{equation}
\begin{split}
f_{\rm grad}={}&\alpha\left(|D_i\Delta_+|^2+|D_i\Delta_-|^2\right)\\
&+\lambda\left(|\Delta_+|^2|D_i\Delta_+|^2+|\Delta_-|^2|D_i\Delta_-|^2\right)\\
&+\eta\left[(\Delta_+^*D_i\Delta_+)(\Delta_-^*D_i\Delta_-)+\mathrm{c.c.}\right]+\cdots .
\end{split}
\label{eq:GL_gradient}
\end{equation}
Here $\lambda$ gives a common quartic-gradient renormalization of the two phase stiffnesses, while $\eta$ distinguishes the superconducting and PDW relative-phase modes~\footnote{Additional symmetry-allowed quartic-gradient terms, evaluated about the symmetry-related state $|\Delta_+|=|\Delta_-|$, reduce to either a common renormalization of the two phase stiffnesses or their relative splitting, and can therefore be absorbed into effective coefficients $\lambda$ and $\eta$.}.

For $\Delta_\pm=\Delta_0e^{i(\phi_{\rm sc}\pm\phi_K)}$, the gradient free energy yields two phase stiffnesses $\rho=4\Delta_0^2[\alpha+(\lambda-\eta)\Delta_0^2]$ and $\kappa=4\Delta_0^2[\alpha+(\lambda+\eta)\Delta_0^2]$.
Hence, within the quartic-gradient truncation,
\begin{equation}
    \frac{\kappa}{\rho}=
    \frac{\alpha+(\lambda+\eta)\Delta_0^2}{\alpha+(\lambda-\eta)\Delta_0^2}=
    1+2\frac{\eta}{\alpha}\Delta_0^2+O(\Delta_0^4).
\label{eq:ratio_GL_exact}
\end{equation}
The last equality follows in the small-$\Delta_0$ limit near the continuous PDW--DSM transition.
Thus $\lambda$ gives only a common stiffness renormalization at leading order, while $\eta$ controls the splitting between $\kappa$ and $\rho$.
As $\Delta_0\rightarrow0$, symmetry between the two PDW components enforces $\kappa/\rho\rightarrow1$, consistent with the emergent $\mathcal N=2$ supersymmetry of the PDW--DSM transition~\cite{Jian2015Emergent}.
Approaching this limit from $\kappa>\rho$, the long-wavelength penalty for vortex fractionalization becomes weak, making the vortex-core energetics decisive.

\textit{Half-vortex energetics and fractionalization}.---
We now ask whether a magnetic full vortex can fractionalize into two half vortices.
Because the magnetic field couples to $\phi_{\rm sc}$ but not to $\phi_K$, the relevant field-induced defect is the full vortex, with phase winding $(\delta\phi_{\rm sc},\delta\phi_K)=(2\pi,0)$.
It can split into two half vortices with $(\delta\phi_{\rm sc},\delta\phi_K)=(\pi,\pm\pi)$.
For two half vortices separated by $d$, the superconducting sector favors their separation, contributing $-(\pi\rho/2)\ln(d/\xi)$, whereas their opposite relative-phase windings contribute $+(\pi\kappa/2)\ln(d/\xi)$~\cite{Chung2010Entropy,How2020Half}.
Relative to the unsplit full vortex,
\begin{equation}
    \Delta E(d)=\frac{\pi}{2}(\kappa-\rho)\ln\frac{d}{\xi}
    +\Delta E_{\rm core},
\label{eq:split_general}
\end{equation}
where $\xi$ is the vortex-core scale and $\Delta E_{\rm core}$ denotes the short-distance energy difference between the two half-vortex cores and the full-vortex core.
For $\kappa<\rho$, the long-wavelength contribution favors fractionalization, whereas for $\kappa>\rho$ it favors the unsplit vortex, requiring a sufficiently negative core contribution for fractionalization.

\begin{figure}[t]
    \centering
    \includegraphics[width=\columnwidth]{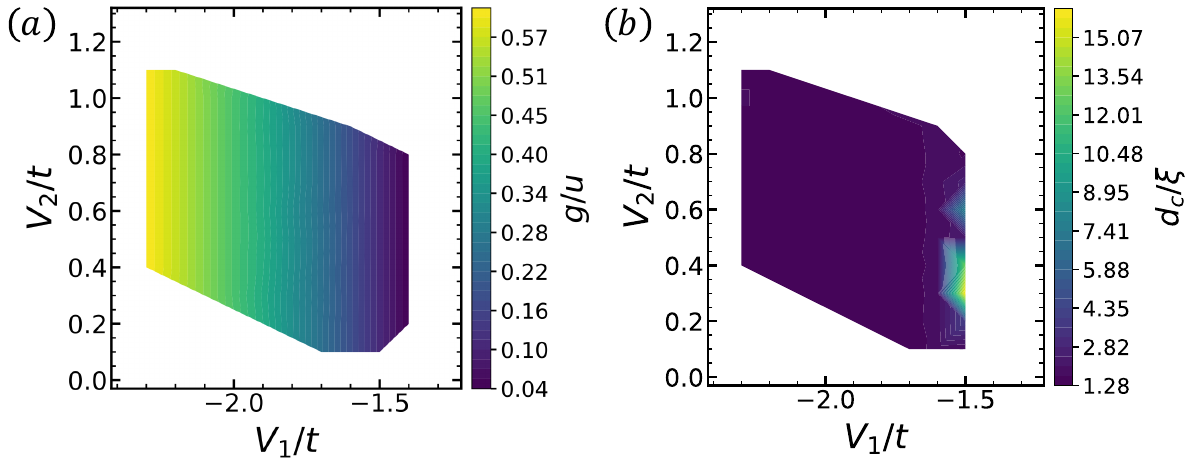}
    \caption{\textbf{Quartic coupling and half-vortex splitting scale.}
    (a) Thermodynamic-limit ratio $g/u$ extracted from the microscopic PDW mean-field energy. 
    (b) Estimated critical separation $d_c/\xi$, obtained by combining $g/u$ with the fully self-consistent stiffness ratio $\kappa/\rho$. 
    Only points with available data in both calculations and $V_1/t\leq-1.5$ are shown.}
\label{fig:ratio}
\end{figure}

To determine the sign and magnitude of $\Delta E_{\rm core}$, we supplement the gradient sector introduced above with the local two-component Ginzburg--Landau potential
\begin{equation}
    f_{\rm pot}=r(|\Delta_+|^2+|\Delta_-|^2)+g|\Delta_+|^2|\Delta_-|^2
    +\frac{u}{2}(|\Delta_+|^4+|\Delta_-|^4).
\label{eq:GL_potential}
\end{equation}
Here $r<0$ in the PDW phase.
The intercomponent coupling $g$ controls the local competition between the two PDW amplitudes.
For $g>0$, suppressing one component allows the other to grow, lowering the half-vortex core energy.
By contrast, $g<0$ favors the two amplitudes varying together.
We extract $u$ and $g$ microscopically by perturbing the two PDW amplitudes around the self-consistent saddle and fitting the interacting energy (see the SM~\cite{SM}).
As shown in Fig.~\ref{fig:ratio}(a), we find $u>0$ and $0<g/u<1$ throughout the studied region, ensuring local stability while favoring half-vortex core formation.
This hierarchy is consistent with the RG structure of the emergent SUSY fixed point~\cite{Jian2015Emergent}, where the intercomponent quartic coupling flows to zero while the intracomponent interaction remains finite.

To quantify the core-energy gain, we compare the local minima of Eq.~\eqref{eq:GL_potential} for the uniform PDW state, an integer-vortex core, and a half-vortex core. 
Allowing one component to relax in the half-vortex core gives $\delta f_{\rm core}=2\delta f_{\rm HV}-\delta f_{\rm IV}=-gr^2/[u(u+g)]$, where $\delta f_{\rm HV}$ ($\delta f_{\rm IV}$) is the energy difference between the half-vortex (integer-vortex) core and the uniform state.
For $g>0$, $\delta f_{\rm core}<0$, showing that the local potential favors splitting the two component cores.
Approximating the core area as $C\pi\xi^2$, with $C$ an order-one factor, we obtain $\Delta E_{\rm core}\simeq-(C\pi/8)(\rho+\kappa)g/u$.
Combining this with Eq.~\eqref{eq:split_general}, for $\kappa>\rho$ the condition $\Delta E(d_c)=0$ gives (see the SM~\cite{SM})
\begin{equation}
    \frac{d_c}{\xi}\simeq
    \exp\left[\frac{C}{4}\frac{1+\kappa/\rho}{\kappa/\rho-1}\frac{g}{u}\right].
\label{eq:dc}
\end{equation}
For $d<d_c$, the core-energy gain overcomes the long-distance elastic penalty, making the split half-vortex configuration energetically favorable.
As $\kappa/\rho\rightarrow1^+$, $d_c/\xi$ grows rapidly.
For $\kappa<\rho$, the long-distance interaction already favors fractionalization, so no finite upper scale $d_c$ arises.
Using the microscopic $g/u$ and $\kappa/\rho$, Fig.~\ref{fig:ratio}(b) shows the strong enhancement of $d_c/\xi$ near the PDW--DSM boundary.
A well-defined half-vortex and Majorana lattice requires $\max(\xi,\xi_M)\ll d<\min(d_c,\ell_{\rm IR})$, where $\ell_{\rm IR}$ is the infrared cutoff of the logarithmic vortex interaction, and $\xi_M$ is the Majorana localization length.
Thus, the rapid growth of $d_c/\xi$ near the PDW--DSM boundary substantially enlarges the accessible window~\footnote{This regime is naturally favored in thin films, where the effective magnetic screening length can greatly exceed $\xi$.}.

The upper bound $d<d_c$ can be translated into a magnetic-field criterion because the vortex spacing is set by the flux density.
Let $a_v$ be the Bravais-lattice spacing, with one half vortex of each flavor per unit cell.
Since the two half vortices together carry one flux quantum, $\Phi_0=B(\sqrt3/2)a_v^2$. 
Defining the upper critical field $H_{c2}$, above which superconductivity is destroyed, we have $\Phi_0=H_{c2}(\sqrt3/2)\xi^2$ for the triangular lattice.
Writing the half-vortex separation as $d=\zeta a_v$ with $\zeta<1$, we find that the condition $d<d_c$ is equivalent to $B>B_c$, where
\begin{equation}
    \frac{B_c}{H_{c2}}
    =\zeta^2\left(\frac{d_c}{\xi}\right)^{-2}.
\end{equation}
Since $\zeta<1$ and $d_c/\xi>1$ in the regime of interest, $B_c<H_{c2}$.
In thin films, the large magnetic screening length can make the lower critical field $H_{c1}$ small, allowing $H_{c1}<B_c<H_{c2}$.
We assume $H_{c1}\ll B_c$ and omit it in Fig.~\ref{fig:phase_B}.
The resulting field evolution is a Meissner state for $B<H_{c1}$, a full-vortex lattice for $H_{c1}<B<B_c$, a half-vortex lattice for $B_c<B<H_{c2}$, and the normal state for $B>H_{c2}$.
As $\kappa/\rho\rightarrow1^+$, $d_c$ grows rapidly and $B_c$ is suppressed, enlarging the half-vortex lattice regime.

\begin{figure}[t]
    \centering
    \includegraphics[width=\columnwidth]{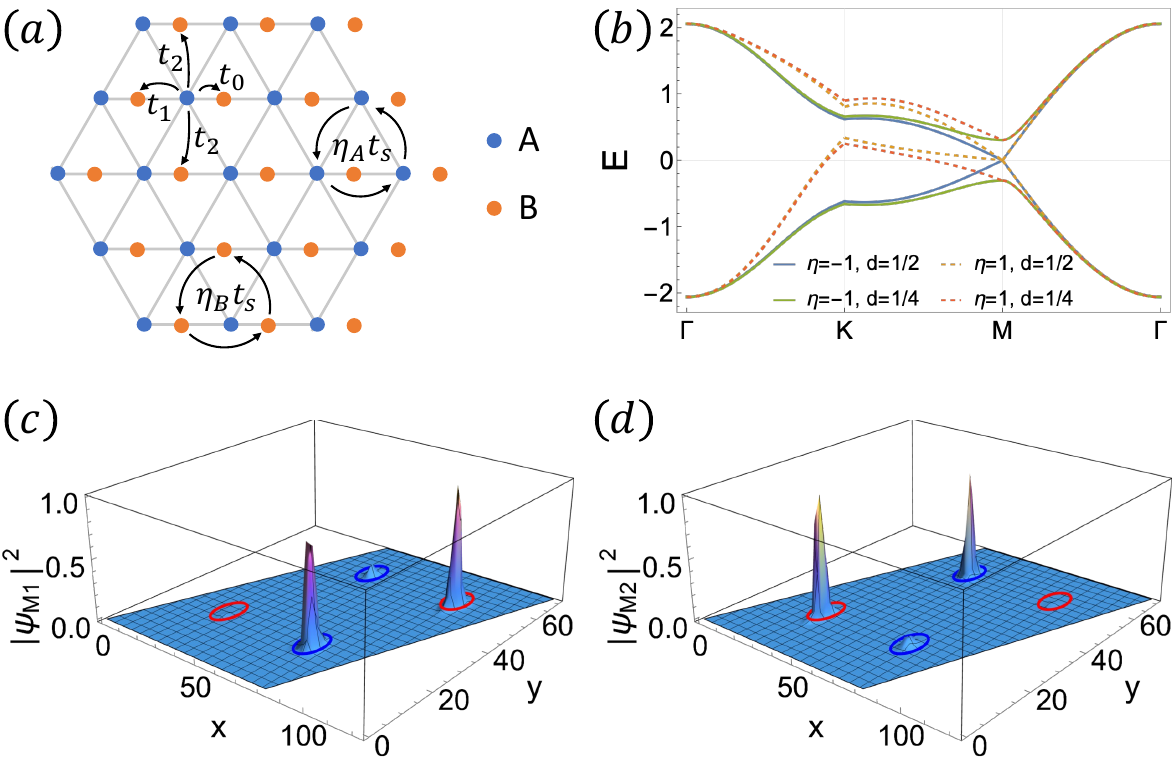}
    \caption{\textbf{Half-vortex Majorana lattice and near-zero-energy states.}
    (a) Effective Majorana hopping model for the two half-vortex flavors $A$ and $B$. 
    Same-flavor Majoranas are coupled by $t_s$ with hopping orientations $\eta_A$ and $\eta_B$, while $t_0$, $t_1$, and $t_2$ denote the inequivalent opposite-flavor hoppings.
    (b) Majorana bands along $\Gamma$--$K$--$M$--$\Gamma$ for representative $(\eta,d)=(-1,1/2)$, $(1,1/2)$, $(-1,1/4)$, and $(1,1/4)$, with $a=1$, $t_c^{(0)}=1$, $t_s^{(0)}=0.3$, and $\xi_M=1$.
    (c,d) Spatial probability densities $|\psi_{M1}|^2$ and $|\psi_{M2}|^2$ of the two near-zero complex BdG eigenstates in a four-half-vortex configuration. 
    Each state has weight on two half vortices of opposite flavor.
    With their particle-hole partners, they span the four-dimensional near-zero subspace corresponding to one Majorana zero mode bound to each half vortex. 
    Red and blue circles denote vortices in the $\Delta_+$ and $\Delta_-$ components, respectively.}
\label{fig:majorana}
\end{figure}

\textit{Half-vortex Majorana lattice}.---
We now determine the structure of the field-induced half-vortex lattice in the regime $|\kappa-\rho|\ll\kappa+\rho$.
Assuming the isotropic logarithmic interaction and a rotationally invariant short-distance core contribution, the same-flavor interaction fixes each vortex flavor, to leading order, into a triangular lattice with primitive vectors $\mathbf a_1=a(1,0)$ and $\mathbf a_2=a(1/2,\sqrt3/2)$, while the weaker interflavor interaction selects their relative displacement $\boldsymbol\tau=d(\cos\theta,\sin\theta)$ (see the SM~\cite{SM}).
Exploiting the sixfold rotational symmetry of the triangular lattice, we find that for any $0<d/a<1$ within the triangular-lattice ansatz,
\begin{equation}
    \theta_{\min}=
    \begin{cases}
        n\pi/3, & \kappa>\rho,\\
        \pi/6+n\pi/3, & \kappa<\rho,
    \end{cases}
    \qquad n\in\mathbb Z.
\label{eq:main_lattice_angle}
\end{equation}
Thus, for $\kappa>\rho$, the two triangular sublattices are displaced along a primitive direction, $\boldsymbol{\tau}\parallel\mathbf a_1$, whereas for $\kappa<\rho$ the displacement is rotated by $\pi/6$, with $\boldsymbol{\tau}\parallel\mathbf a_1+\mathbf a_2$.
These directions are consistent with Ref.~\cite{Chung2010Entropy}, while the analysis here establishes their angular selection for arbitrary $d$ within the triangular-lattice ansatz (see the SM~\cite{SM}).
Near $\kappa\simeq\rho$, relaxation of the triangular Bravais lattice produces only an $O(\frac{|\kappa-\rho|}{\kappa+\rho})$ deformation, so this ansatz remains controlled.

The radial stability differs in the two cases.
For $\kappa<\rho$, the honeycomb displacement $\boldsymbol{\tau}=(\mathbf a_1+\mathbf a_2)/3$ is already a local minimum of the logarithmic lattice interaction, whereas for $\kappa>\rho$ the midpoint $\boldsymbol{\tau}=\mathbf a_1/2$ remains radially unstable.
Short-distance core energetics can reverse this curvature and stabilize the interlaced-rhombus configuration.
Writing $R=a/\xi$, we find that for $\kappa>\rho$ the equilibrium displacement satisfies $d/a<1/2$ in the dilute regime.
As $R$ is reduced below a critical value $R_*$, the midpoint becomes locally stable and the relative displacement locks to $d/a=1/2$ in the well-separated-vortex regime (see the SM~\cite{SM}).
For the Ginzburg--Landau core profile, taking the representative microscopic splitting scale $d_c/\xi\simeq15.7$~\footnote{We set $C=1$ in the numerical estimates.} gives $R_*\simeq16$, corresponding to $d\simeq8\xi$ at the locking point.
Since $d<d_c$, this symmetric configuration remains within the regime where half-vortex splitting is energetically favorable.
The corresponding lattice structure in the $\kappa>\rho$ regime relevant to our microscopic model is shown in Fig.~\ref{fig:majorana}(a).

We verify microscopically that both half-vortex flavors bind Majorana zero modes.
Since $\phi_\pm=\phi_{\rm sc}\pm\phi_K$, the defects $(\delta\phi_{\rm sc},\delta\phi_K)=(\pi,\pm\pi)$ correspond to unit vortices in $\Delta_\pm$.
We construct a gauge-consistent real-space BdG Hamiltonian for four well-separated half vortices, each carrying flux $\Phi_0/2$.
The spectrum contains a four-dimensional near-zero subspace, corresponding to one Majorana zero mode per half vortex (see the SM~\cite{SM}).
Figures~\ref{fig:majorana}(c) and (d) show the two near-zero complex BdG eigenstates, with energies $E_1/t=2.16\times10^{-12}$ and $E_2/t=2.25\times10^{-12}$.
The Majorana manifold is separated from conventional vortex-core excitations~\cite{caroli1964bound} by a gap $\Delta E/t\simeq0.81$.

The Majorana zero modes localized at the half vortices hybridize into dispersive bands.
For the two triangular sublattices $A$ and $B$, we retain the same-flavor nearest-neighbor hopping $t_s$ and three inequivalent opposite-flavor hoppings $t_0$, $t_1$, and $t_2$, as shown in Fig.~\ref{fig:majorana}(a).
The effective Majorana Hamiltonian is
\begin{equation}
    H_M=
    i\sum_{\alpha,\bm R,\bm\delta}t_{s,\alpha}
    \gamma_{\alpha,\bm R}\gamma_{\alpha,\bm R+\bm\delta}
    +i\sum_{\bm R,\bm\iota}t_{\bm\iota}\,
    \gamma_{A,\bm R}\gamma_{B,\bm R+\bm\iota},
\label{eq:majorana_real}
\end{equation}
where $\bm\delta\in\{\bm a_1,-\bm a_2,\bm a_2-\bm a_1\}$ and
$\bm\iota\in\{\bm 0,-\bm a_1,-\bm a_2,\bm a_2-\bm a_1\}$.
The same-flavor hopping is $t_{s,\alpha}=t_s\eta_\alpha$, with $\eta_A=1$ and $\eta_B=\eta=\pm1$ specifying the relative hopping orientation and hence the Majorana flux sector, while $t_{\bm\iota}=(t_0,t_1,t_2,t_2)$ in the listed order.
We model the hopping magnitude as $t(r)=t^{(0)}e^{-r/\xi_M}$, with prefactors $t_s^{(0)}$ and $t_c^{(0)}$ for same- and opposite-flavor bonds, respectively.
With $a=1$, the bond lengths are $1$ for $t_s$ and $d$, $1-d$, and $\sqrt{1-d+d^2}$ for $t_0$, $t_1$, and $t_2$, respectively.

The relative displacement $d$ and flux sector $\eta$ control the nodal structure of the Majorana bands.
For $\eta=-1$, the symmetric configuration $d=1/2$ hosts isolated Majorana Dirac nodes at the symmetry-related $M$ points, while $d\neq1/2$ is fully gapped.
For $\eta=+1$, $d=1/2$ instead exhibits a one-dimensional direct band-crossing locus together with zero-energy Majorana Fermi lines passing through the $M$ points.
Away from $d=1/2$, the direct band crossing is lifted, while the Fermi lines can persist depending on the hopping parameters.
The corresponding spectra are shown in Fig.~\ref{fig:majorana}(b), with the Bloch Hamiltonian and gap analysis in the SM~\cite{SM}.

\textit{Conclusions}.---
Our results establish a microscopic mechanism for field-induced vortex fractionalization in a two-component PDW.
In the spinless honeycomb $t$--$V_1$--$V_2$ model, we find that $\kappa/\rho$ approaches unity near the continuous PDW--DSM transition, strongly suppressing the long-distance cost of fractionalization.
Combined with the microscopic core energetics, this produces a strongly enhanced splitting scale and a field-driven transition from a full-vortex to a half-vortex lattice.
We further determine the two-flavor half-vortex lattice, showing that $\kappa-\rho$ selects the displacement direction, while core energetics can stabilize well-separated half vortices, including at the symmetric midpoint.
Finally, we construct the effective Majorana lattice and show that its hybridization produces geometry- and flux-dependent Dirac nodes and zero-energy Fermi lines.
Overall, our results establish a microscopic route to field-induced half-vortex and Majorana lattices in PDW superconductors.

Experimentally, this mechanism motivates searching for distinctive Majorana signatures in PDW systems under an applied magnetic field~\cite{gu2023detection}.
Half-flux vortices and the accompanying PDW dislocations can provide experimental signatures of vortex fractionalization~\cite{du2020imaging}, while local spectroscopic probes can detect the associated zero-energy bound states~\cite{Zhang2018Observation,Wang2018Evidence,machida2019zero,kong2019half,Zhu2020Nearly,li2022ordered}.
The additional half-vortex flavor provides an extra degree of control over Majorana hybridization and coupling networks, potentially offering new opportunities for the manipulation and encoding of Majorana-based qubits and for topological quantum computation~\cite{Kitaev2001Unpaired,Bonderson2008Measurement,Nayak2008Non,alicea2011non,Alicea2012New,Beenakker2013Search,sarma2015majorana,Aasen2016Milestones,Karzig2017Scalable}.
It would also be interesting to explore these phenomena in a broader class of microscopic PDW realizations and to understand how interactions, disorder, and lattice geometry reshape the resulting Majorana phases~\cite{Chiu2015Strongly,Christian2021Robustness,Vedangi2021Majorana}.

\textit{Acknowledgments}.---We would like to thank Yi-Fan Jiang, Shuo Liu and Zhengzhi Wu for helpful discussions. This work is supported in part by the NSFC under Grant Nos.~12347107 and 12334003 (X.S. and H.Y.), and the New Cornerstone Science Foundation through the Xplorer Prize (H.Y.).


\let\oldaddcontentsline\addcontentsline
\renewcommand{\addcontentsline}[3]{}
\bibliography{refs.bib}

\begin{thebibliography}{110}%
\makeatletter
\providecommand \@ifxundefined [1]{%
 \@ifx{#1\undefined}
}%
\providecommand \@ifnum [1]{%
 \ifnum #1\expandafter \@firstoftwo
 \else \expandafter \@secondoftwo
 \fi
}%
\providecommand \@ifx [1]{%
 \ifx #1\expandafter \@firstoftwo
 \else \expandafter \@secondoftwo
 \fi
}%
\providecommand \natexlab [1]{#1}%
\providecommand \enquote  [1]{``#1''}%
\providecommand \bibnamefont  [1]{#1}%
\providecommand \bibfnamefont [1]{#1}%
\providecommand \citenamefont [1]{#1}%
\providecommand \href@noop [0]{\@secondoftwo}%
\providecommand \href [0]{\begingroup \@sanitize@url \@href}%
\providecommand \@href[1]{\@@startlink{#1}\@@href}%
\providecommand \@@href[1]{\endgroup#1\@@endlink}%
\providecommand \@sanitize@url [0]{\catcode `\\12\catcode `\$12\catcode
  `\&12\catcode `\#12\catcode `\^12\catcode `\_12\catcode `\%12\relax}%
\providecommand \@@startlink[1]{}%
\providecommand \@@endlink[0]{}%
\providecommand \url  [0]{\begingroup\@sanitize@url \@url }%
\providecommand \@url [1]{\endgroup\@href {#1}{\urlprefix }}%
\providecommand \urlprefix  [0]{URL }%
\providecommand \Eprint [0]{\href }%
\providecommand \doibase [0]{https://doi.org/}%
\providecommand \selectlanguage [0]{\@gobble}%
\providecommand \bibinfo  [0]{\@secondoftwo}%
\providecommand \bibfield  [0]{\@secondoftwo}%
\providecommand \translation [1]{[#1]}%
\providecommand \BibitemOpen [0]{}%
\providecommand \bibitemStop [0]{}%
\providecommand \bibitemNoStop [0]{.\EOS\space}%
\providecommand \EOS [0]{\spacefactor3000\relax}%
\providecommand \BibitemShut  [1]{\csname bibitem#1\endcsname}%
\let\auto@bib@innerbib\@empty
\bibitem [{\citenamefont {Fulde}\ and\ \citenamefont
  {Ferrell}(1964)}]{fulde1964superconductivity}%
  \BibitemOpen
  \bibfield  {author} {\bibinfo {author} {\bibfnamefont {P.}~\bibnamefont
  {Fulde}}\ and\ \bibinfo {author} {\bibfnamefont {R.~A.}\ \bibnamefont
  {Ferrell}},\ }\bibfield  {title} {\bibinfo {title} {Superconductivity in a
  strong spin-exchange field},\ }\href
  {https://journals.aps.org/pr/abstract/10.1103/PhysRev.135.A550} {\bibfield
  {journal} {\bibinfo  {journal} {Physical Review}\ }\textbf {\bibinfo {volume}
  {135}},\ \bibinfo {pages} {A550} (\bibinfo {year} {1964})}\BibitemShut
  {NoStop}%
\bibitem [{\citenamefont {Larkin}\ and\ \citenamefont
  {Ovchinnikov}(1965)}]{larkin1965nonuniform}%
  \BibitemOpen
  \bibfield  {author} {\bibinfo {author} {\bibfnamefont {A.}~\bibnamefont
  {Larkin}}\ and\ \bibinfo {author} {\bibfnamefont {Y.~N.}\ \bibnamefont
  {Ovchinnikov}},\ }\bibfield  {title} {\bibinfo {title} {Nonuniform state of
  superconductors},\ }\href@noop {} {\bibfield  {journal} {\bibinfo  {journal}
  {Soviet Physics-JETP}\ }\textbf {\bibinfo {volume} {20}},\ \bibinfo {pages}
  {762} (\bibinfo {year} {1965})}\BibitemShut {NoStop}%
\bibitem [{\citenamefont {Berg}\ \emph {et~al.}(2007)\citenamefont {Berg},
  \citenamefont {Fradkin}, \citenamefont {Kim}, \citenamefont {Kivelson},
  \citenamefont {Oganesyan}, \citenamefont {Tranquada},\ and\ \citenamefont
  {Zhang}}]{Berg2007Dynamical}%
  \BibitemOpen
  \bibfield  {author} {\bibinfo {author} {\bibfnamefont {E.}~\bibnamefont
  {Berg}}, \bibinfo {author} {\bibfnamefont {E.}~\bibnamefont {Fradkin}},
  \bibinfo {author} {\bibfnamefont {E.-A.}\ \bibnamefont {Kim}}, \bibinfo
  {author} {\bibfnamefont {S.~A.}\ \bibnamefont {Kivelson}}, \bibinfo {author}
  {\bibfnamefont {V.}~\bibnamefont {Oganesyan}}, \bibinfo {author}
  {\bibfnamefont {J.~M.}\ \bibnamefont {Tranquada}},\ and\ \bibinfo {author}
  {\bibfnamefont {S.~C.}\ \bibnamefont {Zhang}},\ }\bibfield  {title} {\bibinfo
  {title} {Dynamical layer decoupling in a stripe-ordered high-${T}_{c}$
  superconductor},\ }\href {https://doi.org/10.1103/PhysRevLett.99.127003}
  {\bibfield  {journal} {\bibinfo  {journal} {Phys. Rev. Lett.}\ }\textbf
  {\bibinfo {volume} {99}},\ \bibinfo {pages} {127003} (\bibinfo {year}
  {2007})}\BibitemShut {NoStop}%
\bibitem [{\citenamefont {Li}\ \emph {et~al.}(2007)\citenamefont {Li},
  \citenamefont {H\"ucker}, \citenamefont {Gu}, \citenamefont {Tsvelik},\ and\
  \citenamefont {Tranquada}}]{Li2007Two}%
  \BibitemOpen
  \bibfield  {author} {\bibinfo {author} {\bibfnamefont {Q.}~\bibnamefont
  {Li}}, \bibinfo {author} {\bibfnamefont {M.}~\bibnamefont {H\"ucker}},
  \bibinfo {author} {\bibfnamefont {G.~D.}\ \bibnamefont {Gu}}, \bibinfo
  {author} {\bibfnamefont {A.~M.}\ \bibnamefont {Tsvelik}},\ and\ \bibinfo
  {author} {\bibfnamefont {J.~M.}\ \bibnamefont {Tranquada}},\ }\bibfield
  {title} {\bibinfo {title} {Two-dimensional superconducting fluctuations in
  stripe-ordered
  {$\mathrm{{L}a}_{1.875}\mathrm{{B}a}_{0.125}\mathrm{{C}u{O}}_4$}},\ }\href
  {https://doi.org/10.1103/PhysRevLett.99.067001} {\bibfield  {journal}
  {\bibinfo  {journal} {Phys. Rev. Lett.}\ }\textbf {\bibinfo {volume} {99}},\
  \bibinfo {pages} {067001} (\bibinfo {year} {2007})}\BibitemShut {NoStop}%
\bibitem [{\citenamefont {Berg}\ \emph
  {et~al.}(2009{\natexlab{a}})\citenamefont {Berg}, \citenamefont {Fradkin},
  \citenamefont {Kivelson},\ and\ \citenamefont {Tranquada}}]{Berg2009Striped}%
  \BibitemOpen
  \bibfield  {author} {\bibinfo {author} {\bibfnamefont {E.}~\bibnamefont
  {Berg}}, \bibinfo {author} {\bibfnamefont {E.}~\bibnamefont {Fradkin}},
  \bibinfo {author} {\bibfnamefont {S.~A.}\ \bibnamefont {Kivelson}},\ and\
  \bibinfo {author} {\bibfnamefont {J.~M.}\ \bibnamefont {Tranquada}},\
  }\bibfield  {title} {\bibinfo {title} {Striped superconductors: how spin,
  charge and superconducting orders intertwine in the cuprates},\ }\href
  {https://doi.org/10.1088/1367-2630/11/11/115004} {\bibfield  {journal}
  {\bibinfo  {journal} {New Journal of Physics}\ }\textbf {\bibinfo {volume}
  {11}},\ \bibinfo {pages} {115004} (\bibinfo {year}
  {2009}{\natexlab{a}})}\BibitemShut {NoStop}%
\bibitem [{\citenamefont {Berg}\ \emph
  {et~al.}(2009{\natexlab{b}})\citenamefont {Berg}, \citenamefont {Fradkin},\
  and\ \citenamefont {Kivelson}}]{Berg2009Theory}%
  \BibitemOpen
  \bibfield  {author} {\bibinfo {author} {\bibfnamefont {E.}~\bibnamefont
  {Berg}}, \bibinfo {author} {\bibfnamefont {E.}~\bibnamefont {Fradkin}},\ and\
  \bibinfo {author} {\bibfnamefont {S.~A.}\ \bibnamefont {Kivelson}},\
  }\bibfield  {title} {\bibinfo {title} {Theory of the striped
  superconductor},\ }\href {https://doi.org/10.1103/PhysRevB.79.064515}
  {\bibfield  {journal} {\bibinfo  {journal} {Phys. Rev. B}\ }\textbf {\bibinfo
  {volume} {79}},\ \bibinfo {pages} {064515} (\bibinfo {year}
  {2009}{\natexlab{b}})}\BibitemShut {NoStop}%
\bibitem [{\citenamefont {Berg}\ \emph {et~al.}(2010)\citenamefont {Berg},
  \citenamefont {Fradkin},\ and\ \citenamefont {Kivelson}}]{Berg2010Pair}%
  \BibitemOpen
  \bibfield  {author} {\bibinfo {author} {\bibfnamefont {E.}~\bibnamefont
  {Berg}}, \bibinfo {author} {\bibfnamefont {E.}~\bibnamefont {Fradkin}},\ and\
  \bibinfo {author} {\bibfnamefont {S.~A.}\ \bibnamefont {Kivelson}},\
  }\bibfield  {title} {\bibinfo {title} {Pair-density-wave correlations in the
  kondo-heisenberg model},\ }\href
  {https://doi.org/10.1103/PhysRevLett.105.146403} {\bibfield  {journal}
  {\bibinfo  {journal} {Phys. Rev. Lett.}\ }\textbf {\bibinfo {volume} {105}},\
  \bibinfo {pages} {146403} (\bibinfo {year} {2010})}\BibitemShut {NoStop}%
\bibitem [{\citenamefont {Jaefari}\ and\ \citenamefont
  {Fradkin}(2012)}]{Jaefari2012Pair}%
  \BibitemOpen
  \bibfield  {author} {\bibinfo {author} {\bibfnamefont {A.}~\bibnamefont
  {Jaefari}}\ and\ \bibinfo {author} {\bibfnamefont {E.}~\bibnamefont
  {Fradkin}},\ }\bibfield  {title} {\bibinfo {title} {Pair-density-wave
  superconducting order in two-leg ladders},\ }\href
  {https://doi.org/10.1103/PhysRevB.85.035104} {\bibfield  {journal} {\bibinfo
  {journal} {Phys. Rev. B}\ }\textbf {\bibinfo {volume} {85}},\ \bibinfo
  {pages} {035104} (\bibinfo {year} {2012})}\BibitemShut {NoStop}%
\bibitem [{\citenamefont {Soto-Garrido}\ and\ \citenamefont
  {Fradkin}(2014)}]{Soto2014Pair}%
  \BibitemOpen
  \bibfield  {author} {\bibinfo {author} {\bibfnamefont {R.}~\bibnamefont
  {Soto-Garrido}}\ and\ \bibinfo {author} {\bibfnamefont {E.}~\bibnamefont
  {Fradkin}},\ }\bibfield  {title} {\bibinfo {title} {Pair-density-wave
  superconducting states and electronic liquid-crystal phases},\ }\href
  {https://doi.org/10.1103/PhysRevB.89.165126} {\bibfield  {journal} {\bibinfo
  {journal} {Phys. Rev. B}\ }\textbf {\bibinfo {volume} {89}},\ \bibinfo
  {pages} {165126} (\bibinfo {year} {2014})}\BibitemShut {NoStop}%
\bibitem [{\citenamefont {Lee}(2014)}]{Lee2014Amperean}%
  \BibitemOpen
  \bibfield  {author} {\bibinfo {author} {\bibfnamefont {P.~A.}\ \bibnamefont
  {Lee}},\ }\bibfield  {title} {\bibinfo {title} {Amperean pairing and the
  pseudogap phase of cuprate superconductors},\ }\href
  {https://doi.org/10.1103/PhysRevX.4.031017} {\bibfield  {journal} {\bibinfo
  {journal} {Phys. Rev. X}\ }\textbf {\bibinfo {volume} {4}},\ \bibinfo {pages}
  {031017} (\bibinfo {year} {2014})}\BibitemShut {NoStop}%
\bibitem [{\citenamefont {Fradkin}\ \emph {et~al.}(2015)\citenamefont
  {Fradkin}, \citenamefont {Kivelson},\ and\ \citenamefont
  {Tranquada}}]{Fradkin2015Colloquium}%
  \BibitemOpen
  \bibfield  {author} {\bibinfo {author} {\bibfnamefont {E.}~\bibnamefont
  {Fradkin}}, \bibinfo {author} {\bibfnamefont {S.~A.}\ \bibnamefont
  {Kivelson}},\ and\ \bibinfo {author} {\bibfnamefont {J.~M.}\ \bibnamefont
  {Tranquada}},\ }\bibfield  {title} {\bibinfo {title} {Colloquium: Theory of
  intertwined orders in high temperature superconductors},\ }\href
  {https://doi.org/10.1103/RevModPhys.87.457} {\bibfield  {journal} {\bibinfo
  {journal} {Rev. Mod. Phys.}\ }\textbf {\bibinfo {volume} {87}},\ \bibinfo
  {pages} {457} (\bibinfo {year} {2015})}\BibitemShut {NoStop}%
\bibitem [{\citenamefont {Hamidian}\ \emph {et~al.}(2016)\citenamefont
  {Hamidian}, \citenamefont {Edkins}, \citenamefont {Joo}, \citenamefont
  {Kostin}, \citenamefont {Eisaki}, \citenamefont {Uchida}, \citenamefont
  {Lawler}, \citenamefont {Kim}, \citenamefont {Mackenzie}, \citenamefont
  {Fujita} \emph {et~al.}}]{hamidian2016detection}%
  \BibitemOpen
  \bibfield  {author} {\bibinfo {author} {\bibfnamefont {M.}~\bibnamefont
  {Hamidian}}, \bibinfo {author} {\bibfnamefont {S.~D.}\ \bibnamefont
  {Edkins}}, \bibinfo {author} {\bibfnamefont {S.~H.}\ \bibnamefont {Joo}},
  \bibinfo {author} {\bibfnamefont {A.}~\bibnamefont {Kostin}}, \bibinfo
  {author} {\bibfnamefont {H.}~\bibnamefont {Eisaki}}, \bibinfo {author}
  {\bibfnamefont {S.}~\bibnamefont {Uchida}}, \bibinfo {author} {\bibfnamefont
  {M.}~\bibnamefont {Lawler}}, \bibinfo {author} {\bibfnamefont {E.-A.}\
  \bibnamefont {Kim}}, \bibinfo {author} {\bibfnamefont {A.~P.}\ \bibnamefont
  {Mackenzie}}, \bibinfo {author} {\bibfnamefont {K.}~\bibnamefont {Fujita}},
  \emph {et~al.},\ }\bibfield  {title} {\bibinfo {title} {Detection of a
  cooper-pair density wave in
  $\mathrm{Bi}_2\mathrm{Sr}_2\mathrm{CaCu}_2\mathrm{O}_{8+x}$},\ }\href
  {https://www.nature.com/articles/nature17411} {\bibfield  {journal} {\bibinfo
   {journal} {Nature}\ }\textbf {\bibinfo {volume} {532}},\ \bibinfo {pages}
  {343} (\bibinfo {year} {2016})}\BibitemShut {NoStop}%
\bibitem [{\citenamefont {Ruan}\ \emph {et~al.}(2018)\citenamefont {Ruan},
  \citenamefont {Li}, \citenamefont {Hu}, \citenamefont {Hao}, \citenamefont
  {Li}, \citenamefont {Cai}, \citenamefont {Zhou}, \citenamefont {Lee},\ and\
  \citenamefont {Wang}}]{ruan2018visualization}%
  \BibitemOpen
  \bibfield  {author} {\bibinfo {author} {\bibfnamefont {W.}~\bibnamefont
  {Ruan}}, \bibinfo {author} {\bibfnamefont {X.}~\bibnamefont {Li}}, \bibinfo
  {author} {\bibfnamefont {C.}~\bibnamefont {Hu}}, \bibinfo {author}
  {\bibfnamefont {Z.}~\bibnamefont {Hao}}, \bibinfo {author} {\bibfnamefont
  {H.}~\bibnamefont {Li}}, \bibinfo {author} {\bibfnamefont {P.}~\bibnamefont
  {Cai}}, \bibinfo {author} {\bibfnamefont {X.}~\bibnamefont {Zhou}}, \bibinfo
  {author} {\bibfnamefont {D.-H.}\ \bibnamefont {Lee}},\ and\ \bibinfo {author}
  {\bibfnamefont {Y.}~\bibnamefont {Wang}},\ }\bibfield  {title} {\bibinfo
  {title} {Visualization of the periodic modulation of cooper pairing in a
  cuprate superconductor},\ }\href
  {https://www.nature.com/articles/s41567-018-0276-8} {\bibfield  {journal}
  {\bibinfo  {journal} {Nature Physics}\ }\textbf {\bibinfo {volume} {14}},\
  \bibinfo {pages} {1178} (\bibinfo {year} {2018})}\BibitemShut {NoStop}%
\bibitem [{\citenamefont {Edkins}\ \emph {et~al.}(2019)\citenamefont {Edkins},
  \citenamefont {Kostin}, \citenamefont {Fujita}, \citenamefont {Mackenzie},
  \citenamefont {Eisaki}, \citenamefont {Uchida}, \citenamefont {Sachdev},
  \citenamefont {Lawler}, \citenamefont {Kim}, \citenamefont {Davis},\ and\
  \citenamefont {Hamidian}}]{Edkins2019Magnetic}%
  \BibitemOpen
  \bibfield  {author} {\bibinfo {author} {\bibfnamefont {S.~D.}\ \bibnamefont
  {Edkins}}, \bibinfo {author} {\bibfnamefont {A.}~\bibnamefont {Kostin}},
  \bibinfo {author} {\bibfnamefont {K.}~\bibnamefont {Fujita}}, \bibinfo
  {author} {\bibfnamefont {A.~P.}\ \bibnamefont {Mackenzie}}, \bibinfo {author}
  {\bibfnamefont {H.}~\bibnamefont {Eisaki}}, \bibinfo {author} {\bibfnamefont
  {S.}~\bibnamefont {Uchida}}, \bibinfo {author} {\bibfnamefont
  {S.}~\bibnamefont {Sachdev}}, \bibinfo {author} {\bibfnamefont {M.~J.}\
  \bibnamefont {Lawler}}, \bibinfo {author} {\bibfnamefont {E.-A.}\
  \bibnamefont {Kim}}, \bibinfo {author} {\bibfnamefont {J.~C.~S.}\
  \bibnamefont {Davis}},\ and\ \bibinfo {author} {\bibfnamefont {M.~H.}\
  \bibnamefont {Hamidian}},\ }\bibfield  {title} {\bibinfo {title} {Magnetic
  field–induced pair density wave state in the cuprate vortex halo},\ }\href
  {https://doi.org/10.1126/science.aat1773} {\bibfield  {journal} {\bibinfo
  {journal} {Science}\ }\textbf {\bibinfo {volume} {364}},\ \bibinfo {pages}
  {976} (\bibinfo {year} {2019})}\BibitemShut {NoStop}%
\bibitem [{\citenamefont {Agterberg}\ \emph {et~al.}(2020)\citenamefont
  {Agterberg}, \citenamefont {Davis}, \citenamefont {Edkins}, \citenamefont
  {Fradkin}, \citenamefont {Van~Harlingen}, \citenamefont {Kivelson},
  \citenamefont {Lee}, \citenamefont {Radzihovsky}, \citenamefont {Tranquada},\
  and\ \citenamefont {Wang}}]{Agterberg2020Physics}%
  \BibitemOpen
  \bibfield  {author} {\bibinfo {author} {\bibfnamefont {D.~F.}\ \bibnamefont
  {Agterberg}}, \bibinfo {author} {\bibfnamefont {J.~S.}\ \bibnamefont
  {Davis}}, \bibinfo {author} {\bibfnamefont {S.~D.}\ \bibnamefont {Edkins}},
  \bibinfo {author} {\bibfnamefont {E.}~\bibnamefont {Fradkin}}, \bibinfo
  {author} {\bibfnamefont {D.~J.}\ \bibnamefont {Van~Harlingen}}, \bibinfo
  {author} {\bibfnamefont {S.~A.}\ \bibnamefont {Kivelson}}, \bibinfo {author}
  {\bibfnamefont {P.~A.}\ \bibnamefont {Lee}}, \bibinfo {author} {\bibfnamefont
  {L.}~\bibnamefont {Radzihovsky}}, \bibinfo {author} {\bibfnamefont {J.~M.}\
  \bibnamefont {Tranquada}},\ and\ \bibinfo {author} {\bibfnamefont
  {Y.}~\bibnamefont {Wang}},\ }\bibfield  {title} {\bibinfo {title} {The
  physics of pair-density waves: Cuprate superconductors and beyond},\ }\href
  {https://doi.org/https://doi.org/10.1146/annurev-conmatphys-031119-050711}
  {\bibfield  {journal} {\bibinfo  {journal} {Annual Review of Condensed Matter
  Physics}\ }\textbf {\bibinfo {volume} {11}},\ \bibinfo {pages} {231}
  (\bibinfo {year} {2020})}\BibitemShut {NoStop}%
\bibitem [{\citenamefont {Han}\ \emph {et~al.}(2020)\citenamefont {Han},
  \citenamefont {Kivelson},\ and\ \citenamefont {Yao}}]{Han2020Strong}%
  \BibitemOpen
  \bibfield  {author} {\bibinfo {author} {\bibfnamefont {Z.}~\bibnamefont
  {Han}}, \bibinfo {author} {\bibfnamefont {S.~A.}\ \bibnamefont {Kivelson}},\
  and\ \bibinfo {author} {\bibfnamefont {H.}~\bibnamefont {Yao}},\ }\bibfield
  {title} {\bibinfo {title} {Strong coupling limit of the holstein-hubbard
  model},\ }\href {https://doi.org/10.1103/PhysRevLett.125.167001} {\bibfield
  {journal} {\bibinfo  {journal} {Phys. Rev. Lett.}\ }\textbf {\bibinfo
  {volume} {125}},\ \bibinfo {pages} {167001} (\bibinfo {year}
  {2020})}\BibitemShut {NoStop}%
\bibitem [{\citenamefont {Li}\ \emph {et~al.}(2021)\citenamefont {Li},
  \citenamefont {Zou}, \citenamefont {Ding}, \citenamefont {Yan}, \citenamefont
  {Ye}, \citenamefont {Li}, \citenamefont {Hao}, \citenamefont {Zhao},
  \citenamefont {Zhou},\ and\ \citenamefont {Wang}}]{Li2021Evolution}%
  \BibitemOpen
  \bibfield  {author} {\bibinfo {author} {\bibfnamefont {X.}~\bibnamefont
  {Li}}, \bibinfo {author} {\bibfnamefont {C.}~\bibnamefont {Zou}}, \bibinfo
  {author} {\bibfnamefont {Y.}~\bibnamefont {Ding}}, \bibinfo {author}
  {\bibfnamefont {H.}~\bibnamefont {Yan}}, \bibinfo {author} {\bibfnamefont
  {S.}~\bibnamefont {Ye}}, \bibinfo {author} {\bibfnamefont {H.}~\bibnamefont
  {Li}}, \bibinfo {author} {\bibfnamefont {Z.}~\bibnamefont {Hao}}, \bibinfo
  {author} {\bibfnamefont {L.}~\bibnamefont {Zhao}}, \bibinfo {author}
  {\bibfnamefont {X.}~\bibnamefont {Zhou}},\ and\ \bibinfo {author}
  {\bibfnamefont {Y.}~\bibnamefont {Wang}},\ }\bibfield  {title} {\bibinfo
  {title} {Evolution of charge and pair density modulations in overdoped
  {$\mathrm{Bi}_2\mathrm{Sr}_2\mathrm{CuO}_{6+\delta}$}},\ }\href
  {https://doi.org/10.1103/PhysRevX.11.011007} {\bibfield  {journal} {\bibinfo
  {journal} {Phys. Rev. X}\ }\textbf {\bibinfo {volume} {11}},\ \bibinfo
  {pages} {011007} (\bibinfo {year} {2021})}\BibitemShut {NoStop}%
\bibitem [{\citenamefont {Chen}\ \emph {et~al.}(2021)\citenamefont {Chen},
  \citenamefont {Yang}, \citenamefont {Hu}, \citenamefont {Zhao}, \citenamefont
  {Yuan}, \citenamefont {Xing}, \citenamefont {Qian}, \citenamefont {Huang},
  \citenamefont {Li}, \citenamefont {Ye} \emph {et~al.}}]{chen2021roton}%
  \BibitemOpen
  \bibfield  {author} {\bibinfo {author} {\bibfnamefont {H.}~\bibnamefont
  {Chen}}, \bibinfo {author} {\bibfnamefont {H.}~\bibnamefont {Yang}}, \bibinfo
  {author} {\bibfnamefont {B.}~\bibnamefont {Hu}}, \bibinfo {author}
  {\bibfnamefont {Z.}~\bibnamefont {Zhao}}, \bibinfo {author} {\bibfnamefont
  {J.}~\bibnamefont {Yuan}}, \bibinfo {author} {\bibfnamefont {Y.}~\bibnamefont
  {Xing}}, \bibinfo {author} {\bibfnamefont {G.}~\bibnamefont {Qian}}, \bibinfo
  {author} {\bibfnamefont {Z.}~\bibnamefont {Huang}}, \bibinfo {author}
  {\bibfnamefont {G.}~\bibnamefont {Li}}, \bibinfo {author} {\bibfnamefont
  {Y.}~\bibnamefont {Ye}}, \emph {et~al.},\ }\bibfield  {title} {\bibinfo
  {title} {Roton pair density wave in a strong-coupling kagome
  superconductor},\ }\href {https://www.nature.com/articles/s41586-021-03983-5}
  {\bibfield  {journal} {\bibinfo  {journal} {Nature}\ }\textbf {\bibinfo
  {volume} {599}},\ \bibinfo {pages} {222} (\bibinfo {year}
  {2021})}\BibitemShut {NoStop}%
\bibitem [{\citenamefont {Liu}\ \emph {et~al.}(2021)\citenamefont {Liu},
  \citenamefont {Chong}, \citenamefont {Sharma},\ and\ \citenamefont
  {Davis}}]{Liu2021Discovery}%
  \BibitemOpen
  \bibfield  {author} {\bibinfo {author} {\bibfnamefont {X.}~\bibnamefont
  {Liu}}, \bibinfo {author} {\bibfnamefont {Y.~X.}\ \bibnamefont {Chong}},
  \bibinfo {author} {\bibfnamefont {R.}~\bibnamefont {Sharma}},\ and\ \bibinfo
  {author} {\bibfnamefont {J.~C.~S.}\ \bibnamefont {Davis}},\ }\bibfield
  {title} {\bibinfo {title} {Discovery of a cooper-pair density wave state in a
  transition-metal dichalcogenide},\ }\href
  {https://doi.org/10.1126/science.abd4607} {\bibfield  {journal} {\bibinfo
  {journal} {Science}\ }\textbf {\bibinfo {volume} {372}},\ \bibinfo {pages}
  {1447} (\bibinfo {year} {2021})}\BibitemShut {NoStop}%
\bibitem [{\citenamefont {Huang}\ \emph {et~al.}(2022)\citenamefont {Huang},
  \citenamefont {Han}, \citenamefont {Kivelson},\ and\ \citenamefont
  {Yao}}]{huang2022pair}%
  \BibitemOpen
  \bibfield  {author} {\bibinfo {author} {\bibfnamefont {K.~S.}\ \bibnamefont
  {Huang}}, \bibinfo {author} {\bibfnamefont {Z.}~\bibnamefont {Han}}, \bibinfo
  {author} {\bibfnamefont {S.~A.}\ \bibnamefont {Kivelson}},\ and\ \bibinfo
  {author} {\bibfnamefont {H.}~\bibnamefont {Yao}},\ }\bibfield  {title}
  {\bibinfo {title} {Pair-density-wave in the strong coupling limit of the
  holstein-hubbard model},\ }\href
  {https://www.nature.com/articles/s41535-022-00426-w} {\bibfield  {journal}
  {\bibinfo  {journal} {npj Quantum Materials}\ }\textbf {\bibinfo {volume}
  {7}},\ \bibinfo {pages} {17} (\bibinfo {year} {2022})}\BibitemShut {NoStop}%
\bibitem [{\citenamefont {Han}\ and\ \citenamefont
  {Kivelson}(2022)}]{Han2022Pair}%
  \BibitemOpen
  \bibfield  {author} {\bibinfo {author} {\bibfnamefont {Z.}~\bibnamefont
  {Han}}\ and\ \bibinfo {author} {\bibfnamefont {S.~A.}\ \bibnamefont
  {Kivelson}},\ }\bibfield  {title} {\bibinfo {title} {Pair density wave and
  reentrant superconducting tendencies originating from valley polarization},\
  }\href {https://doi.org/10.1103/PhysRevB.105.L100509} {\bibfield  {journal}
  {\bibinfo  {journal} {Phys. Rev. B}\ }\textbf {\bibinfo {volume} {105}},\
  \bibinfo {pages} {L100509} (\bibinfo {year} {2022})}\BibitemShut {NoStop}%
\bibitem [{\citenamefont {Zhang}\ and\ \citenamefont
  {Vishwanath}(2022)}]{Zhang2022Pair}%
  \BibitemOpen
  \bibfield  {author} {\bibinfo {author} {\bibfnamefont {Y.-H.}\ \bibnamefont
  {Zhang}}\ and\ \bibinfo {author} {\bibfnamefont {A.}~\bibnamefont
  {Vishwanath}},\ }\bibfield  {title} {\bibinfo {title} {Pair-density-wave
  superconductor from doping haldane chain and rung-singlet ladder},\ }\href
  {https://doi.org/10.1103/PhysRevB.106.045103} {\bibfield  {journal} {\bibinfo
   {journal} {Phys. Rev. B}\ }\textbf {\bibinfo {volume} {106}},\ \bibinfo
  {pages} {045103} (\bibinfo {year} {2022})}\BibitemShut {NoStop}%
\bibitem [{\citenamefont {Liu}\ \emph {et~al.}(2023)\citenamefont {Liu},
  \citenamefont {Wei}, \citenamefont {He}, \citenamefont {Zhang}, \citenamefont
  {Wang},\ and\ \citenamefont {Wang}}]{liu2023pair}%
  \BibitemOpen
  \bibfield  {author} {\bibinfo {author} {\bibfnamefont {Y.}~\bibnamefont
  {Liu}}, \bibinfo {author} {\bibfnamefont {T.}~\bibnamefont {Wei}}, \bibinfo
  {author} {\bibfnamefont {G.}~\bibnamefont {He}}, \bibinfo {author}
  {\bibfnamefont {Y.}~\bibnamefont {Zhang}}, \bibinfo {author} {\bibfnamefont
  {Z.}~\bibnamefont {Wang}},\ and\ \bibinfo {author} {\bibfnamefont
  {J.}~\bibnamefont {Wang}},\ }\bibfield  {title} {\bibinfo {title} {Pair
  density wave state in a monolayer high-{$T_c$} iron-based superconductor},\
  }\href {https://doi.org/10.1038/s41586-023-06072-x} {\bibfield  {journal}
  {\bibinfo  {journal} {Nature}\ }\textbf {\bibinfo {volume} {618}},\ \bibinfo
  {pages} {934} (\bibinfo {year} {2023})}\BibitemShut {NoStop}%
\bibitem [{\citenamefont {Zhao}\ \emph {et~al.}(2023)\citenamefont {Zhao},
  \citenamefont {Blackwell}, \citenamefont {Thinel}, \citenamefont {Handa},
  \citenamefont {Ishida}, \citenamefont {Zhu}, \citenamefont {Iyo},
  \citenamefont {Eisaki}, \citenamefont {Pasupathy},\ and\ \citenamefont
  {Fujita}}]{zhao2023smectic}%
  \BibitemOpen
  \bibfield  {author} {\bibinfo {author} {\bibfnamefont {H.}~\bibnamefont
  {Zhao}}, \bibinfo {author} {\bibfnamefont {R.}~\bibnamefont {Blackwell}},
  \bibinfo {author} {\bibfnamefont {M.}~\bibnamefont {Thinel}}, \bibinfo
  {author} {\bibfnamefont {T.}~\bibnamefont {Handa}}, \bibinfo {author}
  {\bibfnamefont {S.}~\bibnamefont {Ishida}}, \bibinfo {author} {\bibfnamefont
  {X.}~\bibnamefont {Zhu}}, \bibinfo {author} {\bibfnamefont {A.}~\bibnamefont
  {Iyo}}, \bibinfo {author} {\bibfnamefont {H.}~\bibnamefont {Eisaki}},
  \bibinfo {author} {\bibfnamefont {A.~N.}\ \bibnamefont {Pasupathy}},\ and\
  \bibinfo {author} {\bibfnamefont {K.}~\bibnamefont {Fujita}},\ }\bibfield
  {title} {\bibinfo {title} {Smectic pair-density-wave order in
  {$\mathrm{EuRbFe}_4\mathrm{As}_4$}},\ }\href
  {https://doi.org/10.1038/s41586-023-06103-7} {\bibfield  {journal} {\bibinfo
  {journal} {Nature}\ }\textbf {\bibinfo {volume} {618}},\ \bibinfo {pages}
  {940} (\bibinfo {year} {2023})}\BibitemShut {NoStop}%
\bibitem [{\citenamefont {Wu}\ \emph {et~al.}(2023{\natexlab{a}})\citenamefont
  {Wu}, \citenamefont {Wu},\ and\ \citenamefont {Yao}}]{Wu2023Pair}%
  \BibitemOpen
  \bibfield  {author} {\bibinfo {author} {\bibfnamefont {Y.-M.}\ \bibnamefont
  {Wu}}, \bibinfo {author} {\bibfnamefont {Z.}~\bibnamefont {Wu}},\ and\
  \bibinfo {author} {\bibfnamefont {H.}~\bibnamefont {Yao}},\ }\bibfield
  {title} {\bibinfo {title} {Pair-density-wave and chiral superconductivity in
  twisted bilayer transition metal dichalcogenides},\ }\href
  {https://doi.org/10.1103/PhysRevLett.130.126001} {\bibfield  {journal}
  {\bibinfo  {journal} {Phys. Rev. Lett.}\ }\textbf {\bibinfo {volume} {130}},\
  \bibinfo {pages} {126001} (\bibinfo {year} {2023}{\natexlab{a}})}\BibitemShut
  {NoStop}%
\bibitem [{\citenamefont {Wu}\ \emph {et~al.}(2023{\natexlab{b}})\citenamefont
  {Wu}, \citenamefont {Nosov}, \citenamefont {Patel},\ and\ \citenamefont
  {Raghu}}]{Wuym2023Pair}%
  \BibitemOpen
  \bibfield  {author} {\bibinfo {author} {\bibfnamefont {Y.-M.}\ \bibnamefont
  {Wu}}, \bibinfo {author} {\bibfnamefont {P.~A.}\ \bibnamefont {Nosov}},
  \bibinfo {author} {\bibfnamefont {A.~A.}\ \bibnamefont {Patel}},\ and\
  \bibinfo {author} {\bibfnamefont {S.}~\bibnamefont {Raghu}},\ }\bibfield
  {title} {\bibinfo {title} {Pair density wave order from electron repulsion},\
  }\href {https://doi.org/10.1103/PhysRevLett.130.026001} {\bibfield  {journal}
  {\bibinfo  {journal} {Phys. Rev. Lett.}\ }\textbf {\bibinfo {volume} {130}},\
  \bibinfo {pages} {026001} (\bibinfo {year} {2023}{\natexlab{b}})}\BibitemShut
  {NoStop}%
\bibitem [{\citenamefont {Wang}\ \emph {et~al.}(2025)\citenamefont {Wang},
  \citenamefont {Sun}, \citenamefont {Wang}, \citenamefont {Han}, \citenamefont
  {Kivelson},\ and\ \citenamefont {Yao}}]{Wang2025Pair}%
  \BibitemOpen
  \bibfield  {author} {\bibinfo {author} {\bibfnamefont {J.}~\bibnamefont
  {Wang}}, \bibinfo {author} {\bibfnamefont {W.}~\bibnamefont {Sun}}, \bibinfo
  {author} {\bibfnamefont {H.-X.}\ \bibnamefont {Wang}}, \bibinfo {author}
  {\bibfnamefont {Z.}~\bibnamefont {Han}}, \bibinfo {author} {\bibfnamefont
  {S.~A.}\ \bibnamefont {Kivelson}},\ and\ \bibinfo {author} {\bibfnamefont
  {H.}~\bibnamefont {Yao}},\ }\bibfield  {title} {\bibinfo {title}
  {Pair-density-wave phase of strongly interacting electrons on the triangular
  lattice: A variational monte carlo study},\ }\href
  {https://doi.org/10.1103/gvw1-xk98} {\bibfield  {journal} {\bibinfo
  {journal} {Phys. Rev. B}\ }\textbf {\bibinfo {volume} {112}},\ \bibinfo
  {pages} {L140505} (\bibinfo {year} {2025})}\BibitemShut {NoStop}%
\bibitem [{\citenamefont {Wang}\ \emph {et~al.}(2015)\citenamefont {Wang},
  \citenamefont {Agterberg},\ and\ \citenamefont
  {Chubukov}}]{Wang2015Coexistence}%
  \BibitemOpen
  \bibfield  {author} {\bibinfo {author} {\bibfnamefont {Y.}~\bibnamefont
  {Wang}}, \bibinfo {author} {\bibfnamefont {D.~F.}\ \bibnamefont
  {Agterberg}},\ and\ \bibinfo {author} {\bibfnamefont {A.}~\bibnamefont
  {Chubukov}},\ }\bibfield  {title} {\bibinfo {title} {Coexistence of
  charge-density-wave and pair-density-wave orders in underdoped cuprates},\
  }\href {https://doi.org/10.1103/PhysRevLett.114.197001} {\bibfield  {journal}
  {\bibinfo  {journal} {Phys. Rev. Lett.}\ }\textbf {\bibinfo {volume} {114}},\
  \bibinfo {pages} {197001} (\bibinfo {year} {2015})}\BibitemShut {NoStop}%
\bibitem [{\citenamefont {Jian}\ \emph {et~al.}(2020)\citenamefont {Jian},
  \citenamefont {Scherer},\ and\ \citenamefont {Yao}}]{Jian2020Mass}%
  \BibitemOpen
  \bibfield  {author} {\bibinfo {author} {\bibfnamefont {S.-K.}\ \bibnamefont
  {Jian}}, \bibinfo {author} {\bibfnamefont {M.~M.}\ \bibnamefont {Scherer}},\
  and\ \bibinfo {author} {\bibfnamefont {H.}~\bibnamefont {Yao}},\ }\bibfield
  {title} {\bibinfo {title} {Mass hierarchy in collective modes of
  pair-density-wave superconductors},\ }\href
  {https://doi.org/10.1103/PhysRevResearch.2.013034} {\bibfield  {journal}
  {\bibinfo  {journal} {Phys. Rev. Res.}\ }\textbf {\bibinfo {volume} {2}},\
  \bibinfo {pages} {013034} (\bibinfo {year} {2020})}\BibitemShut {NoStop}%
\bibitem [{\citenamefont {Wu}\ \emph {et~al.}(2025)\citenamefont {Wu},
  \citenamefont {Chubukov}, \citenamefont {Wang},\ and\ \citenamefont
  {Kivelson}}]{wu2025time}%
  \BibitemOpen
  \bibfield  {author} {\bibinfo {author} {\bibfnamefont {Y.-M.}\ \bibnamefont
  {Wu}}, \bibinfo {author} {\bibfnamefont {A.~V.}\ \bibnamefont {Chubukov}},
  \bibinfo {author} {\bibfnamefont {Y.}~\bibnamefont {Wang}},\ and\ \bibinfo
  {author} {\bibfnamefont {S.~A.}\ \bibnamefont {Kivelson}},\ }\bibfield
  {title} {\bibinfo {title} {Time-reversal symmetry breaking, collective modes,
  and raman spectrum in pair-density-wave states},\ }\href
  {https://www.nature.com/articles/s41535-025-00808-w} {\bibfield  {journal}
  {\bibinfo  {journal} {npj Quantum Materials}\ }\textbf {\bibinfo {volume}
  {10}},\ \bibinfo {pages} {84} (\bibinfo {year} {2025})}\BibitemShut {NoStop}%
\bibitem [{\citenamefont {Nagashima}\ \emph {et~al.}(2025)\citenamefont
  {Nagashima}, \citenamefont {Mouilleron},\ and\ \citenamefont
  {Tsuji}}]{Nagashima2025Optically}%
  \BibitemOpen
  \bibfield  {author} {\bibinfo {author} {\bibfnamefont {R.}~\bibnamefont
  {Nagashima}}, \bibinfo {author} {\bibfnamefont {T.}~\bibnamefont
  {Mouilleron}},\ and\ \bibinfo {author} {\bibfnamefont {N.}~\bibnamefont
  {Tsuji}},\ }\bibfield  {title} {\bibinfo {title} {Optically active higgs and
  leggett modes in multiband pair-density-wave superconductors with lifshitz
  invariant},\ }\href {https://doi.org/10.1103/cgyd-2g11} {\bibfield  {journal}
  {\bibinfo  {journal} {Phys. Rev. B}\ }\textbf {\bibinfo {volume} {112}},\
  \bibinfo {pages} {024503} (\bibinfo {year} {2025})}\BibitemShut {NoStop}%
\bibitem [{\citenamefont {Wang}\ \emph {et~al.}(2026)\citenamefont {Wang},
  \citenamefont {Chen}, \citenamefont {Boyack},\ and\ \citenamefont
  {Levin}}]{wang2026anomalous}%
  \BibitemOpen
  \bibfield  {author} {\bibinfo {author} {\bibfnamefont {K.}~\bibnamefont
  {Wang}}, \bibinfo {author} {\bibfnamefont {Q.}~\bibnamefont {Chen}}, \bibinfo
  {author} {\bibfnamefont {R.}~\bibnamefont {Boyack}},\ and\ \bibinfo {author}
  {\bibfnamefont {K.}~\bibnamefont {Levin}},\ }\bibfield  {title} {\bibinfo
  {title} {Anomalous superfluid density in pair-density-wave superconductors},\
  }\href {https://www.nature.com/articles/s41535-025-00843-7} {\bibfield
  {journal} {\bibinfo  {journal} {npj Quantum Materials}\ }\textbf {\bibinfo
  {volume} {11}},\ \bibinfo {pages} {13} (\bibinfo {year} {2026})}\BibitemShut
  {NoStop}%
\bibitem [{\citenamefont {Agterberg}\ \emph {et~al.}(2011)\citenamefont
  {Agterberg}, \citenamefont {Geracie},\ and\ \citenamefont
  {Tsunetsugu}}]{Agterberg2011Conventional}%
  \BibitemOpen
  \bibfield  {author} {\bibinfo {author} {\bibfnamefont {D.~F.}\ \bibnamefont
  {Agterberg}}, \bibinfo {author} {\bibfnamefont {M.}~\bibnamefont {Geracie}},\
  and\ \bibinfo {author} {\bibfnamefont {H.}~\bibnamefont {Tsunetsugu}},\
  }\bibfield  {title} {\bibinfo {title} {Conventional and charge-six
  superfluids from melting hexagonal fulde-ferrell-larkin-ovchinnikov phases in
  two dimensions},\ }\href {https://doi.org/10.1103/PhysRevB.84.014513}
  {\bibfield  {journal} {\bibinfo  {journal} {Phys. Rev. B}\ }\textbf {\bibinfo
  {volume} {84}},\ \bibinfo {pages} {014513} (\bibinfo {year}
  {2011})}\BibitemShut {NoStop}%
\bibitem [{\citenamefont {Lesser}\ \emph {et~al.}(2026)\citenamefont {Lesser},
  \citenamefont {Huang}, \citenamefont {Sethna},\ and\ \citenamefont
  {Kim}}]{Lesser2026Emblems}%
  \BibitemOpen
  \bibfield  {author} {\bibinfo {author} {\bibfnamefont {O.}~\bibnamefont
  {Lesser}}, \bibinfo {author} {\bibfnamefont {C.}~\bibnamefont {Huang}},
  \bibinfo {author} {\bibfnamefont {J.~P.}\ \bibnamefont {Sethna}},\ and\
  \bibinfo {author} {\bibfnamefont {E.-A.}\ \bibnamefont {Kim}},\ }\bibfield
  {title} {\bibinfo {title} {Emblems of pair density waves: Dual identity of
  topological defects and their transport signatures},\ }\href
  {https://doi.org/10.1103/1gyj-r5fr} {\bibfield  {journal} {\bibinfo
  {journal} {Phys. Rev. B}\ }\textbf {\bibinfo {volume} {113}},\ \bibinfo
  {pages} {214505} (\bibinfo {year} {2026})}\BibitemShut {NoStop}%
\bibitem [{\citenamefont {Agterberg}\ \emph {et~al.}(2015)\citenamefont
  {Agterberg}, \citenamefont {Melchert},\ and\ \citenamefont
  {Kashyap}}]{Agterberg2015Emergent}%
  \BibitemOpen
  \bibfield  {author} {\bibinfo {author} {\bibfnamefont {D.~F.}\ \bibnamefont
  {Agterberg}}, \bibinfo {author} {\bibfnamefont {D.~S.}\ \bibnamefont
  {Melchert}},\ and\ \bibinfo {author} {\bibfnamefont {M.~K.}\ \bibnamefont
  {Kashyap}},\ }\bibfield  {title} {\bibinfo {title} {Emergent loop current
  order from pair density wave superconductivity},\ }\href
  {https://doi.org/10.1103/PhysRevB.91.054502} {\bibfield  {journal} {\bibinfo
  {journal} {Phys. Rev. B}\ }\textbf {\bibinfo {volume} {91}},\ \bibinfo
  {pages} {054502} (\bibinfo {year} {2015})}\BibitemShut {NoStop}%
\bibitem [{\citenamefont {Zhou}\ and\ \citenamefont
  {Wang}(2022)}]{zhou2022chern}%
  \BibitemOpen
  \bibfield  {author} {\bibinfo {author} {\bibfnamefont {S.}~\bibnamefont
  {Zhou}}\ and\ \bibinfo {author} {\bibfnamefont {Z.}~\bibnamefont {Wang}},\
  }\bibfield  {title} {\bibinfo {title} {Chern fermi pocket, topological pair
  density wave, and charge-4e and charge-6e superconductivity in kagom{\'e}
  superconductors},\ }\href {https://doi.org/10.1038/s41467-022-34832-2}
  {\bibfield  {journal} {\bibinfo  {journal} {Nature Communications}\ }\textbf
  {\bibinfo {volume} {13}},\ \bibinfo {pages} {7288} (\bibinfo {year}
  {2022})}\BibitemShut {NoStop}%
\bibitem [{\citenamefont {W\aa{}rdh}\ and\ \citenamefont
  {Granath}(2023)}]{Jonatan2023Nematic}%
  \BibitemOpen
  \bibfield  {author} {\bibinfo {author} {\bibfnamefont {J.}~\bibnamefont
  {W\aa{}rdh}}\ and\ \bibinfo {author} {\bibfnamefont {M.}~\bibnamefont
  {Granath}},\ }\bibfield  {title} {\bibinfo {title} {Nematic single-component
  superconductivity and loop-current order from pair-density wave
  instability},\ }\href {https://doi.org/10.1103/PhysRevB.107.134504}
  {\bibfield  {journal} {\bibinfo  {journal} {Phys. Rev. B}\ }\textbf {\bibinfo
  {volume} {107}},\ \bibinfo {pages} {134504} (\bibinfo {year}
  {2023})}\BibitemShut {NoStop}%
\bibitem [{\citenamefont {Wu}\ and\ \citenamefont {Wang}(2024)}]{wu2024d}%
  \BibitemOpen
  \bibfield  {author} {\bibinfo {author} {\bibfnamefont {Y.-M.}\ \bibnamefont
  {Wu}}\ and\ \bibinfo {author} {\bibfnamefont {Y.}~\bibnamefont {Wang}},\
  }\bibfield  {title} {\bibinfo {title} {d-wave charge-4e superconductivity
  from fluctuating pair density waves},\ }\href
  {https://www.nature.com/articles/s41535-024-00674-y} {\bibfield  {journal}
  {\bibinfo  {journal} {npj Quantum Materials}\ }\textbf {\bibinfo {volume}
  {9}},\ \bibinfo {pages} {66} (\bibinfo {year} {2024})}\BibitemShut {NoStop}%
\bibitem [{\citenamefont {Huecker}\ and\ \citenamefont
  {Wang}(2026)}]{Huecker2026Vestigial}%
  \BibitemOpen
  \bibfield  {author} {\bibinfo {author} {\bibfnamefont {E.}~\bibnamefont
  {Huecker}}\ and\ \bibinfo {author} {\bibfnamefont {Y.}~\bibnamefont {Wang}},\
  }\bibfield  {title} {\bibinfo {title} {Vestigial $d$-wave charge-$4e$
  superconductivity from bidirectional pair density waves},\ }\href
  {https://doi.org/10.1103/7lbj-tzw3} {\bibfield  {journal} {\bibinfo
  {journal} {Phys. Rev. B}\ }\textbf {\bibinfo {volume} {113}},\ \bibinfo
  {pages} {045121} (\bibinfo {year} {2026})}\BibitemShut {NoStop}%
\bibitem [{\citenamefont {Berg}\ \emph
  {et~al.}(2009{\natexlab{c}})\citenamefont {Berg}, \citenamefont {Fradkin},\
  and\ \citenamefont {Kivelson}}]{berg2009charge}%
  \BibitemOpen
  \bibfield  {author} {\bibinfo {author} {\bibfnamefont {E.}~\bibnamefont
  {Berg}}, \bibinfo {author} {\bibfnamefont {E.}~\bibnamefont {Fradkin}},\ and\
  \bibinfo {author} {\bibfnamefont {S.~A.}\ \bibnamefont {Kivelson}},\
  }\bibfield  {title} {\bibinfo {title} {Charge-4e superconductivity from
  pair-density-wave order in certain high-temperature superconductors},\ }\href
  {https://www.nature.com/articles/nphys1389} {\bibfield  {journal} {\bibinfo
  {journal} {Nature Physics}\ }\textbf {\bibinfo {volume} {5}},\ \bibinfo
  {pages} {830} (\bibinfo {year} {2009}{\natexlab{c}})}\BibitemShut {NoStop}%
\bibitem [{\citenamefont {Jiang}\ \emph {et~al.}(2017)\citenamefont {Jiang},
  \citenamefont {Li}, \citenamefont {Kivelson},\ and\ \citenamefont
  {Yao}}]{Jiang2017Charge}%
  \BibitemOpen
  \bibfield  {author} {\bibinfo {author} {\bibfnamefont {Y.-F.}\ \bibnamefont
  {Jiang}}, \bibinfo {author} {\bibfnamefont {Z.-X.}\ \bibnamefont {Li}},
  \bibinfo {author} {\bibfnamefont {S.~A.}\ \bibnamefont {Kivelson}},\ and\
  \bibinfo {author} {\bibfnamefont {H.}~\bibnamefont {Yao}},\ }\bibfield
  {title} {\bibinfo {title} {Charge-$4e$ superconductors: A {Majorana} quantum
  monte carlo study},\ }\href {https://doi.org/10.1103/PhysRevB.95.241103}
  {\bibfield  {journal} {\bibinfo  {journal} {Phys. Rev. B}\ }\textbf {\bibinfo
  {volume} {95}},\ \bibinfo {pages} {241103(R)} (\bibinfo {year}
  {2017})}\BibitemShut {NoStop}%
\bibitem [{\citenamefont {Jian}\ \emph {et~al.}(2021)\citenamefont {Jian},
  \citenamefont {Huang},\ and\ \citenamefont {Yao}}]{Jian2021Charge}%
  \BibitemOpen
  \bibfield  {author} {\bibinfo {author} {\bibfnamefont {S.-K.}\ \bibnamefont
  {Jian}}, \bibinfo {author} {\bibfnamefont {Y.}~\bibnamefont {Huang}},\ and\
  \bibinfo {author} {\bibfnamefont {H.}~\bibnamefont {Yao}},\ }\bibfield
  {title} {\bibinfo {title} {Charge-$4e$ superconductivity from nematic
  superconductors in two and three dimensions},\ }\href
  {https://doi.org/10.1103/PhysRevLett.127.227001} {\bibfield  {journal}
  {\bibinfo  {journal} {Phys. Rev. Lett.}\ }\textbf {\bibinfo {volume} {127}},\
  \bibinfo {pages} {227001} (\bibinfo {year} {2021})}\BibitemShut {NoStop}%
\bibitem [{\citenamefont {Ge}\ \emph {et~al.}(2024)\citenamefont {Ge},
  \citenamefont {Wang}, \citenamefont {Xing}, \citenamefont {Yin},
  \citenamefont {Wang}, \citenamefont {Shen}, \citenamefont {Lei},
  \citenamefont {Wang},\ and\ \citenamefont {Wang}}]{Ge2024Charge}%
  \BibitemOpen
  \bibfield  {author} {\bibinfo {author} {\bibfnamefont {J.}~\bibnamefont
  {Ge}}, \bibinfo {author} {\bibfnamefont {P.}~\bibnamefont {Wang}}, \bibinfo
  {author} {\bibfnamefont {Y.}~\bibnamefont {Xing}}, \bibinfo {author}
  {\bibfnamefont {Q.}~\bibnamefont {Yin}}, \bibinfo {author} {\bibfnamefont
  {A.}~\bibnamefont {Wang}}, \bibinfo {author} {\bibfnamefont {J.}~\bibnamefont
  {Shen}}, \bibinfo {author} {\bibfnamefont {H.}~\bibnamefont {Lei}}, \bibinfo
  {author} {\bibfnamefont {Z.}~\bibnamefont {Wang}},\ and\ \bibinfo {author}
  {\bibfnamefont {J.}~\bibnamefont {Wang}},\ }\bibfield  {title} {\bibinfo
  {title} {Charge-$4e$ and charge-$6e$ flux quantization and higher charge
  superconductivity in kagome superconductor ring devices},\ }\href
  {https://doi.org/10.1103/PhysRevX.14.021025} {\bibfield  {journal} {\bibinfo
  {journal} {Phys. Rev. X}\ }\textbf {\bibinfo {volume} {14}},\ \bibinfo
  {pages} {021025} (\bibinfo {year} {2024})}\BibitemShut {NoStop}%
\bibitem [{\citenamefont {Zou}\ \emph {et~al.}(2026)\citenamefont {Zou},
  \citenamefont {Wan},\ and\ \citenamefont {Yao}}]{Zou2026Emergence}%
  \BibitemOpen
  \bibfield  {author} {\bibinfo {author} {\bibfnamefont {X.}~\bibnamefont
  {Zou}}, \bibinfo {author} {\bibfnamefont {Z.-Q.}\ \bibnamefont {Wan}},\ and\
  \bibinfo {author} {\bibfnamefont {H.}~\bibnamefont {Yao}},\ }\bibfield
  {title} {\bibinfo {title} {Emergence of charge-$4e$ superconductivity from 2d
  nematic superconductors},\ }\href {https://doi.org/10.1103/j7nh-jzdd}
  {\bibfield  {journal} {\bibinfo  {journal} {Phys. Rev. Lett.}\ }\textbf
  {\bibinfo {volume} {137}},\ \bibinfo {pages} {066505} (\bibinfo {year}
  {2026})}\BibitemShut {NoStop}%
\bibitem [{\citenamefont {Agterberg}\ and\ \citenamefont
  {Tsunetsugu}(2008)}]{agterberg2008dislocations}%
  \BibitemOpen
  \bibfield  {author} {\bibinfo {author} {\bibfnamefont {D.}~\bibnamefont
  {Agterberg}}\ and\ \bibinfo {author} {\bibfnamefont {H.}~\bibnamefont
  {Tsunetsugu}},\ }\bibfield  {title} {\bibinfo {title} {Dislocations and
  vortices in pair-density-wave superconductors},\ }\href
  {https://www.nature.com/articles/nphys999} {\bibfield  {journal} {\bibinfo
  {journal} {Nature Physics}\ }\textbf {\bibinfo {volume} {4}},\ \bibinfo
  {pages} {639} (\bibinfo {year} {2008})}\BibitemShut {NoStop}%
\bibitem [{\citenamefont {Radzihovsky}\ and\ \citenamefont
  {Vishwanath}(2009)}]{Radzihovsky2009Quantum}%
  \BibitemOpen
  \bibfield  {author} {\bibinfo {author} {\bibfnamefont {L.}~\bibnamefont
  {Radzihovsky}}\ and\ \bibinfo {author} {\bibfnamefont {A.}~\bibnamefont
  {Vishwanath}},\ }\bibfield  {title} {\bibinfo {title} {Quantum liquid
  crystals in an imbalanced fermi gas: Fluctuations and fractional vortices in
  larkin-ovchinnikov states},\ }\href
  {https://doi.org/10.1103/PhysRevLett.103.010404} {\bibfield  {journal}
  {\bibinfo  {journal} {Phys. Rev. Lett.}\ }\textbf {\bibinfo {volume} {103}},\
  \bibinfo {pages} {010404} (\bibinfo {year} {2009})}\BibitemShut {NoStop}%
\bibitem [{\citenamefont {Mross}\ and\ \citenamefont
  {Senthil}(2015)}]{Mross2015Spin}%
  \BibitemOpen
  \bibfield  {author} {\bibinfo {author} {\bibfnamefont {D.~F.}\ \bibnamefont
  {Mross}}\ and\ \bibinfo {author} {\bibfnamefont {T.}~\bibnamefont
  {Senthil}},\ }\bibfield  {title} {\bibinfo {title} {Spin- and
  pair-density-wave glasses},\ }\href
  {https://doi.org/10.1103/PhysRevX.5.031008} {\bibfield  {journal} {\bibinfo
  {journal} {Phys. Rev. X}\ }\textbf {\bibinfo {volume} {5}},\ \bibinfo {pages}
  {031008} (\bibinfo {year} {2015})}\BibitemShut {NoStop}%
\bibitem [{\citenamefont {Rosales}\ and\ \citenamefont
  {Fradkin}(2024)}]{Rosales2024Electronic}%
  \BibitemOpen
  \bibfield  {author} {\bibinfo {author} {\bibfnamefont {M.}~\bibnamefont
  {Rosales}}\ and\ \bibinfo {author} {\bibfnamefont {E.}~\bibnamefont
  {Fradkin}},\ }\bibfield  {title} {\bibinfo {title} {Electronic structure of
  topological defects in the pair density wave superconductor},\ }\href
  {https://doi.org/10.1103/PhysRevB.110.214508} {\bibfield  {journal} {\bibinfo
   {journal} {Phys. Rev. B}\ }\textbf {\bibinfo {volume} {110}},\ \bibinfo
  {pages} {214508} (\bibinfo {year} {2024})}\BibitemShut {NoStop}%
\bibitem [{\citenamefont {Salomaa}\ and\ \citenamefont
  {Volovik}(1985)}]{Salomaa1985Half}%
  \BibitemOpen
  \bibfield  {author} {\bibinfo {author} {\bibfnamefont {M.~M.}\ \bibnamefont
  {Salomaa}}\ and\ \bibinfo {author} {\bibfnamefont {G.~E.}\ \bibnamefont
  {Volovik}},\ }\bibfield  {title} {\bibinfo {title} {Half-quantum vortices in
  superfluid $^{3}\mathrm{He}$-$a$},\ }\href
  {https://doi.org/10.1103/PhysRevLett.55.1184} {\bibfield  {journal} {\bibinfo
   {journal} {Phys. Rev. Lett.}\ }\textbf {\bibinfo {volume} {55}},\ \bibinfo
  {pages} {1184} (\bibinfo {year} {1985})}\BibitemShut {NoStop}%
\bibitem [{\citenamefont {Babaev}(2002)}]{Babaev2002Vortices}%
  \BibitemOpen
  \bibfield  {author} {\bibinfo {author} {\bibfnamefont {E.}~\bibnamefont
  {Babaev}},\ }\bibfield  {title} {\bibinfo {title} {Vortices with fractional
  flux in two-gap superconductors and in extended faddeev model},\ }\href
  {https://doi.org/10.1103/PhysRevLett.89.067001} {\bibfield  {journal}
  {\bibinfo  {journal} {Phys. Rev. Lett.}\ }\textbf {\bibinfo {volume} {89}},\
  \bibinfo {pages} {067001} (\bibinfo {year} {2002})}\BibitemShut {NoStop}%
\bibitem [{\citenamefont {Chung}\ \emph {et~al.}(2007)\citenamefont {Chung},
  \citenamefont {Bluhm},\ and\ \citenamefont {Kim}}]{Chung2007Stability}%
  \BibitemOpen
  \bibfield  {author} {\bibinfo {author} {\bibfnamefont {S.~B.}\ \bibnamefont
  {Chung}}, \bibinfo {author} {\bibfnamefont {H.}~\bibnamefont {Bluhm}},\ and\
  \bibinfo {author} {\bibfnamefont {E.-A.}\ \bibnamefont {Kim}},\ }\bibfield
  {title} {\bibinfo {title} {Stability of half-quantum vortices in
  ${p}_{x}+i{p}_{y}$ superconductors},\ }\href
  {https://doi.org/10.1103/PhysRevLett.99.197002} {\bibfield  {journal}
  {\bibinfo  {journal} {Phys. Rev. Lett.}\ }\textbf {\bibinfo {volume} {99}},\
  \bibinfo {pages} {197002} (\bibinfo {year} {2007})}\BibitemShut {NoStop}%
\bibitem [{\citenamefont {Jang}\ \emph {et~al.}(2011)\citenamefont {Jang},
  \citenamefont {Ferguson}, \citenamefont {Vakaryuk}, \citenamefont {Budakian},
  \citenamefont {Chung}, \citenamefont {Goldbart},\ and\ \citenamefont
  {Maeno}}]{Jang2011Observation}%
  \BibitemOpen
  \bibfield  {author} {\bibinfo {author} {\bibfnamefont {J.}~\bibnamefont
  {Jang}}, \bibinfo {author} {\bibfnamefont {D.~G.}\ \bibnamefont {Ferguson}},
  \bibinfo {author} {\bibfnamefont {V.}~\bibnamefont {Vakaryuk}}, \bibinfo
  {author} {\bibfnamefont {R.}~\bibnamefont {Budakian}}, \bibinfo {author}
  {\bibfnamefont {S.~B.}\ \bibnamefont {Chung}}, \bibinfo {author}
  {\bibfnamefont {P.~M.}\ \bibnamefont {Goldbart}},\ and\ \bibinfo {author}
  {\bibfnamefont {Y.}~\bibnamefont {Maeno}},\ }\bibfield  {title} {\bibinfo
  {title} {Observation of half-height magnetization steps in {Sr$_2$RuO$_4$}},\
  }\href {https://doi.org/10.1126/science.1193839} {\bibfield  {journal}
  {\bibinfo  {journal} {Science}\ }\textbf {\bibinfo {volume} {331}},\ \bibinfo
  {pages} {186} (\bibinfo {year} {2011})}\BibitemShut {NoStop}%
\bibitem [{\citenamefont {Chung}\ \emph {et~al.}(2009)\citenamefont {Chung},
  \citenamefont {Agterberg},\ and\ \citenamefont {Kim}}]{Chung2009Fractional}%
  \BibitemOpen
  \bibfield  {author} {\bibinfo {author} {\bibfnamefont {S.~B.}\ \bibnamefont
  {Chung}}, \bibinfo {author} {\bibfnamefont {D.~F.}\ \bibnamefont
  {Agterberg}},\ and\ \bibinfo {author} {\bibfnamefont {E.-A.}\ \bibnamefont
  {Kim}},\ }\bibfield  {title} {\bibinfo {title} {Fractional vortex lattice
  structures in spin-triplet superconductors},\ }\href
  {https://doi.org/10.1088/1367-2630/11/8/085004} {\bibfield  {journal}
  {\bibinfo  {journal} {New Journal of Physics}\ }\textbf {\bibinfo {volume}
  {11}},\ \bibinfo {pages} {085004} (\bibinfo {year} {2009})}\BibitemShut
  {NoStop}%
\bibitem [{\citenamefont {Chung}\ and\ \citenamefont
  {Kivelson}(2010)}]{Chung2010Entropy}%
  \BibitemOpen
  \bibfield  {author} {\bibinfo {author} {\bibfnamefont {S.~B.}\ \bibnamefont
  {Chung}}\ and\ \bibinfo {author} {\bibfnamefont {S.~A.}\ \bibnamefont
  {Kivelson}},\ }\bibfield  {title} {\bibinfo {title} {Entropy-driven formation
  of a half-quantum vortex lattice},\ }\href
  {https://doi.org/10.1103/PhysRevB.82.214512} {\bibfield  {journal} {\bibinfo
  {journal} {Phys. Rev. B}\ }\textbf {\bibinfo {volume} {82}},\ \bibinfo
  {pages} {214512} (\bibinfo {year} {2010})}\BibitemShut {NoStop}%
\bibitem [{\citenamefont {Cho}\ \emph {et~al.}(2012)\citenamefont {Cho},
  \citenamefont {Bardarson}, \citenamefont {Lu},\ and\ \citenamefont
  {Moore}}]{Cho2012Superconductivity}%
  \BibitemOpen
  \bibfield  {author} {\bibinfo {author} {\bibfnamefont {G.~Y.}\ \bibnamefont
  {Cho}}, \bibinfo {author} {\bibfnamefont {J.~H.}\ \bibnamefont {Bardarson}},
  \bibinfo {author} {\bibfnamefont {Y.-M.}\ \bibnamefont {Lu}},\ and\ \bibinfo
  {author} {\bibfnamefont {J.~E.}\ \bibnamefont {Moore}},\ }\bibfield  {title}
  {\bibinfo {title} {Superconductivity of doped weyl semimetals:
  Finite-momentum pairing and electronic analog of the ${}^{3}$he-$a$ phase},\
  }\href {https://doi.org/10.1103/PhysRevB.86.214514} {\bibfield  {journal}
  {\bibinfo  {journal} {Phys. Rev. B}\ }\textbf {\bibinfo {volume} {86}},\
  \bibinfo {pages} {214514} (\bibinfo {year} {2012})}\BibitemShut {NoStop}%
\bibitem [{\citenamefont {Cho}\ \emph {et~al.}(2014)\citenamefont {Cho},
  \citenamefont {Soto-Garrido},\ and\ \citenamefont
  {Fradkin}}]{Cho2014Topological}%
  \BibitemOpen
  \bibfield  {author} {\bibinfo {author} {\bibfnamefont {G.~Y.}\ \bibnamefont
  {Cho}}, \bibinfo {author} {\bibfnamefont {R.}~\bibnamefont {Soto-Garrido}},\
  and\ \bibinfo {author} {\bibfnamefont {E.}~\bibnamefont {Fradkin}},\
  }\bibfield  {title} {\bibinfo {title} {Topological pair-density-wave
  superconducting states},\ }\href
  {https://doi.org/10.1103/PhysRevLett.113.256405} {\bibfield  {journal}
  {\bibinfo  {journal} {Phys. Rev. Lett.}\ }\textbf {\bibinfo {volume} {113}},\
  \bibinfo {pages} {256405} (\bibinfo {year} {2014})}\BibitemShut {NoStop}%
\bibitem [{\citenamefont {Chan}\ and\ \citenamefont {Liu}(2017)}]{Chan2017Non}%
  \BibitemOpen
  \bibfield  {author} {\bibinfo {author} {\bibfnamefont {C.}~\bibnamefont
  {Chan}}\ and\ \bibinfo {author} {\bibfnamefont {X.-J.}\ \bibnamefont {Liu}},\
  }\bibfield  {title} {\bibinfo {title} {Non-abelian {Majorana} modes protected
  by an emergent second chern number},\ }\href
  {https://doi.org/10.1103/PhysRevLett.118.207002} {\bibfield  {journal}
  {\bibinfo  {journal} {Phys. Rev. Lett.}\ }\textbf {\bibinfo {volume} {118}},\
  \bibinfo {pages} {207002} (\bibinfo {year} {2017})}\BibitemShut {NoStop}%
\bibitem [{\citenamefont {Santos}\ \emph {et~al.}(2019)\citenamefont {Santos},
  \citenamefont {Wang},\ and\ \citenamefont {Fradkin}}]{Santos2019Pair}%
  \BibitemOpen
  \bibfield  {author} {\bibinfo {author} {\bibfnamefont {L.~H.}\ \bibnamefont
  {Santos}}, \bibinfo {author} {\bibfnamefont {Y.}~\bibnamefont {Wang}},\ and\
  \bibinfo {author} {\bibfnamefont {E.}~\bibnamefont {Fradkin}},\ }\bibfield
  {title} {\bibinfo {title} {Pair-density-wave order and paired fractional
  quantum hall fluids},\ }\href {https://doi.org/10.1103/PhysRevX.9.021047}
  {\bibfield  {journal} {\bibinfo  {journal} {Phys. Rev. X}\ }\textbf {\bibinfo
  {volume} {9}},\ \bibinfo {pages} {021047} (\bibinfo {year}
  {2019})}\BibitemShut {NoStop}%
\bibitem [{\citenamefont {Read}\ and\ \citenamefont
  {Green}(2000)}]{Read2000Paired}%
  \BibitemOpen
  \bibfield  {author} {\bibinfo {author} {\bibfnamefont {N.}~\bibnamefont
  {Read}}\ and\ \bibinfo {author} {\bibfnamefont {D.}~\bibnamefont {Green}},\
  }\bibfield  {title} {\bibinfo {title} {Paired states of fermions in two
  dimensions with breaking of parity and time-reversal symmetries and the
  fractional quantum hall effect},\ }\href
  {https://doi.org/10.1103/PhysRevB.61.10267} {\bibfield  {journal} {\bibinfo
  {journal} {Phys. Rev. B}\ }\textbf {\bibinfo {volume} {61}},\ \bibinfo
  {pages} {10267} (\bibinfo {year} {2000})}\BibitemShut {NoStop}%
\bibitem [{\citenamefont {Ivanov}(2001)}]{Ivanov2001Non}%
  \BibitemOpen
  \bibfield  {author} {\bibinfo {author} {\bibfnamefont {D.~A.}\ \bibnamefont
  {Ivanov}},\ }\bibfield  {title} {\bibinfo {title} {Non-abelian statistics of
  half-quantum vortices in $\mathit{p}$-wave superconductors},\ }\href
  {https://doi.org/10.1103/PhysRevLett.86.268} {\bibfield  {journal} {\bibinfo
  {journal} {Phys. Rev. Lett.}\ }\textbf {\bibinfo {volume} {86}},\ \bibinfo
  {pages} {268} (\bibinfo {year} {2001})}\BibitemShut {NoStop}%
\bibitem [{\citenamefont {Stone}\ and\ \citenamefont
  {Chung}(2006)}]{Stone2006Fusion}%
  \BibitemOpen
  \bibfield  {author} {\bibinfo {author} {\bibfnamefont {M.}~\bibnamefont
  {Stone}}\ and\ \bibinfo {author} {\bibfnamefont {S.-B.}\ \bibnamefont
  {Chung}},\ }\bibfield  {title} {\bibinfo {title} {Fusion rules and vortices
  in ${p}_{x}+i{p}_{y}$ superconductors},\ }\href
  {https://doi.org/10.1103/PhysRevB.73.014505} {\bibfield  {journal} {\bibinfo
  {journal} {Phys. Rev. B}\ }\textbf {\bibinfo {volume} {73}},\ \bibinfo
  {pages} {014505} (\bibinfo {year} {2006})}\BibitemShut {NoStop}%
\bibitem [{\citenamefont {Tewari}\ \emph {et~al.}(2007)\citenamefont {Tewari},
  \citenamefont {Das~Sarma},\ and\ \citenamefont {Lee}}]{Tewari2007Index}%
  \BibitemOpen
  \bibfield  {author} {\bibinfo {author} {\bibfnamefont {S.}~\bibnamefont
  {Tewari}}, \bibinfo {author} {\bibfnamefont {S.}~\bibnamefont {Das~Sarma}},\
  and\ \bibinfo {author} {\bibfnamefont {D.-H.}\ \bibnamefont {Lee}},\
  }\bibfield  {title} {\bibinfo {title} {Index theorem for the zero modes of
  {Majorana} fermion vortices in chiral $p$-wave superconductors},\ }\href
  {https://doi.org/10.1103/PhysRevLett.99.037001} {\bibfield  {journal}
  {\bibinfo  {journal} {Phys. Rev. Lett.}\ }\textbf {\bibinfo {volume} {99}},\
  \bibinfo {pages} {037001} (\bibinfo {year} {2007})}\BibitemShut {NoStop}%
\bibitem [{\citenamefont {Tewari}\ \emph {et~al.}(2008)\citenamefont {Tewari},
  \citenamefont {Zhang}, \citenamefont {Das~Sarma}, \citenamefont {Nayak},\
  and\ \citenamefont {Lee}}]{Tewari2008Testable}%
  \BibitemOpen
  \bibfield  {author} {\bibinfo {author} {\bibfnamefont {S.}~\bibnamefont
  {Tewari}}, \bibinfo {author} {\bibfnamefont {C.}~\bibnamefont {Zhang}},
  \bibinfo {author} {\bibfnamefont {S.}~\bibnamefont {Das~Sarma}}, \bibinfo
  {author} {\bibfnamefont {C.}~\bibnamefont {Nayak}},\ and\ \bibinfo {author}
  {\bibfnamefont {D.-H.}\ \bibnamefont {Lee}},\ }\bibfield  {title} {\bibinfo
  {title} {Testable signatures of quantum nonlocality in a two-dimensional
  chiral $p$-wave superconductor},\ }\href
  {https://doi.org/10.1103/PhysRevLett.100.027001} {\bibfield  {journal}
  {\bibinfo  {journal} {Phys. Rev. Lett.}\ }\textbf {\bibinfo {volume} {100}},\
  \bibinfo {pages} {027001} (\bibinfo {year} {2008})}\BibitemShut {NoStop}%
\bibitem [{\citenamefont {Fu}\ and\ \citenamefont
  {Kane}(2008)}]{Fu2008Superconducting}%
  \BibitemOpen
  \bibfield  {author} {\bibinfo {author} {\bibfnamefont {L.}~\bibnamefont
  {Fu}}\ and\ \bibinfo {author} {\bibfnamefont {C.~L.}\ \bibnamefont {Kane}},\
  }\bibfield  {title} {\bibinfo {title} {Superconducting proximity effect and
  {Majorana} fermions at the surface of a topological insulator},\ }\href
  {https://doi.org/10.1103/PhysRevLett.100.096407} {\bibfield  {journal}
  {\bibinfo  {journal} {Phys. Rev. Lett.}\ }\textbf {\bibinfo {volume} {100}},\
  \bibinfo {pages} {096407} (\bibinfo {year} {2008})}\BibitemShut {NoStop}%
\bibitem [{\citenamefont {Teo}\ and\ \citenamefont
  {Kane}(2010)}]{Teo2010Topological}%
  \BibitemOpen
  \bibfield  {author} {\bibinfo {author} {\bibfnamefont {J.~C.~Y.}\
  \bibnamefont {Teo}}\ and\ \bibinfo {author} {\bibfnamefont {C.~L.}\
  \bibnamefont {Kane}},\ }\bibfield  {title} {\bibinfo {title} {Topological
  defects and gapless modes in insulators and superconductors},\ }\href
  {https://doi.org/10.1103/PhysRevB.82.115120} {\bibfield  {journal} {\bibinfo
  {journal} {Phys. Rev. B}\ }\textbf {\bibinfo {volume} {82}},\ \bibinfo
  {pages} {115120} (\bibinfo {year} {2010})}\BibitemShut {NoStop}%
\bibitem [{\citenamefont {Wang}\ \emph {et~al.}(2012)\citenamefont {Wang},
  \citenamefont {Liu}, \citenamefont {Xu}, \citenamefont {Yang}, \citenamefont
  {Miao}, \citenamefont {Yao}, \citenamefont {Gao}, \citenamefont {Shen},
  \citenamefont {Ma}, \citenamefont {Chen}, \citenamefont {Xu}, \citenamefont
  {Liu}, \citenamefont {Zhang}, \citenamefont {Qian}, \citenamefont {Jia},\
  and\ \citenamefont {Xue}}]{Wang2012The}%
  \BibitemOpen
  \bibfield  {author} {\bibinfo {author} {\bibfnamefont {M.-X.}\ \bibnamefont
  {Wang}}, \bibinfo {author} {\bibfnamefont {C.}~\bibnamefont {Liu}}, \bibinfo
  {author} {\bibfnamefont {J.-P.}\ \bibnamefont {Xu}}, \bibinfo {author}
  {\bibfnamefont {F.}~\bibnamefont {Yang}}, \bibinfo {author} {\bibfnamefont
  {L.}~\bibnamefont {Miao}}, \bibinfo {author} {\bibfnamefont {M.-Y.}\
  \bibnamefont {Yao}}, \bibinfo {author} {\bibfnamefont {C.~L.}\ \bibnamefont
  {Gao}}, \bibinfo {author} {\bibfnamefont {C.}~\bibnamefont {Shen}}, \bibinfo
  {author} {\bibfnamefont {X.}~\bibnamefont {Ma}}, \bibinfo {author}
  {\bibfnamefont {X.}~\bibnamefont {Chen}}, \bibinfo {author} {\bibfnamefont
  {Z.-A.}\ \bibnamefont {Xu}}, \bibinfo {author} {\bibfnamefont
  {Y.}~\bibnamefont {Liu}}, \bibinfo {author} {\bibfnamefont {S.-C.}\
  \bibnamefont {Zhang}}, \bibinfo {author} {\bibfnamefont {D.}~\bibnamefont
  {Qian}}, \bibinfo {author} {\bibfnamefont {J.-F.}\ \bibnamefont {Jia}},\ and\
  \bibinfo {author} {\bibfnamefont {Q.-K.}\ \bibnamefont {Xue}},\ }\bibfield
  {title} {\bibinfo {title} {The coexistence of superconductivity and
  topological order in the $\mathrm{Bi_2Se_3}$ thin films},\ }\href
  {https://doi.org/10.1126/science.1216466} {\bibfield  {journal} {\bibinfo
  {journal} {Science}\ }\textbf {\bibinfo {volume} {336}},\ \bibinfo {pages}
  {52} (\bibinfo {year} {2012})}\BibitemShut {NoStop}%
\bibitem [{\citenamefont {Xu}\ \emph {et~al.}(2014)\citenamefont {Xu},
  \citenamefont {Liu}, \citenamefont {Wang}, \citenamefont {Ge}, \citenamefont
  {Liu}, \citenamefont {Yang}, \citenamefont {Chen}, \citenamefont {Liu},
  \citenamefont {Xu}, \citenamefont {Gao}, \citenamefont {Qian}, \citenamefont
  {Zhang},\ and\ \citenamefont {Jia}}]{Xu2014Artificial}%
  \BibitemOpen
  \bibfield  {author} {\bibinfo {author} {\bibfnamefont {J.-P.}\ \bibnamefont
  {Xu}}, \bibinfo {author} {\bibfnamefont {C.}~\bibnamefont {Liu}}, \bibinfo
  {author} {\bibfnamefont {M.-X.}\ \bibnamefont {Wang}}, \bibinfo {author}
  {\bibfnamefont {J.}~\bibnamefont {Ge}}, \bibinfo {author} {\bibfnamefont
  {Z.-L.}\ \bibnamefont {Liu}}, \bibinfo {author} {\bibfnamefont
  {X.}~\bibnamefont {Yang}}, \bibinfo {author} {\bibfnamefont {Y.}~\bibnamefont
  {Chen}}, \bibinfo {author} {\bibfnamefont {Y.}~\bibnamefont {Liu}}, \bibinfo
  {author} {\bibfnamefont {Z.-A.}\ \bibnamefont {Xu}}, \bibinfo {author}
  {\bibfnamefont {C.-L.}\ \bibnamefont {Gao}}, \bibinfo {author} {\bibfnamefont
  {D.}~\bibnamefont {Qian}}, \bibinfo {author} {\bibfnamefont {F.-C.}\
  \bibnamefont {Zhang}},\ and\ \bibinfo {author} {\bibfnamefont {J.-F.}\
  \bibnamefont {Jia}},\ }\bibfield  {title} {\bibinfo {title} {Artificial
  topological superconductor by the proximity effect},\ }\href
  {https://doi.org/10.1103/PhysRevLett.112.217001} {\bibfield  {journal}
  {\bibinfo  {journal} {Phys. Rev. Lett.}\ }\textbf {\bibinfo {volume} {112}},\
  \bibinfo {pages} {217001} (\bibinfo {year} {2014})}\BibitemShut {NoStop}%
\bibitem [{\citenamefont {Xu}\ \emph {et~al.}(2015)\citenamefont {Xu},
  \citenamefont {Wang}, \citenamefont {Liu}, \citenamefont {Ge}, \citenamefont
  {Yang}, \citenamefont {Liu}, \citenamefont {Xu}, \citenamefont {Guan},
  \citenamefont {Gao}, \citenamefont {Qian}, \citenamefont {Liu}, \citenamefont
  {Wang}, \citenamefont {Zhang}, \citenamefont {Xue},\ and\ \citenamefont
  {Jia}}]{Xu2015Experimental}%
  \BibitemOpen
  \bibfield  {author} {\bibinfo {author} {\bibfnamefont {J.-P.}\ \bibnamefont
  {Xu}}, \bibinfo {author} {\bibfnamefont {M.-X.}\ \bibnamefont {Wang}},
  \bibinfo {author} {\bibfnamefont {Z.~L.}\ \bibnamefont {Liu}}, \bibinfo
  {author} {\bibfnamefont {J.-F.}\ \bibnamefont {Ge}}, \bibinfo {author}
  {\bibfnamefont {X.}~\bibnamefont {Yang}}, \bibinfo {author} {\bibfnamefont
  {C.}~\bibnamefont {Liu}}, \bibinfo {author} {\bibfnamefont {Z.~A.}\
  \bibnamefont {Xu}}, \bibinfo {author} {\bibfnamefont {D.}~\bibnamefont
  {Guan}}, \bibinfo {author} {\bibfnamefont {C.~L.}\ \bibnamefont {Gao}},
  \bibinfo {author} {\bibfnamefont {D.}~\bibnamefont {Qian}}, \bibinfo {author}
  {\bibfnamefont {Y.}~\bibnamefont {Liu}}, \bibinfo {author} {\bibfnamefont
  {Q.-H.}\ \bibnamefont {Wang}}, \bibinfo {author} {\bibfnamefont {F.-C.}\
  \bibnamefont {Zhang}}, \bibinfo {author} {\bibfnamefont {Q.-K.}\ \bibnamefont
  {Xue}},\ and\ \bibinfo {author} {\bibfnamefont {J.-F.}\ \bibnamefont {Jia}},\
  }\bibfield  {title} {\bibinfo {title} {Experimental detection of a {Majorana}
  mode in the core of a magnetic vortex inside a topological
  insulator-superconductor $\mathrm{Bi}_2\mathrm{Te}_3/\mathrm{NbSe}_2$
  heterostructure},\ }\href {https://doi.org/10.1103/PhysRevLett.114.017001}
  {\bibfield  {journal} {\bibinfo  {journal} {Phys. Rev. Lett.}\ }\textbf
  {\bibinfo {volume} {114}},\ \bibinfo {pages} {017001} (\bibinfo {year}
  {2015})}\BibitemShut {NoStop}%
\bibitem [{\citenamefont {Sun}\ \emph {et~al.}(2016)\citenamefont {Sun},
  \citenamefont {Zhang}, \citenamefont {Hu}, \citenamefont {Li}, \citenamefont
  {Wang}, \citenamefont {Ma}, \citenamefont {Xu}, \citenamefont {Gao},
  \citenamefont {Guan}, \citenamefont {Li}, \citenamefont {Liu}, \citenamefont
  {Qian}, \citenamefont {Zhou}, \citenamefont {Fu}, \citenamefont {Li},
  \citenamefont {Zhang},\ and\ \citenamefont {Jia}}]{Sun2016Majorana}%
  \BibitemOpen
  \bibfield  {author} {\bibinfo {author} {\bibfnamefont {H.-H.}\ \bibnamefont
  {Sun}}, \bibinfo {author} {\bibfnamefont {K.-W.}\ \bibnamefont {Zhang}},
  \bibinfo {author} {\bibfnamefont {L.-H.}\ \bibnamefont {Hu}}, \bibinfo
  {author} {\bibfnamefont {C.}~\bibnamefont {Li}}, \bibinfo {author}
  {\bibfnamefont {G.-Y.}\ \bibnamefont {Wang}}, \bibinfo {author}
  {\bibfnamefont {H.-Y.}\ \bibnamefont {Ma}}, \bibinfo {author} {\bibfnamefont
  {Z.-A.}\ \bibnamefont {Xu}}, \bibinfo {author} {\bibfnamefont {C.-L.}\
  \bibnamefont {Gao}}, \bibinfo {author} {\bibfnamefont {D.-D.}\ \bibnamefont
  {Guan}}, \bibinfo {author} {\bibfnamefont {Y.-Y.}\ \bibnamefont {Li}},
  \bibinfo {author} {\bibfnamefont {C.}~\bibnamefont {Liu}}, \bibinfo {author}
  {\bibfnamefont {D.}~\bibnamefont {Qian}}, \bibinfo {author} {\bibfnamefont
  {Y.}~\bibnamefont {Zhou}}, \bibinfo {author} {\bibfnamefont {L.}~\bibnamefont
  {Fu}}, \bibinfo {author} {\bibfnamefont {S.-C.}\ \bibnamefont {Li}}, \bibinfo
  {author} {\bibfnamefont {F.-C.}\ \bibnamefont {Zhang}},\ and\ \bibinfo
  {author} {\bibfnamefont {J.-F.}\ \bibnamefont {Jia}},\ }\bibfield  {title}
  {\bibinfo {title} {Majorana zero mode detected with spin selective {Andreev}
  reflection in the vortex of a topological superconductor},\ }\href
  {https://doi.org/10.1103/PhysRevLett.116.257003} {\bibfield  {journal}
  {\bibinfo  {journal} {Phys. Rev. Lett.}\ }\textbf {\bibinfo {volume} {116}},\
  \bibinfo {pages} {257003} (\bibinfo {year} {2016})}\BibitemShut {NoStop}%
\bibitem [{\citenamefont {Xu}\ \emph {et~al.}(2016)\citenamefont {Xu},
  \citenamefont {Lian}, \citenamefont {Tang}, \citenamefont {Qi},\ and\
  \citenamefont {Zhang}}]{Xu2016Topological}%
  \BibitemOpen
  \bibfield  {author} {\bibinfo {author} {\bibfnamefont {G.}~\bibnamefont
  {Xu}}, \bibinfo {author} {\bibfnamefont {B.}~\bibnamefont {Lian}}, \bibinfo
  {author} {\bibfnamefont {P.}~\bibnamefont {Tang}}, \bibinfo {author}
  {\bibfnamefont {X.-L.}\ \bibnamefont {Qi}},\ and\ \bibinfo {author}
  {\bibfnamefont {S.-C.}\ \bibnamefont {Zhang}},\ }\bibfield  {title} {\bibinfo
  {title} {Topological superconductivity on the surface of {Fe}-based
  superconductors},\ }\href {https://doi.org/10.1103/PhysRevLett.117.047001}
  {\bibfield  {journal} {\bibinfo  {journal} {Phys. Rev. Lett.}\ }\textbf
  {\bibinfo {volume} {117}},\ \bibinfo {pages} {047001} (\bibinfo {year}
  {2016})}\BibitemShut {NoStop}%
\bibitem [{\citenamefont {Biswas}(2013)}]{Biswas2013Majorana}%
  \BibitemOpen
  \bibfield  {author} {\bibinfo {author} {\bibfnamefont {R.~R.}\ \bibnamefont
  {Biswas}},\ }\bibfield  {title} {\bibinfo {title} {Majorana fermions in
  vortex lattices},\ }\href {https://doi.org/10.1103/PhysRevLett.111.136401}
  {\bibfield  {journal} {\bibinfo  {journal} {Phys. Rev. Lett.}\ }\textbf
  {\bibinfo {volume} {111}},\ \bibinfo {pages} {136401} (\bibinfo {year}
  {2013})}\BibitemShut {NoStop}%
\bibitem [{\citenamefont {Murray}\ and\ \citenamefont
  {Vafek}(2015)}]{Murray2015Majorana}%
  \BibitemOpen
  \bibfield  {author} {\bibinfo {author} {\bibfnamefont {J.~M.}\ \bibnamefont
  {Murray}}\ and\ \bibinfo {author} {\bibfnamefont {O.}~\bibnamefont {Vafek}},\
  }\bibfield  {title} {\bibinfo {title} {Majorana bands, berry curvature, and
  thermal hall conductivity in the vortex state of a chiral $p$-wave
  superconductor},\ }\href {https://doi.org/10.1103/PhysRevB.92.134520}
  {\bibfield  {journal} {\bibinfo  {journal} {Phys. Rev. B}\ }\textbf {\bibinfo
  {volume} {92}},\ \bibinfo {pages} {134520} (\bibinfo {year}
  {2015})}\BibitemShut {NoStop}%
\bibitem [{\citenamefont {Liu}\ and\ \citenamefont
  {Franz}(2015)}]{Liu2015Electronic}%
  \BibitemOpen
  \bibfield  {author} {\bibinfo {author} {\bibfnamefont {T.}~\bibnamefont
  {Liu}}\ and\ \bibinfo {author} {\bibfnamefont {M.}~\bibnamefont {Franz}},\
  }\bibfield  {title} {\bibinfo {title} {Electronic structure of topological
  superconductors in the presence of a vortex lattice},\ }\href
  {https://doi.org/10.1103/PhysRevB.92.134519} {\bibfield  {journal} {\bibinfo
  {journal} {Phys. Rev. B}\ }\textbf {\bibinfo {volume} {92}},\ \bibinfo
  {pages} {134519} (\bibinfo {year} {2015})}\BibitemShut {NoStop}%
\bibitem [{\citenamefont {Yoshida}\ and\ \citenamefont
  {Udagawa}(2016)}]{Yoshida2016Generic}%
  \BibitemOpen
  \bibfield  {author} {\bibinfo {author} {\bibfnamefont {T.}~\bibnamefont
  {Yoshida}}\ and\ \bibinfo {author} {\bibfnamefont {M.}~\bibnamefont
  {Udagawa}},\ }\bibfield  {title} {\bibinfo {title} {Generic weyl phase in the
  vortex state of quasi-two-dimensional chiral superconductors},\ }\href
  {https://doi.org/10.1103/PhysRevB.94.060507} {\bibfield  {journal} {\bibinfo
  {journal} {Phys. Rev. B}\ }\textbf {\bibinfo {volume} {94}},\ \bibinfo
  {pages} {060507(R)} (\bibinfo {year} {2016})}\BibitemShut {NoStop}%
\bibitem [{\citenamefont {Jian}\ \emph {et~al.}(2015)\citenamefont {Jian},
  \citenamefont {Jiang},\ and\ \citenamefont {Yao}}]{Jian2015Emergent}%
  \BibitemOpen
  \bibfield  {author} {\bibinfo {author} {\bibfnamefont {S.-K.}\ \bibnamefont
  {Jian}}, \bibinfo {author} {\bibfnamefont {Y.-F.}\ \bibnamefont {Jiang}},\
  and\ \bibinfo {author} {\bibfnamefont {H.}~\bibnamefont {Yao}},\ }\bibfield
  {title} {\bibinfo {title} {Emergent spacetime supersymmetry in 3d weyl
  semimetals and 2d dirac semimetals},\ }\href
  {https://doi.org/10.1103/PhysRevLett.114.237001} {\bibfield  {journal}
  {\bibinfo  {journal} {Phys. Rev. Lett.}\ }\textbf {\bibinfo {volume} {114}},\
  \bibinfo {pages} {237001} (\bibinfo {year} {2015})}\BibitemShut {NoStop}%
\bibitem [{\citenamefont {Jian}\ \emph {et~al.}(2017)\citenamefont {Jian},
  \citenamefont {Lin}, \citenamefont {Maciejko},\ and\ \citenamefont
  {Yao}}]{Jian2017Emergence}%
  \BibitemOpen
  \bibfield  {author} {\bibinfo {author} {\bibfnamefont {S.-K.}\ \bibnamefont
  {Jian}}, \bibinfo {author} {\bibfnamefont {C.-H.}\ \bibnamefont {Lin}},
  \bibinfo {author} {\bibfnamefont {J.}~\bibnamefont {Maciejko}},\ and\
  \bibinfo {author} {\bibfnamefont {H.}~\bibnamefont {Yao}},\ }\bibfield
  {title} {\bibinfo {title} {Emergence of supersymmetric quantum
  electrodynamics},\ }\href {https://doi.org/10.1103/PhysRevLett.118.166802}
  {\bibfield  {journal} {\bibinfo  {journal} {Phys. Rev. Lett.}\ }\textbf
  {\bibinfo {volume} {118}},\ \bibinfo {pages} {166802} (\bibinfo {year}
  {2017})}\BibitemShut {NoStop}%
\bibitem [{\citenamefont {Jiang}\ and\ \citenamefont
  {Yao}(2024)}]{Jiang2024Pair}%
  \BibitemOpen
  \bibfield  {author} {\bibinfo {author} {\bibfnamefont {Y.-F.}\ \bibnamefont
  {Jiang}}\ and\ \bibinfo {author} {\bibfnamefont {H.}~\bibnamefont {Yao}},\
  }\bibfield  {title} {\bibinfo {title} {Pair-density-wave superconductivity: A
  microscopic model on the 2d honeycomb lattice},\ }\href
  {https://doi.org/10.1103/PhysRevLett.133.176501} {\bibfield  {journal}
  {\bibinfo  {journal} {Phys. Rev. Lett.}\ }\textbf {\bibinfo {volume} {133}},\
  \bibinfo {pages} {176501} (\bibinfo {year} {2024})}\BibitemShut {NoStop}%
\bibitem [{\citenamefont {Raghu}\ \emph {et~al.}(2008)\citenamefont {Raghu},
  \citenamefont {Qi}, \citenamefont {Honerkamp},\ and\ \citenamefont
  {Zhang}}]{Raghu2008Topological}%
  \BibitemOpen
  \bibfield  {author} {\bibinfo {author} {\bibfnamefont {S.}~\bibnamefont
  {Raghu}}, \bibinfo {author} {\bibfnamefont {X.-L.}\ \bibnamefont {Qi}},
  \bibinfo {author} {\bibfnamefont {C.}~\bibnamefont {Honerkamp}},\ and\
  \bibinfo {author} {\bibfnamefont {S.-C.}\ \bibnamefont {Zhang}},\ }\bibfield
  {title} {\bibinfo {title} {Topological mott insulators},\ }\href
  {https://doi.org/10.1103/PhysRevLett.100.156401} {\bibfield  {journal}
  {\bibinfo  {journal} {Phys. Rev. Lett.}\ }\textbf {\bibinfo {volume} {100}},\
  \bibinfo {pages} {156401} (\bibinfo {year} {2008})}\BibitemShut {NoStop}%
\bibitem [{\citenamefont {Radzihovsky}(2011)}]{Radzihovsky2011Fluctuations}%
  \BibitemOpen
  \bibfield  {author} {\bibinfo {author} {\bibfnamefont {L.}~\bibnamefont
  {Radzihovsky}},\ }\bibfield  {title} {\bibinfo {title} {Fluctuations and
  phase transitions in larkin-ovchinnikov liquid-crystal states of a
  population-imbalanced resonant fermi gas},\ }\href
  {https://doi.org/10.1103/PhysRevA.84.023611} {\bibfield  {journal} {\bibinfo
  {journal} {Phys. Rev. A}\ }\textbf {\bibinfo {volume} {84}},\ \bibinfo
  {pages} {023611} (\bibinfo {year} {2011})}\BibitemShut {NoStop}%
\bibitem [{\citenamefont {W\aa{}rdh}\ \emph {et~al.}(2018)\citenamefont
  {W\aa{}rdh}, \citenamefont {Andersen},\ and\ \citenamefont
  {Granath}}]{Waardh2018Suppression}%
  \BibitemOpen
  \bibfield  {author} {\bibinfo {author} {\bibfnamefont {J.}~\bibnamefont
  {W\aa{}rdh}}, \bibinfo {author} {\bibfnamefont {B.~M.}\ \bibnamefont
  {Andersen}},\ and\ \bibinfo {author} {\bibfnamefont {M.}~\bibnamefont
  {Granath}},\ }\bibfield  {title} {\bibinfo {title} {Suppression of superfluid
  stiffness near a lifshitz-point instability to finite-momentum
  superconductivity},\ }\href {https://doi.org/10.1103/PhysRevB.98.224501}
  {\bibfield  {journal} {\bibinfo  {journal} {Phys. Rev. B}\ }\textbf {\bibinfo
  {volume} {98}},\ \bibinfo {pages} {224501} (\bibinfo {year}
  {2018})}\BibitemShut {NoStop}%
\bibitem [{Note1()}]{Note1}%
  \BibitemOpen
  \bibinfo {note} {In the fully self-consistent calculation, the HFB fields are
  reoptimized at each imposed deformation. In the auxiliary-BdG scheme, the
  deformed BdG Hamiltonian is constructed from the undeformed mean-field saddle
  and diagonalized once, while in the rigid calculation the undeformed Gaussian
  state itself is kept fixed. The rigid construction applies only to the
  superconducting phase twist. See the SM~\cite {SM} for details.}\BibitemShut
  {Stop}%
\bibitem [{SM()}]{SM}%
  \BibitemOpen
  \href@noop {} {}\bibinfo {note} {See Supplemental Material for details of the
  mean-field calculation, phase-stiffness extraction, Ginzburg--Landau
  couplings and vortex-core energetics, gauge-consistent BdG calculation of
  half-vortex {Majorana} modes, half-vortex lattice geometry and shape
  relaxation, and effective {Majorana} band structure.}\BibitemShut {Stop}%
\bibitem [{Note2()}]{Note2}%
  \BibitemOpen
  \bibinfo {note} {Additional symmetry-allowed quartic-gradient terms,
  evaluated about the symmetry-related state $|\Delta _+|=|\Delta _-|$, reduce
  to either a common renormalization of the two phase stiffnesses or their
  relative splitting, and can therefore be absorbed into effective coefficients
  $\lambda $ and $\eta $.}\BibitemShut {Stop}%
\bibitem [{\citenamefont {How}\ and\ \citenamefont {Yip}(2020)}]{How2020Half}%
  \BibitemOpen
  \bibfield  {author} {\bibinfo {author} {\bibfnamefont {P.~T.}\ \bibnamefont
  {How}}\ and\ \bibinfo {author} {\bibfnamefont {S.-K.}\ \bibnamefont {Yip}},\
  }\bibfield  {title} {\bibinfo {title} {Half quantum vortices in a nematic
  superconductor},\ }\href {https://doi.org/10.1103/PhysRevResearch.2.043192}
  {\bibfield  {journal} {\bibinfo  {journal} {Phys. Rev. Res.}\ }\textbf
  {\bibinfo {volume} {2}},\ \bibinfo {pages} {043192} (\bibinfo {year}
  {2020})}\BibitemShut {NoStop}%
\bibitem [{Note3()}]{Note3}%
  \BibitemOpen
  \bibinfo {note} {This regime is naturally favored in thin films, where the
  effective magnetic screening length can greatly exceed $\xi $.}\BibitemShut
  {Stop}%
\bibitem [{Note4()}]{Note4}%
  \BibitemOpen
  \bibinfo {note} {We set $C=1$ in the numerical estimates.}\BibitemShut
  {Stop}%
\bibitem [{\citenamefont {Caroli}\ \emph {et~al.}(1964)\citenamefont {Caroli},
  \citenamefont {De~Gennes},\ and\ \citenamefont {Matricon}}]{caroli1964bound}%
  \BibitemOpen
  \bibfield  {author} {\bibinfo {author} {\bibfnamefont {C.}~\bibnamefont
  {Caroli}}, \bibinfo {author} {\bibfnamefont {P.}~\bibnamefont {De~Gennes}},\
  and\ \bibinfo {author} {\bibfnamefont {J.}~\bibnamefont {Matricon}},\
  }\bibfield  {title} {\bibinfo {title} {Bound fermion states on a vortex line
  in a type ii superconductor},\ }\href@noop {} {\bibfield  {journal} {\bibinfo
   {journal} {Physics Letters}\ }\textbf {\bibinfo {volume} {9}},\ \bibinfo
  {pages} {307} (\bibinfo {year} {1964})}\BibitemShut {NoStop}%
\bibitem [{\citenamefont {Gu}\ \emph {et~al.}(2023)\citenamefont {Gu},
  \citenamefont {Carroll}, \citenamefont {Wang}, \citenamefont {Ran},
  \citenamefont {Broyles}, \citenamefont {Siddiquee}, \citenamefont {Butch},
  \citenamefont {Saha}, \citenamefont {Paglione}, \citenamefont {Davis} \emph
  {et~al.}}]{gu2023detection}%
  \BibitemOpen
  \bibfield  {author} {\bibinfo {author} {\bibfnamefont {Q.}~\bibnamefont
  {Gu}}, \bibinfo {author} {\bibfnamefont {J.~P.}\ \bibnamefont {Carroll}},
  \bibinfo {author} {\bibfnamefont {S.}~\bibnamefont {Wang}}, \bibinfo {author}
  {\bibfnamefont {S.}~\bibnamefont {Ran}}, \bibinfo {author} {\bibfnamefont
  {C.}~\bibnamefont {Broyles}}, \bibinfo {author} {\bibfnamefont
  {H.}~\bibnamefont {Siddiquee}}, \bibinfo {author} {\bibfnamefont {N.~P.}\
  \bibnamefont {Butch}}, \bibinfo {author} {\bibfnamefont {S.~R.}\ \bibnamefont
  {Saha}}, \bibinfo {author} {\bibfnamefont {J.}~\bibnamefont {Paglione}},
  \bibinfo {author} {\bibfnamefont {J.~S.}\ \bibnamefont {Davis}}, \emph
  {et~al.},\ }\bibfield  {title} {\bibinfo {title} {Detection of a pair density
  wave state in ute2},\ }\href {https://doi.org/10.1038/s41586-023-05919-7}
  {\bibfield  {journal} {\bibinfo  {journal} {Nature}\ }\textbf {\bibinfo
  {volume} {618}},\ \bibinfo {pages} {921} (\bibinfo {year}
  {2023})}\BibitemShut {NoStop}%
\bibitem [{\citenamefont {Du}\ \emph {et~al.}(2020)\citenamefont {Du},
  \citenamefont {Li}, \citenamefont {Joo}, \citenamefont {Donoway},
  \citenamefont {Lee}, \citenamefont {Davis}, \citenamefont {Gu}, \citenamefont
  {Johnson},\ and\ \citenamefont {Fujita}}]{du2020imaging}%
  \BibitemOpen
  \bibfield  {author} {\bibinfo {author} {\bibfnamefont {Z.}~\bibnamefont
  {Du}}, \bibinfo {author} {\bibfnamefont {H.}~\bibnamefont {Li}}, \bibinfo
  {author} {\bibfnamefont {S.~H.}\ \bibnamefont {Joo}}, \bibinfo {author}
  {\bibfnamefont {E.~P.}\ \bibnamefont {Donoway}}, \bibinfo {author}
  {\bibfnamefont {J.}~\bibnamefont {Lee}}, \bibinfo {author} {\bibfnamefont
  {J.~S.}\ \bibnamefont {Davis}}, \bibinfo {author} {\bibfnamefont
  {G.}~\bibnamefont {Gu}}, \bibinfo {author} {\bibfnamefont {P.~D.}\
  \bibnamefont {Johnson}},\ and\ \bibinfo {author} {\bibfnamefont
  {K.}~\bibnamefont {Fujita}},\ }\bibfield  {title} {\bibinfo {title} {Imaging
  the energy gap modulations of the cuprate pair-density-wave state},\ }\href
  {https://doi.org/10.1038/s41586-020-2143-x} {\bibfield  {journal} {\bibinfo
  {journal} {Nature}\ }\textbf {\bibinfo {volume} {580}},\ \bibinfo {pages}
  {65} (\bibinfo {year} {2020})}\BibitemShut {NoStop}%
\bibitem [{\citenamefont {Zhang}\ \emph {et~al.}(2018)\citenamefont {Zhang},
  \citenamefont {Yaji}, \citenamefont {Hashimoto}, \citenamefont {Ota},
  \citenamefont {Kondo}, \citenamefont {Okazaki}, \citenamefont {Wang},
  \citenamefont {Wen}, \citenamefont {Gu}, \citenamefont {Ding},\ and\
  \citenamefont {Shin}}]{Zhang2018Observation}%
  \BibitemOpen
  \bibfield  {author} {\bibinfo {author} {\bibfnamefont {P.}~\bibnamefont
  {Zhang}}, \bibinfo {author} {\bibfnamefont {K.}~\bibnamefont {Yaji}},
  \bibinfo {author} {\bibfnamefont {T.}~\bibnamefont {Hashimoto}}, \bibinfo
  {author} {\bibfnamefont {Y.}~\bibnamefont {Ota}}, \bibinfo {author}
  {\bibfnamefont {T.}~\bibnamefont {Kondo}}, \bibinfo {author} {\bibfnamefont
  {K.}~\bibnamefont {Okazaki}}, \bibinfo {author} {\bibfnamefont
  {Z.}~\bibnamefont {Wang}}, \bibinfo {author} {\bibfnamefont {J.}~\bibnamefont
  {Wen}}, \bibinfo {author} {\bibfnamefont {G.~D.}\ \bibnamefont {Gu}},
  \bibinfo {author} {\bibfnamefont {H.}~\bibnamefont {Ding}},\ and\ \bibinfo
  {author} {\bibfnamefont {S.}~\bibnamefont {Shin}},\ }\bibfield  {title}
  {\bibinfo {title} {Observation of topological superconductivity on the
  surface of an iron-based superconductor},\ }\href
  {https://doi.org/10.1126/science.aan4596} {\bibfield  {journal} {\bibinfo
  {journal} {Science}\ }\textbf {\bibinfo {volume} {360}},\ \bibinfo {pages}
  {182} (\bibinfo {year} {2018})}\BibitemShut {NoStop}%
\bibitem [{\citenamefont {Wang}\ \emph {et~al.}(2018)\citenamefont {Wang},
  \citenamefont {Kong}, \citenamefont {Fan}, \citenamefont {Chen},
  \citenamefont {Zhu}, \citenamefont {Liu}, \citenamefont {Cao}, \citenamefont
  {Sun}, \citenamefont {Du}, \citenamefont {Schneeloch}, \citenamefont {Zhong},
  \citenamefont {Gu}, \citenamefont {Fu}, \citenamefont {Ding},\ and\
  \citenamefont {Gao}}]{Wang2018Evidence}%
  \BibitemOpen
  \bibfield  {author} {\bibinfo {author} {\bibfnamefont {D.}~\bibnamefont
  {Wang}}, \bibinfo {author} {\bibfnamefont {L.}~\bibnamefont {Kong}}, \bibinfo
  {author} {\bibfnamefont {P.}~\bibnamefont {Fan}}, \bibinfo {author}
  {\bibfnamefont {H.}~\bibnamefont {Chen}}, \bibinfo {author} {\bibfnamefont
  {S.}~\bibnamefont {Zhu}}, \bibinfo {author} {\bibfnamefont {W.}~\bibnamefont
  {Liu}}, \bibinfo {author} {\bibfnamefont {L.}~\bibnamefont {Cao}}, \bibinfo
  {author} {\bibfnamefont {Y.}~\bibnamefont {Sun}}, \bibinfo {author}
  {\bibfnamefont {S.}~\bibnamefont {Du}}, \bibinfo {author} {\bibfnamefont
  {J.}~\bibnamefont {Schneeloch}}, \bibinfo {author} {\bibfnamefont
  {R.}~\bibnamefont {Zhong}}, \bibinfo {author} {\bibfnamefont
  {G.}~\bibnamefont {Gu}}, \bibinfo {author} {\bibfnamefont {L.}~\bibnamefont
  {Fu}}, \bibinfo {author} {\bibfnamefont {H.}~\bibnamefont {Ding}},\ and\
  \bibinfo {author} {\bibfnamefont {H.-J.}\ \bibnamefont {Gao}},\ }\bibfield
  {title} {\bibinfo {title} {Evidence for {Majorana} bound states in an
  iron-based superconductor},\ }\href {https://doi.org/10.1126/science.aao1797}
  {\bibfield  {journal} {\bibinfo  {journal} {Science}\ }\textbf {\bibinfo
  {volume} {362}},\ \bibinfo {pages} {333} (\bibinfo {year}
  {2018})}\BibitemShut {NoStop}%
\bibitem [{\citenamefont {Machida}\ \emph {et~al.}(2019)\citenamefont
  {Machida}, \citenamefont {Sun}, \citenamefont {Pyon}, \citenamefont {Takeda},
  \citenamefont {Kohsaka}, \citenamefont {Hanaguri}, \citenamefont {Sasagawa},\
  and\ \citenamefont {Tamegai}}]{machida2019zero}%
  \BibitemOpen
  \bibfield  {author} {\bibinfo {author} {\bibfnamefont {T.}~\bibnamefont
  {Machida}}, \bibinfo {author} {\bibfnamefont {Y.}~\bibnamefont {Sun}},
  \bibinfo {author} {\bibfnamefont {S.}~\bibnamefont {Pyon}}, \bibinfo {author}
  {\bibfnamefont {S.}~\bibnamefont {Takeda}}, \bibinfo {author} {\bibfnamefont
  {Y.}~\bibnamefont {Kohsaka}}, \bibinfo {author} {\bibfnamefont
  {T.}~\bibnamefont {Hanaguri}}, \bibinfo {author} {\bibfnamefont
  {T.}~\bibnamefont {Sasagawa}},\ and\ \bibinfo {author} {\bibfnamefont
  {T.}~\bibnamefont {Tamegai}},\ }\bibfield  {title} {\bibinfo {title}
  {Zero-energy vortex bound state in the superconducting topological surface
  state of {Fe(Se,Te)}},\ }\href
  {https://www.nature.com/articles/s41563-019-0397-1} {\bibfield  {journal}
  {\bibinfo  {journal} {Nature materials}\ }\textbf {\bibinfo {volume} {18}},\
  \bibinfo {pages} {811} (\bibinfo {year} {2019})}\BibitemShut {NoStop}%
\bibitem [{\citenamefont {Kong}\ \emph {et~al.}(2019)\citenamefont {Kong},
  \citenamefont {Zhu}, \citenamefont {Papaj}, \citenamefont {Chen},
  \citenamefont {Cao}, \citenamefont {Isobe}, \citenamefont {Xing},
  \citenamefont {Liu}, \citenamefont {Wang}, \citenamefont {Fan} \emph
  {et~al.}}]{kong2019half}%
  \BibitemOpen
  \bibfield  {author} {\bibinfo {author} {\bibfnamefont {L.}~\bibnamefont
  {Kong}}, \bibinfo {author} {\bibfnamefont {S.}~\bibnamefont {Zhu}}, \bibinfo
  {author} {\bibfnamefont {M.}~\bibnamefont {Papaj}}, \bibinfo {author}
  {\bibfnamefont {H.}~\bibnamefont {Chen}}, \bibinfo {author} {\bibfnamefont
  {L.}~\bibnamefont {Cao}}, \bibinfo {author} {\bibfnamefont {H.}~\bibnamefont
  {Isobe}}, \bibinfo {author} {\bibfnamefont {Y.}~\bibnamefont {Xing}},
  \bibinfo {author} {\bibfnamefont {W.}~\bibnamefont {Liu}}, \bibinfo {author}
  {\bibfnamefont {D.}~\bibnamefont {Wang}}, \bibinfo {author} {\bibfnamefont
  {P.}~\bibnamefont {Fan}}, \emph {et~al.},\ }\bibfield  {title} {\bibinfo
  {title} {Half-integer level shift of vortex bound states in an iron-based
  superconductor},\ }\href {https://www.nature.com/articles/s41567-019-0630-5}
  {\bibfield  {journal} {\bibinfo  {journal} {Nature Physics}\ }\textbf
  {\bibinfo {volume} {15}},\ \bibinfo {pages} {1181} (\bibinfo {year}
  {2019})}\BibitemShut {NoStop}%
\bibitem [{\citenamefont {Zhu}\ \emph {et~al.}(2020)\citenamefont {Zhu},
  \citenamefont {Kong}, \citenamefont {Cao}, \citenamefont {Chen},
  \citenamefont {Papaj}, \citenamefont {Du}, \citenamefont {Xing},
  \citenamefont {Liu}, \citenamefont {Wang}, \citenamefont {Shen},
  \citenamefont {Yang}, \citenamefont {Schneeloch}, \citenamefont {Zhong},
  \citenamefont {Gu}, \citenamefont {Fu}, \citenamefont {Zhang}, \citenamefont
  {Ding},\ and\ \citenamefont {Gao}}]{Zhu2020Nearly}%
  \BibitemOpen
  \bibfield  {author} {\bibinfo {author} {\bibfnamefont {S.}~\bibnamefont
  {Zhu}}, \bibinfo {author} {\bibfnamefont {L.}~\bibnamefont {Kong}}, \bibinfo
  {author} {\bibfnamefont {L.}~\bibnamefont {Cao}}, \bibinfo {author}
  {\bibfnamefont {H.}~\bibnamefont {Chen}}, \bibinfo {author} {\bibfnamefont
  {M.}~\bibnamefont {Papaj}}, \bibinfo {author} {\bibfnamefont
  {S.}~\bibnamefont {Du}}, \bibinfo {author} {\bibfnamefont {Y.}~\bibnamefont
  {Xing}}, \bibinfo {author} {\bibfnamefont {W.}~\bibnamefont {Liu}}, \bibinfo
  {author} {\bibfnamefont {D.}~\bibnamefont {Wang}}, \bibinfo {author}
  {\bibfnamefont {C.}~\bibnamefont {Shen}}, \bibinfo {author} {\bibfnamefont
  {F.}~\bibnamefont {Yang}}, \bibinfo {author} {\bibfnamefont {J.}~\bibnamefont
  {Schneeloch}}, \bibinfo {author} {\bibfnamefont {R.}~\bibnamefont {Zhong}},
  \bibinfo {author} {\bibfnamefont {G.}~\bibnamefont {Gu}}, \bibinfo {author}
  {\bibfnamefont {L.}~\bibnamefont {Fu}}, \bibinfo {author} {\bibfnamefont
  {Y.-Y.}\ \bibnamefont {Zhang}}, \bibinfo {author} {\bibfnamefont
  {H.}~\bibnamefont {Ding}},\ and\ \bibinfo {author} {\bibfnamefont {H.-J.}\
  \bibnamefont {Gao}},\ }\bibfield  {title} {\bibinfo {title} {Nearly quantized
  conductance plateau of vortex zero mode in an iron-based superconductor},\
  }\href {https://doi.org/10.1126/science.aax0274} {\bibfield  {journal}
  {\bibinfo  {journal} {Science}\ }\textbf {\bibinfo {volume} {367}},\ \bibinfo
  {pages} {189} (\bibinfo {year} {2020})}\BibitemShut {NoStop}%
\bibitem [{\citenamefont {Li}\ \emph {et~al.}(2022)\citenamefont {Li},
  \citenamefont {Li}, \citenamefont {Cao}, \citenamefont {Zhou}, \citenamefont
  {Wang}, \citenamefont {Jin}, \citenamefont {Chiu}, \citenamefont {Pennycook},
  \citenamefont {Wang},\ and\ \citenamefont {Gao}}]{li2022ordered}%
  \BibitemOpen
  \bibfield  {author} {\bibinfo {author} {\bibfnamefont {M.}~\bibnamefont
  {Li}}, \bibinfo {author} {\bibfnamefont {G.}~\bibnamefont {Li}}, \bibinfo
  {author} {\bibfnamefont {L.}~\bibnamefont {Cao}}, \bibinfo {author}
  {\bibfnamefont {X.}~\bibnamefont {Zhou}}, \bibinfo {author} {\bibfnamefont
  {X.}~\bibnamefont {Wang}}, \bibinfo {author} {\bibfnamefont {C.}~\bibnamefont
  {Jin}}, \bibinfo {author} {\bibfnamefont {C.-K.}\ \bibnamefont {Chiu}},
  \bibinfo {author} {\bibfnamefont {S.~J.}\ \bibnamefont {Pennycook}}, \bibinfo
  {author} {\bibfnamefont {Z.}~\bibnamefont {Wang}},\ and\ \bibinfo {author}
  {\bibfnamefont {H.-J.}\ \bibnamefont {Gao}},\ }\bibfield  {title} {\bibinfo
  {title} {Ordered and tunable {Majorana}-zero-mode lattice in naturally
  strained {LiFeAs}},\ }\href
  {https://www.nature.com/articles/s41586-022-04744-8} {\bibfield  {journal}
  {\bibinfo  {journal} {Nature}\ }\textbf {\bibinfo {volume} {606}},\ \bibinfo
  {pages} {890} (\bibinfo {year} {2022})}\BibitemShut {NoStop}%
\bibitem [{\citenamefont {Kitaev}(2001)}]{Kitaev2001Unpaired}%
  \BibitemOpen
  \bibfield  {author} {\bibinfo {author} {\bibfnamefont {A.~Y.}\ \bibnamefont
  {Kitaev}},\ }\bibfield  {title} {\bibinfo {title} {Unpaired {Majorana}
  fermions in quantum wires},\ }\href
  {https://doi.org/10.1070/1063-7869/44/10s/s29} {\bibfield  {journal}
  {\bibinfo  {journal} {Physics-Uspekhi}\ }\textbf {\bibinfo {volume} {44}},\
  \bibinfo {pages} {131–136} (\bibinfo {year} {2001})}\BibitemShut {NoStop}%
\bibitem [{\citenamefont {Bonderson}\ \emph {et~al.}(2008)\citenamefont
  {Bonderson}, \citenamefont {Freedman},\ and\ \citenamefont
  {Nayak}}]{Bonderson2008Measurement}%
  \BibitemOpen
  \bibfield  {author} {\bibinfo {author} {\bibfnamefont {P.}~\bibnamefont
  {Bonderson}}, \bibinfo {author} {\bibfnamefont {M.}~\bibnamefont
  {Freedman}},\ and\ \bibinfo {author} {\bibfnamefont {C.}~\bibnamefont
  {Nayak}},\ }\bibfield  {title} {\bibinfo {title} {Measurement-only
  topological quantum computation},\ }\href
  {https://doi.org/10.1103/PhysRevLett.101.010501} {\bibfield  {journal}
  {\bibinfo  {journal} {Phys. Rev. Lett.}\ }\textbf {\bibinfo {volume} {101}},\
  \bibinfo {pages} {010501} (\bibinfo {year} {2008})}\BibitemShut {NoStop}%
\bibitem [{\citenamefont {Nayak}\ \emph {et~al.}(2008)\citenamefont {Nayak},
  \citenamefont {Simon}, \citenamefont {Stern}, \citenamefont {Freedman},\ and\
  \citenamefont {Das~Sarma}}]{Nayak2008Non}%
  \BibitemOpen
  \bibfield  {author} {\bibinfo {author} {\bibfnamefont {C.}~\bibnamefont
  {Nayak}}, \bibinfo {author} {\bibfnamefont {S.~H.}\ \bibnamefont {Simon}},
  \bibinfo {author} {\bibfnamefont {A.}~\bibnamefont {Stern}}, \bibinfo
  {author} {\bibfnamefont {M.}~\bibnamefont {Freedman}},\ and\ \bibinfo
  {author} {\bibfnamefont {S.}~\bibnamefont {Das~Sarma}},\ }\bibfield  {title}
  {\bibinfo {title} {Non-abelian anyons and topological quantum computation},\
  }\href {https://doi.org/10.1103/RevModPhys.80.1083} {\bibfield  {journal}
  {\bibinfo  {journal} {Rev. Mod. Phys.}\ }\textbf {\bibinfo {volume} {80}},\
  \bibinfo {pages} {1083} (\bibinfo {year} {2008})}\BibitemShut {NoStop}%
\bibitem [{\citenamefont {Alicea}\ \emph {et~al.}(2011)\citenamefont {Alicea},
  \citenamefont {Oreg}, \citenamefont {Refael}, \citenamefont {Von~Oppen},\
  and\ \citenamefont {Fisher}}]{alicea2011non}%
  \BibitemOpen
  \bibfield  {author} {\bibinfo {author} {\bibfnamefont {J.}~\bibnamefont
  {Alicea}}, \bibinfo {author} {\bibfnamefont {Y.}~\bibnamefont {Oreg}},
  \bibinfo {author} {\bibfnamefont {G.}~\bibnamefont {Refael}}, \bibinfo
  {author} {\bibfnamefont {F.}~\bibnamefont {Von~Oppen}},\ and\ \bibinfo
  {author} {\bibfnamefont {M.~P.}\ \bibnamefont {Fisher}},\ }\bibfield  {title}
  {\bibinfo {title} {Non-abelian statistics and topological quantum information
  processing in 1d wire networks},\ }\href
  {https://www.nature.com/articles/nphys1915} {\bibfield  {journal} {\bibinfo
  {journal} {Nature Physics}\ }\textbf {\bibinfo {volume} {7}},\ \bibinfo
  {pages} {412} (\bibinfo {year} {2011})}\BibitemShut {NoStop}%
\bibitem [{\citenamefont {Alicea}(2012)}]{Alicea2012New}%
  \BibitemOpen
  \bibfield  {author} {\bibinfo {author} {\bibfnamefont {J.}~\bibnamefont
  {Alicea}},\ }\bibfield  {title} {\bibinfo {title} {New directions in the
  pursuit of {Majorana} fermions in solid state systems},\ }\href
  {https://doi.org/10.1088/0034-4885/75/7/076501} {\bibfield  {journal}
  {\bibinfo  {journal} {Reports on Progress in Physics}\ }\textbf {\bibinfo
  {volume} {75}},\ \bibinfo {pages} {076501} (\bibinfo {year}
  {2012})}\BibitemShut {NoStop}%
\bibitem [{\citenamefont {Beenakker}(2013)}]{Beenakker2013Search}%
  \BibitemOpen
  \bibfield  {author} {\bibinfo {author} {\bibfnamefont {C.}~\bibnamefont
  {Beenakker}},\ }\bibfield  {title} {\bibinfo {title} {Search for {Majorana}
  fermions in superconductors},\ }\href
  {https://doi.org/https://doi.org/10.1146/annurev-conmatphys-030212-184337}
  {\bibfield  {journal} {\bibinfo  {journal} {Annual Review of Condensed Matter
  Physics}\ }\textbf {\bibinfo {volume} {4}},\ \bibinfo {pages} {113} (\bibinfo
  {year} {2013})}\BibitemShut {NoStop}%
\bibitem [{\citenamefont {Das~Sarma}\ \emph {et~al.}(2015)\citenamefont
  {Das~Sarma}, \citenamefont {Freedman},\ and\ \citenamefont
  {Nayak}}]{sarma2015majorana}%
  \BibitemOpen
  \bibfield  {author} {\bibinfo {author} {\bibfnamefont {S.}~\bibnamefont
  {Das~Sarma}}, \bibinfo {author} {\bibfnamefont {M.}~\bibnamefont
  {Freedman}},\ and\ \bibinfo {author} {\bibfnamefont {C.}~\bibnamefont
  {Nayak}},\ }\bibfield  {title} {\bibinfo {title} {Majorana zero modes and
  topological quantum computation},\ }\href
  {https://www.nature.com/articles/npjqi20151} {\bibfield  {journal} {\bibinfo
  {journal} {npj Quantum Information}\ }\textbf {\bibinfo {volume} {1}},\
  \bibinfo {pages} {15001} (\bibinfo {year} {2015})}\BibitemShut {NoStop}%
\bibitem [{\citenamefont {Aasen}\ \emph {et~al.}(2016)\citenamefont {Aasen},
  \citenamefont {Hell}, \citenamefont {Mishmash}, \citenamefont {Higginbotham},
  \citenamefont {Danon}, \citenamefont {Leijnse}, \citenamefont {Jespersen},
  \citenamefont {Folk}, \citenamefont {Marcus}, \citenamefont {Flensberg},\
  and\ \citenamefont {Alicea}}]{Aasen2016Milestones}%
  \BibitemOpen
  \bibfield  {author} {\bibinfo {author} {\bibfnamefont {D.}~\bibnamefont
  {Aasen}}, \bibinfo {author} {\bibfnamefont {M.}~\bibnamefont {Hell}},
  \bibinfo {author} {\bibfnamefont {R.~V.}\ \bibnamefont {Mishmash}}, \bibinfo
  {author} {\bibfnamefont {A.}~\bibnamefont {Higginbotham}}, \bibinfo {author}
  {\bibfnamefont {J.}~\bibnamefont {Danon}}, \bibinfo {author} {\bibfnamefont
  {M.}~\bibnamefont {Leijnse}}, \bibinfo {author} {\bibfnamefont {T.~S.}\
  \bibnamefont {Jespersen}}, \bibinfo {author} {\bibfnamefont {J.~A.}\
  \bibnamefont {Folk}}, \bibinfo {author} {\bibfnamefont {C.~M.}\ \bibnamefont
  {Marcus}}, \bibinfo {author} {\bibfnamefont {K.}~\bibnamefont {Flensberg}},\
  and\ \bibinfo {author} {\bibfnamefont {J.}~\bibnamefont {Alicea}},\
  }\bibfield  {title} {\bibinfo {title} {Milestones toward {Majorana}-based
  quantum computing},\ }\href {https://doi.org/10.1103/PhysRevX.6.031016}
  {\bibfield  {journal} {\bibinfo  {journal} {Phys. Rev. X}\ }\textbf {\bibinfo
  {volume} {6}},\ \bibinfo {pages} {031016} (\bibinfo {year}
  {2016})}\BibitemShut {NoStop}%
\bibitem [{\citenamefont {Karzig}\ \emph {et~al.}(2017)\citenamefont {Karzig},
  \citenamefont {Knapp}, \citenamefont {Lutchyn}, \citenamefont {Bonderson},
  \citenamefont {Hastings}, \citenamefont {Nayak}, \citenamefont {Alicea},
  \citenamefont {Flensberg}, \citenamefont {Plugge}, \citenamefont {Oreg},
  \citenamefont {Marcus},\ and\ \citenamefont {Freedman}}]{Karzig2017Scalable}%
  \BibitemOpen
  \bibfield  {author} {\bibinfo {author} {\bibfnamefont {T.}~\bibnamefont
  {Karzig}}, \bibinfo {author} {\bibfnamefont {C.}~\bibnamefont {Knapp}},
  \bibinfo {author} {\bibfnamefont {R.~M.}\ \bibnamefont {Lutchyn}}, \bibinfo
  {author} {\bibfnamefont {P.}~\bibnamefont {Bonderson}}, \bibinfo {author}
  {\bibfnamefont {M.~B.}\ \bibnamefont {Hastings}}, \bibinfo {author}
  {\bibfnamefont {C.}~\bibnamefont {Nayak}}, \bibinfo {author} {\bibfnamefont
  {J.}~\bibnamefont {Alicea}}, \bibinfo {author} {\bibfnamefont
  {K.}~\bibnamefont {Flensberg}}, \bibinfo {author} {\bibfnamefont
  {S.}~\bibnamefont {Plugge}}, \bibinfo {author} {\bibfnamefont
  {Y.}~\bibnamefont {Oreg}}, \bibinfo {author} {\bibfnamefont {C.~M.}\
  \bibnamefont {Marcus}},\ and\ \bibinfo {author} {\bibfnamefont {M.~H.}\
  \bibnamefont {Freedman}},\ }\bibfield  {title} {\bibinfo {title} {Scalable
  designs for quasiparticle-poisoning-protected topological quantum computation
  with {Majorana} zero modes},\ }\href
  {https://doi.org/10.1103/PhysRevB.95.235305} {\bibfield  {journal} {\bibinfo
  {journal} {Phys. Rev. B}\ }\textbf {\bibinfo {volume} {95}},\ \bibinfo
  {pages} {235305} (\bibinfo {year} {2017})}\BibitemShut {NoStop}%
\bibitem [{\citenamefont {Chiu}\ \emph {et~al.}(2015)\citenamefont {Chiu},
  \citenamefont {Pikulin},\ and\ \citenamefont {Franz}}]{Chiu2015Strongly}%
  \BibitemOpen
  \bibfield  {author} {\bibinfo {author} {\bibfnamefont {C.-K.}\ \bibnamefont
  {Chiu}}, \bibinfo {author} {\bibfnamefont {D.~I.}\ \bibnamefont {Pikulin}},\
  and\ \bibinfo {author} {\bibfnamefont {M.}~\bibnamefont {Franz}},\ }\bibfield
   {title} {\bibinfo {title} {Strongly interacting {Majorana} fermions},\
  }\href {https://doi.org/10.1103/PhysRevB.91.165402} {\bibfield  {journal}
  {\bibinfo  {journal} {Phys. Rev. B}\ }\textbf {\bibinfo {volume} {91}},\
  \bibinfo {pages} {165402} (\bibinfo {year} {2015})}\BibitemShut {NoStop}%
\bibitem [{\citenamefont {Christian}\ \emph {et~al.}(2021)\citenamefont
  {Christian}, \citenamefont {Dumitrescu},\ and\ \citenamefont
  {Hal\'asz}}]{Christian2021Robustness}%
  \BibitemOpen
  \bibfield  {author} {\bibinfo {author} {\bibfnamefont {C.}~\bibnamefont
  {Christian}}, \bibinfo {author} {\bibfnamefont {E.~F.}\ \bibnamefont
  {Dumitrescu}},\ and\ \bibinfo {author} {\bibfnamefont {G.~B.}\ \bibnamefont
  {Hal\'asz}},\ }\bibfield  {title} {\bibinfo {title} {Robustness of
  vortex-bound {Majorana} zero modes against correlated disorder},\ }\href
  {https://doi.org/10.1103/PhysRevB.104.L020505} {\bibfield  {journal}
  {\bibinfo  {journal} {Phys. Rev. B}\ }\textbf {\bibinfo {volume} {104}},\
  \bibinfo {pages} {L020505} (\bibinfo {year} {2021})}\BibitemShut {NoStop}%
\bibitem [{\citenamefont {Pathak}\ \emph {et~al.}(2021)\citenamefont {Pathak},
  \citenamefont {Plugge},\ and\ \citenamefont {Franz}}]{Vedangi2021Majorana}%
  \BibitemOpen
  \bibfield  {author} {\bibinfo {author} {\bibfnamefont {V.}~\bibnamefont
  {Pathak}}, \bibinfo {author} {\bibfnamefont {S.}~\bibnamefont {Plugge}},\
  and\ \bibinfo {author} {\bibfnamefont {M.}~\bibnamefont {Franz}},\ }\bibfield
   {title} {\bibinfo {title} {Majorana bound states in vortex lattices on
  iron-based superconductors},\ }\href
  {https://doi.org/https://doi.org/10.1016/j.aop.2021.168431} {\bibfield
  {journal} {\bibinfo  {journal} {Annals of Physics}\ }\textbf {\bibinfo
  {volume} {435}},\ \bibinfo {pages} {168431} (\bibinfo {year} {2021})},\
  \bibinfo {note} {special issue on Philip W. Anderson}\BibitemShut {NoStop}%
\bibitem [{\citenamefont {Ariad}\ \emph {et~al.}(2018)\citenamefont {Ariad},
  \citenamefont {Avishai},\ and\ \citenamefont {Grosfeld}}]{Ariad2018How}%
  \BibitemOpen
  \bibfield  {author} {\bibinfo {author} {\bibfnamefont {D.}~\bibnamefont
  {Ariad}}, \bibinfo {author} {\bibfnamefont {Y.}~\bibnamefont {Avishai}},\
  and\ \bibinfo {author} {\bibfnamefont {E.}~\bibnamefont {Grosfeld}},\
  }\bibfield  {title} {\bibinfo {title} {How vortex bound states affect the
  hall conductivity of a chiral $p\ifmmode\pm\else\textpm\fi{}ip$
  superconductor},\ }\href {https://doi.org/10.1103/PhysRevB.98.104511}
  {\bibfield  {journal} {\bibinfo  {journal} {Phys. Rev. B}\ }\textbf {\bibinfo
  {volume} {98}},\ \bibinfo {pages} {104511} (\bibinfo {year}
  {2018})}\BibitemShut {NoStop}%
\bibitem [{\citenamefont {Sandier}\ and\ \citenamefont
  {Serfaty}(2012)}]{sandier2012ginzburg}%
  \BibitemOpen
  \bibfield  {author} {\bibinfo {author} {\bibfnamefont {E.}~\bibnamefont
  {Sandier}}\ and\ \bibinfo {author} {\bibfnamefont {S.}~\bibnamefont
  {Serfaty}},\ }\bibfield  {title} {\bibinfo {title} {From the ginzburg-landau
  model to vortex lattice problems},\ }\href
  {https://doi.org/10.1007/s00220-012-1508-x} {\bibfield  {journal} {\bibinfo
  {journal} {Communications in Mathematical Physics}\ }\textbf {\bibinfo
  {volume} {313}},\ \bibinfo {pages} {635} (\bibinfo {year}
  {2012})}\BibitemShut {NoStop}%
\bibitem [{\citenamefont {Zhang}(2015)}]{Zhang2015On}%
  \BibitemOpen
  \bibfield  {author} {\bibinfo {author} {\bibfnamefont {P.}~\bibnamefont
  {Zhang}},\ }\bibfield  {title} {\bibinfo {title} {On the minimizer of a
  renormalized energy related to the ginzburg–landau model},\ }\href
  {https://doi.org/https://doi.org/10.1016/j.crma.2015.01.001} {\bibfield
  {journal} {\bibinfo  {journal} {Comptes Rendus Mathematique}\ }\textbf
  {\bibinfo {volume} {353}},\ \bibinfo {pages} {255} (\bibinfo {year}
  {2015})}\BibitemShut {NoStop}%
\end{thebibliography}%
\let\addcontentsline\oldaddcontentsline

\clearpage


\setcounter{section}{0}
\setcounter{subsection}{0}
\setcounter{secnumdepth}{2}
\setcounter{tocdepth}{2}

\renewcommand{\thesection}{\Roman{section}}
\renewcommand{\thesubsection}{\Alph{subsection}}

\setcounter{table}{0}
\renewcommand{\thetable}{S\arabic{table}}
\setcounter{figure}{0}
\renewcommand{\thefigure}{S\arabic{figure}}
\setcounter{equation}{0}
\renewcommand{\theequation}{S\arabic{equation}}

\begin{widetext}

\begin{center}
\textbf{Supplemental Material for ``Majorana zero modes in half-quantum vortices of pair density wave superconductors''}
\end{center}

\setcounter{table}{0}
\renewcommand{\thetable}{S\arabic{table}}
\setcounter{figure}{0}
\renewcommand{\thefigure}{S\arabic{figure}}
\setcounter{equation}{0}
\renewcommand{\theequation}{S\arabic{equation}}


\tableofcontents

\section{Extraction of the phase stiffnesses $\rho$ and $\kappa$}
\label{secS:stiffness}

The two PDW components are parameterized as
\begin{equation}
    \Delta_{\alpha,\pm}=|\Delta_{\alpha,\pm}|e^{i(\phi_{\rm sc}\pm\phi_K)}.
\label{eqS:phase_decomposition}
\end{equation}
The common phase $\phi_{\rm sc}$ is charged under the microscopic $U(1)$ symmetry, whereas $\phi_K$ is the neutral relative phase that translates the PDW modulation.
Because these two deformations have different microscopic implementations, we calculate their energy curvatures separately.

\subsection{Superfluid stiffness}

We impose a uniform superconducting phase gradient $\phi_{\rm sc}(\bm r)=\bm q\cdot\bm r$.
For a bond $(i,j)$ centered at $\bm R_{ij}=(\bm r_i+\bm r_j)/2$, the pairing field therefore acquires $\Delta_{ij}(\bm q)=\Delta_{ij}(0)e^{i\bm q\cdot\bm R_{ij}}$.
A charge gauge transformation assigning half of the pair phase to each fermion removes this spatial dependence from the pairing field and transfers it to the hopping.
In our directed-bond convention, a hopping from $j$ to $i$ becomes
\begin{equation}
    t_{ij}(\bm q)=t_{ij}(0)\exp\left[\frac{i}{2}\bm q\cdot(\bm r_i-\bm r_j)\right].
\label{eqS:twisted_hopping}
\end{equation}
The density interactions themselves are unchanged. The small-$q$ energy density is then fitted to
\begin{equation}
    e_{\rm sc}(\bm q)-e(0)=\frac{\rho}{2}q^2+O(q^4).
\label{eqS:rho_fit}
\end{equation}
Numerically, we use symmetric fermionic twists $\Phi=0,\pm0.02,\pm0.04,\pm0.06$ along one direction, implemented in a uniform gauge with periodic Bloch boundary conditions.
In our lattice convention, these correspond to $q_{\rm sc}=4\Phi/9$, and $\rho$ is extracted from the quadratic fit in Eq.~\eqref{eqS:rho_fit}.
The calculation is repeated for $N_k=8,10,12,14,16$, and the thermodynamic-limit stiffness is obtained from $\rho(N_k)=\rho_\infty+a_\rho N_k^{-2}$.

We use three levels of variational relaxation.
In the fully self-consistent calculation, denoted by $\rho_{\rm scf}$, all HFB fields and the chemical potential are reoptimized for every imposed $\bm q$. 
In the auxiliary-BdG calculation, the undeformed mean-field saddle is used to construct the deformed quadratic Hamiltonian, which is diagonalized once without further self-consistent iteration, and the corresponding interacting energy defines $\rho_{\rm aux}$.
Finally, in the rigid-state calculation, the undeformed Gaussian state itself is kept fixed while the Peierls-twisted interacting Hamiltonian is evaluated, defining $\rho_{\rm rigid}$. 
The increasing amount of variational freedom gives
\begin{equation}
    \rho_{\rm scf}\leq\rho_{\rm aux}\leq\rho_{\rm rigid},
\label{eqS:rho_hierarchy}
\end{equation}
and the inequalities are numerically strict in the resolved PDW regime, as shown in the main text.
The resulting fully self-consistent thermodynamic-limit stiffness $\rho_{\rm scf}$ across the PDW region is shown in Fig.~\ref{figS:stiffness}(a).

\begin{figure}[t]
    \centering
    \includegraphics[width=0.6\columnwidth]{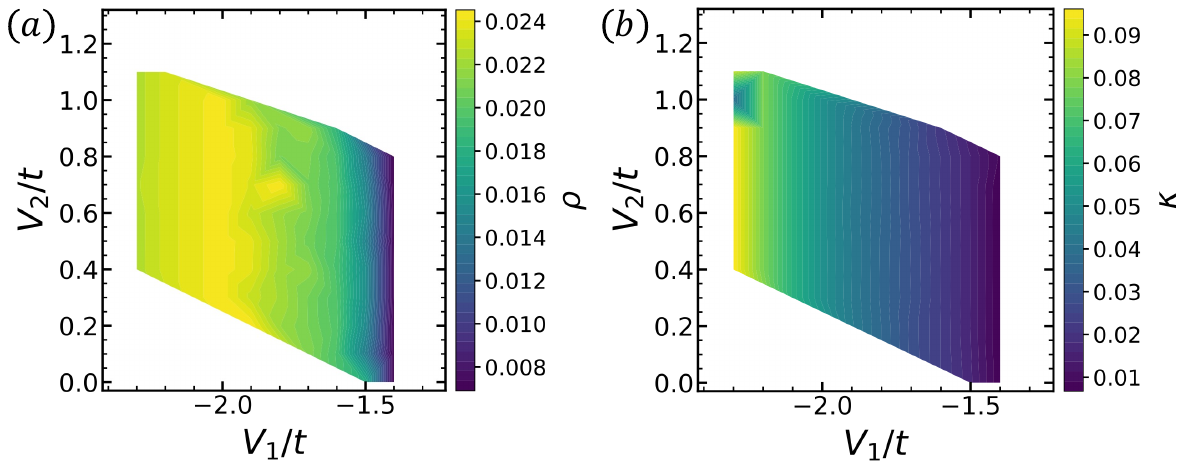}
    \caption{\textbf{Thermodynamic PDW stiffnesses across the PDW phase.}
    Thermodynamic-limit fully self-consistent values of (a) the superfluid stiffness $\rho$ and (b) the PDW relative-phase stiffness $\kappa$.
    Both soften toward the PDW--DSM boundary, with $\kappa$ decreasing more rapidly than $\rho$.}
\label{figS:stiffness}
\end{figure}

\subsection{PDW relative-phase stiffness}

A uniform relative-phase gradient $\phi_K(\bm r)=\bm q\cdot\bm r$ changes the two PDW harmonics from $\pm\bm K$ to $\pm(\bm K+\bm q)$. 
This deformation cannot be removed by a microscopic charge gauge transformation. 
Moreover, a generic $\bm q$ is not compatible with the original $3\times3$ periodicity.
We therefore keep real space along the direction of the imposed modulation and momentum space in the transverse direction.

During every self-consistency step, the pairing field on each bond orientation is projected onto the shifted harmonic subspace
\begin{equation}
    \mathcal B_{\bm q}=\operatorname{span}\left\{1,e^{i(\bm K+\bm q)\cdot\bm r},e^{-i(\bm K+\bm q)\cdot\bm r}\right\}.
\label{eqS:constrained_basis}
\end{equation}
The projected pairing texture is mixed into the next iteration, while the remaining density, exchange, and pairing degrees of freedom are allowed to relax subject to the prescribed ordering wave vectors.
The resulting energy curvature defines
\begin{equation}
    e_K(\bm q)-e(0)=\frac{\kappa}{2}q^2+O(q^4).
\label{eqS:kappa_fit}
\end{equation}
This fully relaxed result is denoted $\kappa_{\rm scf}$.
Numerically, we impose the deformation along one direction on a periodic real-space strip of length $L_r=3N_k$. 
The total relative-phase twists are chosen as $\Phi_K=2\pi m$ with $m=0,\pm1,\pm2,\pm3$, corresponding to a modulation $q_{\rm lat}=\Phi_K/L_r$. 
In our lattice convention, the physical gradient entering Eq.~\eqref{eqS:kappa_fit} satisfies $q_K=2\Phi_K/3L_r$.
The calculation is repeated for $N_k=8,10,12,14,16$, and the thermodynamic-limit stiffness is obtained from $\kappa(N_k)=\kappa_\infty+a_\kappa N_k^{-2}$.

For comparison, $\kappa_{\rm aux}$ is obtained by imposing the shifted PDW texture on the undeformed saddle, diagonalizing the resulting BdG Hamiltonian once, and evaluating the corresponding interacting energy without further relaxation. 
There is no useful rigid-state analogue of $\kappa_{\rm rigid}$ because the relative-phase deformation is not generated by a microscopic gauge transformation of the bare interacting Hamiltonian.
We find
\begin{equation}
    \kappa_{\rm scf}\leq\kappa_{\rm aux},
\label{eqS:kappa_hierarchy}
\end{equation}
with a sizable reduction from internal mean-field relaxation, as illustrated in the main text.
The resulting fully self-consistent thermodynamic-limit stiffness $\kappa_{\rm scf}$ across the PDW region is shown in Fig.~\ref{figS:stiffness}(b).

\section{Microscopic Ginzburg--Landau couplings and vortex-core energetics}
\label{secS:quartic_core}

\subsection{Microscopic extraction of the quartic couplings $u$ and $g$}
\label{secS:quartic}

The local two-component Ginzburg--Landau potential used in the main text is
\begin{equation}
f_{\rm pot}=r\left(|\Delta_+|^2+|\Delta_-|^2\right)
+\frac{u}{2}\left(|\Delta_+|^4+|\Delta_-|^4\right)
+g|\Delta_+|^2|\Delta_-|^2.
\label{eqS:GL_potential}
\end{equation}
We extract the local quartic curvatures directly from the microscopic mean-field state by a fixed-background amplitude-response calculation.

For each converged PDW saddle, we define $A_+$ and $A_-$ by projecting the nearest-neighbor anomalous expectation values $F_{ij}=\langle c_i c_j\rangle$ onto the two symmetry-related PDW harmonics at $+\bm K$ and $-\bm K$.
The absolute normalization of $A_\pm$ fixes the units of $u$ and $g$, while their ratio $g/u$ is independent of an overall rescaling of the projected amplitudes.

Starting from the converged saddle, we rescale the two PDW components while keeping the remaining mean-field background fixed.
We use symmetric amplitude deformations $\epsilon=0,\pm0.025,\pm0.050,\pm0.075,\pm0.100$. 
For the common-amplitude channel, both PDW components are rescaled as $A_\pm\rightarrow(1+\epsilon)A_\pm$, while for the relative-amplitude channel we use $A_\pm\rightarrow(1\pm\epsilon)A_\pm$.
For each deformation, the modified BdG Hamiltonian is diagonalized once, without a new self-consistent relaxation. 
From the resulting Gaussian state we measure the actual projected amplitudes $A_\pm$ and evaluate the expectation value of the original interacting Hamiltonian.
Thus, this procedure measures the local amplitude curvature around the microscopic saddle while avoiding contamination from relaxation of other HFB channels.

We define $x=|A_+|^2,\ y=|A_-|^2,\ \delta x=x-x_0,\ \delta y=y-y_0$, where $(x_0,y_0)$ are the values at the undeformed saddle. 
The interacting energy density is fitted to
\begin{equation}
    e-e_0=c+\ell(\delta x+\delta y)+\frac{u}{2}\left(\delta x^2+\delta y^2\right)+g\,\delta x\delta y.
\label{eqS:ug_fit}
\end{equation}
The constant $c$ and the small linear coefficient $\ell$ absorb residual numerical offsets of the one-shot projected saddle and do not affect the Hessian that determines $u$ and $g$. 
The symmetric deformations predominantly measure the $u+g$ curvature, while the antisymmetric deformations measure $u-g$.
Fitting both sets simultaneously separates the self- and intercomponent quartic couplings.

\begin{figure}[t]
    \centering
    \includegraphics[width=0.6\textwidth]{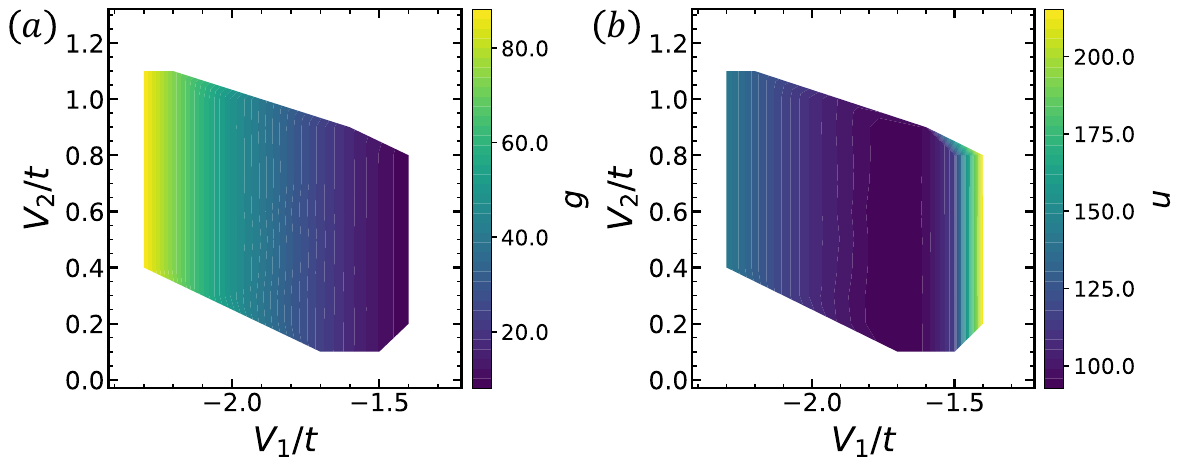}
    \caption{\textbf{Microscopic Ginzburg--Landau quartic couplings.}
    Thermodynamic-limit values of (a) the intercomponent coupling $g$ and (b) the self-interaction $u$, extracted from the fixed-background amplitude response in Eq.~\eqref{eqS:ug_fit}.
    The absolute values correspond to the convention in which the PDW amplitudes are obtained directly from the nearest-neighbor expectation values $\langle c_i c_j\rangle$. 
    The normalization-independent ratio $g/u$ is shown in the main text.}
\label{figS:gu}
\end{figure}

We perform the amplitude-response calculation for $N_k=12,16,20,24,28$, and extrapolate the two coefficients separately according to $u(N_k)=u_\infty+a_u N_k^{-2}$ and $g(N_k)=g_\infty+a_g N_k^{-2}$.
Figure~\ref{figS:gu} shows the resulting thermodynamic-limit values.
The microscopic fits give $u>0$ and $0<g/u<1$ throughout the region used in the main text. 
Consequently, the positive intercomponent coupling $g$ favors an imbalance between the two components when one of them is suppressed.
This is the microscopic origin of the negative half-vortex core contribution discussed in the main text.
We next use these couplings to estimate the corresponding half-vortex core-energy gain and critical splitting scale.

\subsection{Core-energy estimate and critical separation}
\label{secS:core_energy}

We now derive the core-energy estimate used in the main text.
For the uniform symmetric PDW state, minimizing Eq.~\eqref{eqS:GL_potential} gives
\begin{equation}
    |\Delta_+|^2=|\Delta_-|^2=\Delta_0^2=-\frac{r}{u+g},\qquad
    f_0=-\frac{r^2}{u+g}.
\label{eqS:uniform_saddle}
\end{equation}
At the center of an integer vortex, both PDW components are suppressed, so that $f_{\rm IV}=0$.
For a half vortex, only one component vanishes, while the nonsingular component can relax locally. Minimizing the potential with, for example, $\Delta_+=0$ gives
\begin{equation}
    |\Delta_-|^2=-\frac{r}{u},\qquad
    f_{\rm HV}=-\frac{r^2}{2u}.
\end{equation}
Defining the core-energy densities relative to the uniform state as $\delta f_{\rm IV}=f_{\rm IV}-f_0$ and $\delta f_{\rm HV}=f_{\rm HV}-f_0$, the energy-density difference between two half-vortex cores and one integer-vortex core is
\begin{equation}
    \delta f_{\rm core}
    =2\delta f_{\rm HV}-\delta f_{\rm IV}
    =-\frac{gr^2}{u(u+g)}.
\label{eqS:core_density}
\end{equation}
Thus $g>0$ gives $\delta f_{\rm core}<0$, so the local potential favors separating the two component cores.

To convert this local energy-density gain into a vortex-core energy, we approximate the effective core area by $C\pi\xi^2$, where $C$ is an order-one factor that accounts for the detailed spatial profile. 
For the symmetric PDW state, the common gradient coefficient entering the two phase modes is $\alpha_{\rm eff}=\alpha+\lambda\Delta_0^2$, and the stiffnesses satisfy $\rho+\kappa=8\alpha_{\rm eff}\Delta_0^2$.
Using $\Delta_0^2=|r|/(u+g)$ and estimating the coherence length from the balance of gradient and condensation energies, $\xi^2\sim\alpha_{\rm eff}/|r|$, Eq.~\eqref{eqS:core_density} gives
\begin{equation}
    \Delta E_{\rm core}\simeq C\pi\xi^2\delta f_{\rm core}\simeq-\frac{C\pi}{8}(\rho+\kappa)\frac{g}{u}.
\label{eqS:core_energy}
\end{equation}
All order-one corrections associated with the detailed vortex profile are absorbed into $C$.

For two half vortices separated by $d$, the long-distance phase contribution relative to an unsplit integer vortex is
\begin{equation}
    \Delta E(d)
    =\frac{\pi}{2}(\kappa-\rho)\ln\frac{d}{\xi}
    +\Delta E_{\rm core}.
\label{eqS:split_energy}
\end{equation}
For $\kappa>\rho$, the logarithmic term opposes separation, while the negative core contribution favors it.
Defining the critical separation by $\Delta E(d_c)=0$ and using Eq.~\eqref{eqS:core_energy}, we obtain
\begin{equation}
    \frac{d_c}{\xi}
    \simeq
    \exp\left[
    \frac{C}{4}
    \frac{\rho+\kappa}{\kappa-\rho}
    \frac{g}{u}
    \right]
    =
    \exp\left[
    \frac{C}{4}
    \frac{1+\kappa/\rho}{\kappa/\rho-1}
    \frac{g}{u}
    \right].
\label{eqS:dc}
\end{equation}
Hence the split configuration is favored for $d<d_c$ within this estimate.
As $\kappa/\rho\rightarrow1^+$, the logarithmic penalty becomes parametrically weak and $d_c/\xi$ grows rapidly.
For $\kappa<\rho$, the long-distance contribution already favors separation, so no finite upper scale $d_c$ of this form arises.



\section{Gauge-consistent BdG calculation of half-vortex Majorana modes}
\label{secS:majorana}

For the real-space Majorana calculation, we consider a periodic torus containing four half-quantum vortices, two of each flavor.
The two PDW component phases are $\phi_\pm=\phi_{\rm sc}\pm\phi_K$, so a half vortex with $(\delta\phi_{\rm sc},\delta\phi_K)=(\pi,\pi)$ carries a $2\pi$ winding only in $\phi_+$, while $(\pi,-\pi)$ carries a $2\pi$ winding only in $\phi_-$. 
We therefore divide the four defects into two sets, $\mathcal I_+$ and $\mathcal I_-$, containing two vortices of the $\Delta_+$ and $\Delta_-$ components, respectively. 
The four-HQV configuration carries two superconducting flux quanta in total and is compatible with magnetic periodic boundary conditions on the torus. 
For each component, the amplitude is suppressed smoothly near its own vortex cores according to
\begin{equation}
    F_\pm(\bm r)=\prod_{j\in\mathcal I_\pm}\tanh\left[\frac{d_T(\bm r,\bm R_j)}{\ell_0}\right],
\label{eqS:vortex_profile}
\end{equation}
where $d_T$ is the shortest distance on the torus.
For the calculation shown in the main text, we use $L_x=L_y=48$, $t=1$, $\mu=0$, $\Delta_0=0.6$, and $\ell_0=1.5$, with the vortex centers placed away from lattice sites and separated by approximately half a torus period.

Following the vortex-phase construction of Ref.~\cite{Ariad2018How}, we use the Jacobi theta function to describe vortices on the torus. 
Let $\tau_j=T_{jx}+iT_{jy}$, $j=1,2$, be the complex representations of the two torus periods. 
For a unit vortex centered at $z_j=R_{jx}+iR_{jy}$, we define
\begin{equation}
    \theta(z-z_j)=\operatorname{Im}\left\{\ln\left[i\vartheta_1\left(\frac{z-z_j}{\tau_2},-\frac{\tau_1}{\tau_2}\right)\right]-\frac{2i(z-z_j)^2}{\tau_1\tau_2}\arctan\left(i\frac{\tau_1}{\tau_2}\right)\right\},
\label{eqS:theta_vortex}
\end{equation}
where $z=r_x+ir_y$. 
This phase has the required $2\pi$ winding around the vortex and the quasi-periodicity appropriate to a periodic vortex array~\cite{Ariad2018How}. 
We apply it independently to the two PDW components,
\begin{equation}
    \phi_\pm(\bm r)=\sum_{j\in\mathcal I_\pm}\theta(z-z_j),\qquad
    \phi_{\rm sc}=\frac{\phi_++\phi_-}{2},\qquad
    \phi_K=\frac{\phi_+-\phi_-}{2}.
\label{eqS:component_vortex_phase}
\end{equation}
Thus the electromagnetic gauge field is determined only by the common phase $\phi_{\rm sc}$, whereas $\phi_K$ is the neutral relative phase.

We implement this texture in the real-space mean-field Hamiltonian
\begin{equation}
    H_{\rm MF}=-t\sum_{\langle ij\rangle}\left(e^{ia_{ij}}c_i^\dagger c_j+\mathrm{H.c.}\right)-\mu\sum_i n_i+\sum_{\langle ij\rangle}\left(\Delta_{ij}c_i^\dagger c_j^\dagger+\mathrm{H.c.}\right).
\label{eqS:realspace_bdg}
\end{equation}
The superconducting phase entering the pairing texture and the Peierls phase entering the hopping are generated from the same vortex phase field. 
For a hopping link from $j$ to $i$, the numerical calculation uses
\begin{equation}
    a_{ij}=-\frac12\operatorname{Arg}\left[e^{i\phi_{\rm sc}(\bm r_j)}e^{-i\phi_{\rm sc}(\bm r_i)}\right].
\label{eqS:local_peierls}
\end{equation}
Equivalently, this is the lattice London relation $a_{ij}=(e/\hbar c)\int_{\bm r_j}^{\bm r_i}\bm A\cdot d\bm l$, with $\bm A=(\hbar c/2e)\bm\nabla\phi_{\rm sc}$ locally away from the vortex singularities. 
Thus the phase winding of each half vortex is accompanied by the corresponding half-quantum magnetic flux.

For each nearest-neighbor bond oriented from an $A$-sublattice site $i$ to its $B$-sublattice neighbor $j=i+\bm e_m$, we denote the bond midpoint by $\bm r_m=\bm r_i+\bm e_m/2$. 
The two PDW contributions are
\begin{align}
    \Delta_{+,m}(\bm r_i)&=\Delta_0F_+(\bm r_m)e^{i\phi_+(\bm r_m)}e^{+i\bm K\cdot(\bm r_i-\bm e_m)},\nonumber\\
    \Delta_{-,m}(\bm r_i)&=\Delta_0F_-(\bm r_m)e^{i\phi_-(\bm r_m)}e^{-i\bm K\cdot(\bm r_i-\bm e_m)},
\label{eqS:bdg_pdw_components}
\end{align}
and the pairing matrix element is
\begin{equation}
    \Delta_{ij}=\frac12\left[\Delta_{+,m}(\bm r_i)+\Delta_{-,m}(\bm r_i)\right],\qquad \Delta_{ji}=-\Delta_{ij}.
\label{eqS:bdg_pairing}
\end{equation}
In the vortex-free equal-amplitude limit this reduces to
\begin{equation}
    \Delta_{ij}=\Delta_0e^{i\phi_{\rm sc}}\cos\left[\bm K\cdot(\bm r_i-\bm e_m)+\phi_K\right],
\label{eqS:bdg_pairing_cos}
\end{equation}
which is the lattice PDW form used in the real-space calculation.

Because the torus carries nonzero net magnetic flux, the vector potential cannot in general be chosen as a globally periodic function.
However, the physical magnetic field is periodic.
So for a torus translation $\bm S=n_1\bm T_1+n_2\bm T_2$, the difference $\bm A(\bm r+\bm S)-\bm A(\bm r)$ has vanishing curl and can therefore be written on the covering space as a gauge transformation,
\begin{equation}
    \bm A(\bm r+\bm S)=\bm A(\bm r)+\bm\nabla\Lambda_{\bm S}(\bm r).
\label{eqS:A_transition}
\end{equation}
Correspondingly, the electron operator transforms as $c(\bm r)\rightarrow e^{ie\Lambda(\bm r)/\hbar c}c(\bm r)$, while the superconducting phase transforms as $\phi_{\rm sc}(\bm r)\rightarrow\phi_{\rm sc}(\bm r)+2e\Lambda(\bm r)/\hbar c$.
The quasi-periodic jump of the same theta-function phase therefore determines the magnetic boundary condition,
\begin{equation}
    c(\bm r+\bm S)=G_{\bm S}(\bm r)c(\bm r),\qquad
    G_{\bm S}(\bm r)=\exp\left[\frac{i}{2}\delta_{\bm S}\phi_{\rm sc}(\bm r)\right].
\label{eqS:transition_function}
\end{equation}
Thus the pairing texture, hopping Peierls phases, and magnetic boundary conditions are all generated from the same superconducting vortex phase, yielding a globally well-defined BdG Hamiltonian on the torus.

Diagonalizing Eq.~\eqref{eqS:realspace_bdg} for $L_x=L_y=48$, we find the two smallest positive energies $E_1/t=2.16\times10^{-12}$ and $E_2/t=2.25\times10^{-12}$, together with their particle-hole partners, forming a four-dimensional near-zero subspace. 
The next positive-energy vortex-core levels occur at approximately $E/t=0.813,\ 0.825,\ 0.838,\ 0.849,\ldots$, so the Majorana manifold is separated from conventional core excitations by a gap of order $0.81t$.
In the main text, we show the spatial profiles of the two near-zero complex BdG eigenstates, each of which has weight on two half vortices of opposite flavor, while Fig.~\ref{figS:Majorana_Ex} shows the four lowest finite-energy core states, each localized at an individual half-vortex core.
For a normalized BdG eigenvector $\Psi_n=(u_n,v_n)^T$, we characterize its spatial profile by $P_n(i)=|u_{n,i}|^2+|v_{n,i}|^2$.
Because the four near-zero modes are exponentially close in energy, the complex eigenvectors returned by diagonalization need not be individually localized at single vortex cores.
Within the four-dimensional near-zero subspace, one can instead choose a particle-hole-symmetric Majorana basis whose four wave packets are localized at the four half-vortex cores.


\begin{figure}[t]
    \centering
    \includegraphics[width=\textwidth]{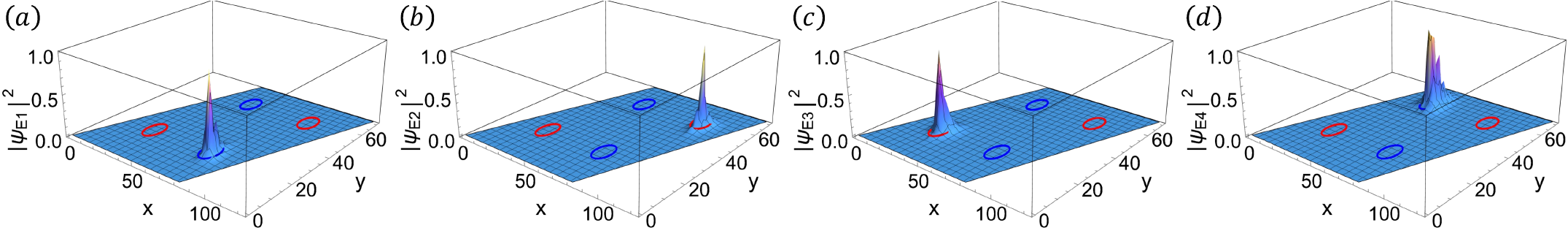}
    \caption{\textbf{Lowest finite-energy vortex-core states.}
    Spatial probability densities of the four lowest excited BdG states above the Majorana zero-mode manifold. 
    Panels (a--d) show that each state is strongly localized at an individual half-vortex core. 
    Red and blue circles denote the two half-vortex flavors associated with vortices in the $\Delta_+$ and $\Delta_-$ components, respectively.}
\label{figS:Majorana_Ex}
\end{figure}

\section{Half-vortex lattice geometry and shape relaxation}
\label{sec:sm_vortex_lattice}

\subsection{Triangular reference lattice}

We consider equal densities of the two half-vortex flavors, whose phase windings are $(1/2,s/2)$ with $s=\pm1$.
In the isotropic logarithmic regime, their pair interaction is
\begin{equation}
    V_{ss'}(r)=-\frac{\pi}{2}(\rho+\kappa ss')\ln\frac{r}{\xi}.
\label{eq:sm_pair_interaction}
\end{equation}
Here $\rho,\kappa>0$, and $\xi$ denotes a short-distance reference scale.
We work at fixed vortex density and initially restrict the variational space to commensurate configurations containing one vortex of each flavor per common Bravais cell. 
The two sublattices are related by a uniform displacement $\boldsymbol\tau$.

A triangular reference shape is motivated by first minimizing the same-flavor interaction. 
At fixed density, the triangular lattice minimizes the regularized one-component logarithmic energy among Bravais lattices~\cite{sandier2012ginzburg,Zhang2015On}.
A useful perturbative regime is
\begin{equation}
    \delta=\frac{|\kappa-\rho|}{\kappa+\rho}\ll1,
\label{eq:sm_weak_interflavor}
\end{equation}
in which the same-flavor energy determines the leading sublattice shape, while the weaker interflavor interaction selects the relative displacement. 
We therefore neglect the interflavor coupling only when choosing the zeroth-order shape and retain it when minimizing the displacement energy. 
Setting this coupling identically to zero would leave the relative displacement undetermined by the logarithmic interaction.

We choose the reference vectors
\begin{equation}
    \mathbf a_1=a(1,0),\qquad
    \mathbf a_2=a\left(\frac12,\frac{\sqrt3}{2}\right),\qquad
    A=\frac{\sqrt3}{2}a^2,
\label{eq:sm_triangular_vectors}
\end{equation}
and write $\boldsymbol\tau=d(\cos\theta,\sin\theta)$. 
The length $a$ is fixed by the density and remains a reference length when shape relaxation is considered below. 
Short-distance physics is represented by an additive core contribution $U_{\rm core}(d)$.
Its assumed rotational invariance means that it does not affect angular minimization at fixed $d$.
This form describes an internal pair contribution.
It does not replace a full lattice sum of core interactions when overlap with several neighboring vortices is appreciable.

\subsection{Angular selection at arbitrary displacement within the convergence disk}

For the fixed triangular lattice, the same-flavor contribution is independent of the relative displacement $\bm{\tau}$.
The displacement-dependent energy per cell can therefore be written as
\begin{equation}
    E_{\triangle}(d,\theta)=E_{\triangle,0}+\frac{\pi}{2}(\kappa-\rho)G_{\triangle}(d,\theta)+U_{\rm core}(d).
\end{equation}
Here $G_{\triangle}$ denotes the regularized logarithmic interaction between the two translated triangular sublattices.
Since $U_{\rm core}(d)$ is assumed to be rotationally invariant, it does not affect the angular minimization at fixed $d$.

For the angular minimization, only the $\theta$-dependent part of $G_{\triangle}$ is needed.
Set $x=d/a$, $\omega=m+n\omega_0$, and $\omega_0=e^{i\pi/3}$. 
For $x<1$,
\begin{equation}
    \ln|a\omega-de^{i\theta}|=\ln(a|\omega|)-\sum_{p=1}^{\infty}\frac{x^p}{p}\operatorname{Re}\left(\frac{e^{ip\theta}}{\omega^p}\right),
\end{equation}
where we used $\ln{|1-z|}=\frac{1}{2}(\ln(1-z)+\ln(1-z^*))=-{\rm Re}(\sum_{n=1}^\infty z^n/n)$ for $|z|<1$. 
All terms independent of $\theta$, including the regularization of the logarithmic lattice sum, are absorbed into $G_{\triangle,0}(d)$.

The angular dependence simplifies strongly because of the sixfold symmetry of the triangular lattice. 
Under $\omega\mapsto e^{i\pi/3}\omega$, each term $\omega^{-p}$ acquires a factor $e^{-ip\pi/3}$. 
Summing over each sixfold orbit therefore gives $\sum_{m=0}^{5}e^{-imp\pi/3}=6\,\delta_{p\,{\rm mod}\,6,0}$, so only $p=6k$ survives. 
Reflection symmetry makes the corresponding lattice sums real. 
Hence
\begin{equation}
    G_{\triangle}(d,\theta)=G_{\triangle,0}(d)-\sum_{k=1}^{\infty}\frac{S_{6k}}{6k}x^{6k}\cos(6k\theta),
\label{eq:G_triangle}
\end{equation}
where
\begin{equation}
    S_{6k}=\sum_{\omega\neq0}\omega^{-6k}=\sum_{(m,n)\neq(0,0)}(m+n\omega_0)^{-6k}\in\mathbb R.
\label{eq:S6k}
\end{equation}
The series is absolutely convergent for $0<x<1$.

We next show that $S_{6k}>0$ for every $k\geq1$. 
The six nearest lattice sites have $|\omega|=1$ and satisfy $\omega^{6k}=1$, so together they contribute exactly $6$. 
Writing
\begin{equation}
    S_{6k}=6+R_k,\qquad |R_k|\leq\sum_{|\omega|>1}|\omega|^{-6k}\leq B_6,
\label{eq:approx S6k}
\end{equation}
where $B_6=\sum_{|\omega|>1}|\omega|^{-6}$, it remains only to bound the total contribution of all more distant sites.
Introduce the hexagonal shell $j=\max(|m|,|n|,|m+n|)$. 
The shell $j$ contains $6j$ sites, and for $j\geq2$ one has $|\omega|\geq\sqrt{3}j/2$.
Hence
\begin{align}
    B_6\leq\sum_{j=2}^{\infty}6j\left(\frac{2}{\sqrt{3}j}\right)^6
    =\frac{128}{9}\sum_{j=2}^{\infty}j^{-5}
    <\frac{128}{9}\left(\frac{1}{2^5}+\int_2^\infty t^{-5}dt\right)=\frac{2}{3}.
\label{eq:B6}
\end{align}
It follows immediately that
\begin{equation}
    S_{6k}>6-\frac{2}{3}=\frac{16}{3}>0
\end{equation}
for every $k\geq1$. 
Thus, all coefficients in the angular expansion have the same sign, and $G_{\triangle}$ is minimized when $\cos(6k\theta)=1$ for every $k$, namely at $\theta=n\pi/3$.

To locate the maxima, we show that $G_{\triangle}$ is strictly increasing for $0<\theta<\pi/6$. 
Define $y=x^6\in(0,1)$ and $\phi=6\theta\in(0,\pi)$.
Since $S_{6k}=6+R_k$, differentiation of Eq.~\eqref{eq:G_triangle} gives
\begin{equation}
    \partial_\theta G_{\triangle}=\frac{6y\sin\phi}{1-2y\cos\phi+y^2}+\sum_{k=1}^{\infty}R_ky^k\sin(k\phi).
\end{equation}
The first term is the exact contribution of the six nearest sites, obtained from $\sum_{k\geq1}y^k\sin(k\phi)=y\sin\phi/(1-2y\cos\phi+y^2)$.
Since $1-2y\cos\phi+y^2\leq(1+y)^2<4$,
\begin{equation}
    \frac{6y\sin\phi}{1-2y\cos\phi+y^2}>\frac{3}{2}y\sin\phi.
\end{equation}

The remaining sites give a smaller correction. 
Since $R_k=\sum_{|\omega|>1}\omega^{-6k}$ and $|\sin(k\phi)|\leq k\sin\phi$ for $0<\phi<\pi$,
\begin{equation}
    \left|\sum_{k=1}^{\infty}R_ky^k\sin(k\phi)\right|
    \leq\sin\phi\sum_{|\omega|>1}\sum_{k=1}^{\infty}k(y|\omega|^{-6})^k
    =y\sin\phi\sum_{|\omega|>1}\frac{|\omega|^{-6}}{(1-y|\omega|^{-6})^2}.
\end{equation}
For all non-nearest sites $|\omega|\geq\sqrt3$, so $y|\omega|^{-6}<1/27$. 
Using $B_6<2/3$ then gives
\begin{equation}
    \left|\sum_{k=1}^{\infty}R_ky^k\sin(k\phi)\right|
    <\left(\frac{27}{26}\right)^2B_6y\sin\phi<\frac{3}{4}y\sin\phi.
\end{equation}
Therefore
\begin{equation}
    \partial_\theta G_{\triangle}>\left(\frac{3}{2}-\frac{3}{4}\right)y\sin\phi>0,\qquad 0<\theta<\frac{\pi}{6}.
\end{equation}
Thus $G_{\triangle}$ increases strictly from $\theta=0$ to $\theta=\pi/6$.
By reflection symmetry and sixfold periodicity, $\theta=n\pi/3$ are its minima and $\theta=\pi/6+n\pi/3$ are its maxima.

Since the angular part of the energy is proportional to $(\kappa-\rho)G_{\triangle}$, the fixed-distance energy minima are therefore
\begin{equation}
    \theta_{\min}=
    \begin{cases}
        n\pi/3, & \kappa>\rho,\\
        \pi/6+n\pi/3, & \kappa<\rho,
    \end{cases}
    \qquad n\in\mathbb{Z}.
\label{eq:minimal theta}
\end{equation}
At $\kappa=\rho$, the logarithmic displacement energy is independent of $\theta$. 
Equation~\eqref{eq:minimal theta} is exact within the fixed triangular ansatz for every $0<d<a$.
The proof has a simple interpretation.
The six nearest sites favor the two high-symmetry directions, while the bounds above show that the combined contribution of all more distant sites is too small to change either the minima or the monotonic interpolation between them.

Our angular selection is consistent with the high-symmetry structures considered by Chung and Kivelson~\cite{Chung2010Entropy}.
For $\kappa>\rho$, the selected direction is along a primitive lattice vector, $\boldsymbol\tau\parallel\mathbf a_1$, and the special displacement $d=a/2$ gives their interlaced-rhombus configuration, with one flavor located midway between two vortices of the other flavor. 
For $\kappa<\rho$, the selected direction is along $\mathbf a_1+\mathbf a_2$, and $d=a/\sqrt3$ gives $\boldsymbol\tau=(\mathbf a_1+\mathbf a_2)/3$, corresponding to the honeycomb configuration. 
Ref.~\cite{Chung2010Entropy} compares these two high-symmetry structures and finds the honeycomb structure favored for $\kappa<\rho$ and the interlaced-rhombus structure favored for $\kappa>\rho$.
Our result complements this comparison by allowing an arbitrary displacement angle at fixed $d$ and showing that these two symmetry directions are selected throughout the convergence region $0<d<a$.
We emphasize, however, that the angular minimization does not determine the equilibrium separation $d$.
The two high-symmetry structures occur at different radii, and the preferred $d$ depends on short-distance core physics and other contributions beyond the logarithmic angular interaction.
In particular, the thermal stabilization discussed in Ref.~\cite{Chung2010Entropy} is distinct from the static angular selection established here.

\subsection{Weak relaxation of the triangular lattice}

We finally examine how the triangular reference lattice is modified when the weak interflavor coupling $\kappa-\rho$ is retained.
For $\kappa>\rho$, the angular analysis selects $\boldsymbol\tau=(d,0)$. 
We consider the area-preserving strain
\begin{equation}
    \mathbf b_1(\epsilon)=a(e^\epsilon,0),\qquad
    \mathbf b_2(\epsilon)=a\left(\frac{e^\epsilon}{2},\frac{\sqrt3e^{-\epsilon}}{2}\right),
\label{eq:sm_strained_vectors}
\end{equation}
which compresses the lattice along $\boldsymbol\tau$ for $\epsilon<0$ and expands it in the perpendicular direction.

For small $\epsilon$ and weak interflavor coupling, the energy has the form
\begin{equation}
    E(\epsilon,d)-E(0,d)=\mu_{s}\epsilon^2+(\kappa-\rho)\chi(d)\epsilon+\cdots.
\label{eq:sm_strain_general}
\end{equation}
The first term is the elastic cost of deforming the same-flavor triangular lattices. 
The absence of a linear term follows because the triangular lattice is stationary at fixed area, and its stiffness satisfies $\mu_{s}\propto\kappa+\rho$.

The second term describes the coupling between the strain and the relative displacement. 
Its small-$d$ behavior follows directly from symmetry. 
At $d=0$ there is no distinguished displacement direction, so the linear strain coupling vanishes, $\chi(0)=0$.
Moreover, inversion symmetry $\boldsymbol\tau\rightarrow-\boldsymbol\tau$ forbids terms odd in $\boldsymbol\tau$.
The leading strain-displacement coupling is therefore quadratic and has the form $\epsilon(\tau_x^2-\tau_y^2)$.
For the selected displacement $\boldsymbol\tau=(d,0)$, this implies $\chi(d)=\chi_2(d/a)^2+O\left((d/a)^4\right)$, where $\chi_2$ is a dimensionless constant determined by the logarithmic lattice interaction.

Minimizing Eq.~\eqref{eq:sm_strain_general} with respect to $\epsilon$ gives
\begin{equation}
    \epsilon_*=-\frac{(\kappa-\rho)\chi(d)}{2\mu_s}
    =\mathcal{O}\left[\frac{\kappa-\rho}{\kappa+\rho}\left(\frac{d}{a}\right)^2\right].
\label{eq:sm_optimal_strain}
\end{equation}
The sign of $\epsilon_*$ is not fixed by symmetry and depends on the microscopic lattice sum.
Thus the magnitude of the deformation is linear in the weak interflavor coupling
$|\kappa-\rho|/(\kappa+\rho)$ and quadratic in the relative displacement for $d/a\ll1$.
Hence the triangular lattice remains a controlled zeroth-order description for
$|\kappa-\rho|\ll\kappa+\rho$, with only a small shape relaxation.

\subsection{Radial stability of high-symmetry displacements}

We next examine the radial stability of the high-symmetry displacements selected by the angular analysis.
A direct sixfold-symmetric real-space regularization of the logarithmic lattice sum can be written as
\begin{equation}
    \widetilde G_\triangle(\boldsymbol\tau)
    =
    \sum_{\mathbf R\in\Lambda}^{\rm sym}
    \ln|\mathbf R-\boldsymbol\tau|,
\label{eq:sm_log_shell_sum}
\end{equation}
where the displacement-independent divergence of the logarithmic sum should be subtracted.
This symmetric-shell sum still contains the smooth contribution generated by the average lattice density $1/A$.
For a sixfold-symmetric cutoff, this contribution is
$\pi|\boldsymbol\tau|^2/(2A)$, where
$A=\sqrt3a^2/2$ is the triangular-lattice unit-cell area.
The translationally invariant displacement energy is therefore
\begin{equation}
    G_\triangle(\boldsymbol\tau)=\widetilde G_\triangle(\boldsymbol\tau)-\frac{\pi}{2A}|\boldsymbol\tau|^2.
\label{eq:sm_periodic_log}
\end{equation}
The quadratic term does not represent an additional physical interaction.
It removes the smooth cutoff-dependent part of the divergent real-space sum and does not affect the angular minimization at fixed $|\boldsymbol\tau|$.

For a displacement $\boldsymbol\tau=d\hat{\mathbf e}$, define $r_\parallel=(\mathbf R-d\hat{\mathbf e})\cdot\hat{\mathbf e},\ r_\perp=(\mathbf R-d\hat{\mathbf e})\cdot\hat{\mathbf e}_\perp$.
The first two radial derivatives are
\begin{equation}
    \widetilde G_\triangle'(d)=-\sum_{\mathbf R}^{\rm sym}\frac{r_\parallel}{r_\parallel^2+r_\perp^2},\qquad
    \widetilde G_\triangle''(d)=\sum_{\mathbf R}^{\rm sym}\frac{r_\perp^2-r_\parallel^2}{(r_\parallel^2+r_\perp^2)^2}.
\label{eq:sm_log_derivative}
\end{equation}

For $\kappa>\rho$, the selected direction is
$\hat{\mathbf e}=\hat{\mathbf a}_1$.
At the symmetric midpoint $d=a/2$, ${r_\parallel}/{a}=(2m+n-1)/2,\ {r_\perp}/{a}={\sqrt3 n}/{2}$.
Evaluating the first and second derivative with sixfold-symmetric shells gives
\begin{equation}
    G_\triangle'\left(\frac a2\right)=0,\qquad
    G_\triangle''\left(\frac a2\right)\approx-\frac{9.52594}{a^2}<0
\end{equation}
Since the physical displacement energy is
$\frac{\pi}{2}(\kappa-\rho)G_\triangle$,
the midpoint $d=a/2$ is a radial local maximum for $\kappa>\rho$.
Together with the angular minimum established above, the interlaced-rhombus midpoint is therefore a saddle point of the pure logarithmic interaction.

For $\kappa<\rho$, the selected direction is
$\hat{\mathbf e}_h=(\mathbf a_1+\mathbf a_2)/|\mathbf a_1+\mathbf a_2|$.
At the honeycomb displacement $\boldsymbol\tau_h={(\mathbf a_1+\mathbf a_2)}/{3},\ d_h={a}/{\sqrt3}$, the projected coordinates are ${r_\parallel}/{a}={\sqrt3}(m+n-2/3)/{2},\ {r_\perp}/{a}={(n-m)}/{2}$.
The corresponding sums give
\begin{equation}
    G_\triangle'\left(\frac{a}{\sqrt3}\right)=0,\qquad
    G_\triangle''\left(\frac{a}{\sqrt3}\right)\approx-\frac{3.6276}{a^2}<0.
\label{eq:sm_honeycomb_radial_curvature}
\end{equation}
Because $\kappa-\rho<0$, the physical energy has positive curvature at this point.
The honeycomb configuration is therefore a local minimum of the pure logarithmic interaction.

\subsection{Core-selected separation and lattice locking}

The logarithmic interaction determines the long-distance interaction between two half vortices, while their equilibrium separation also depends on the overlap of the vortex cores.
We first consider an isolated pair and then extend the discussion to the periodic half-vortex lattice.
Let $z=d/\xi$, and write the separation-dependent core contribution as
\begin{equation}
    \Delta E_{\rm core}(d)=-E_c[1-F(z)],
\label{eq:sm_general_core_overlap}
\end{equation}
where $E_c>0$ is the asymptotic core-energy gain, with $F(0)=1$ and $F(z\rightarrow\infty)=0$.
For $\kappa>\rho$, the total energy in the logarithmic regime is
\begin{equation}
    \frac{\Delta E_2(z)}{E_c}=2q\ln z-1+F(z),\qquad
    q=\frac{\pi(\kappa-\rho)}{4E_c}.
\label{eq:sm_general_pair_energy}
\end{equation}
The equilibrium separation $z_0=d_0/\xi$ satisfies $2q/z_0+F'(z_0)=0$.
Defining $Q(z)=-zF'(z)/2$, the stationary condition becomes $q=Q(z_0)$, while at a stationary point $\Delta E_2''(z)/E_c=-2Q'(z)/z$.
The stable solution therefore lies on the branch with $Q'(z)<0$.
For a core interaction whose force vanishes at large separation, $Q(z)\rightarrow0$ as $z\rightarrow\infty$.
Provided that the stable branch extends continuously to large $z$, $q\rightarrow0^+$ therefore implies $d_0/\xi\rightarrow\infty$.
Thus a small stiffness mismatch relative to the core-energy scale naturally favors well-separated half-vortex cores.

We next consider the periodic half-vortex lattice.
For the $\kappa>\rho$ branch, the angular analysis selects a primitive lattice direction, which we take to be $\mathbf a_1$.
We define $R=a/\xi$ and $x=d/a$, with $0<x\leq1/2$, so that $d/\xi=Rx$.
Dropping displacement-independent constants, the lattice energy can be written as
\begin{equation}
    \frac{E_R(x)}{E_c}=2q\,G_\triangle(x)+\mathcal F_R(x),\qquad
    \mathcal F_R(x)=\sum_{m,n\in\mathbb Z}\left[F\left(Rr_{mn}(x)\right)-F\left(Rr_{mn}(0)\right)\right],
\label{eq:sm_general_core_lattice_sum}
\end{equation}
where $r_{mn}(x)=[(m+n/2-x)^2+3n^2/4]^{1/2}$.
The lattice energy is symmetric under $x\rightarrow1-x$, so $x=1/2$ is always a stationary point.
Its curvature is
\begin{equation}
    \frac{1}{E_c}E_R''\left(\frac12\right)= 2q\,G_\triangle''\left(\frac12\right)+\mathcal F_R''\left(\frac12\right).
\label{eq:sm_general_midpoint_curvature}
\end{equation}
As shown above, $G_\triangle''(1/2)\approx-9.52594$, so the logarithmic interaction alone destabilizes the midpoint.
For a decaying core interaction, $\mathcal F_R''(1/2)\rightarrow0$ in the dilute limit $R\rightarrow\infty$, and the midpoint is therefore unstable at sufficiently large $R$.
As $R$ decreases, the core contribution can become large enough to stabilize the symmetric configuration.
We define $R_*$ as the first curvature crossing encountered upon decreasing $R$ from the dilute limit,
\begin{equation}
    2q\,G_\triangle''\left(\frac12\right)+\mathcal F_{R_*}''\left(\frac12\right)=0.
\label{eq:sm_general_Rstar}
\end{equation}
Starting from the dilute limit and decreasing $R$, the midpoint is unstable for $R>R_*$ and becomes locally stable for $R<R_*$ in the well-separated-vortex regime.
Accordingly, the equilibrium displacement satisfies $d_0/a<1/2$ on the dilute side and locks to $d_0/a=1/2$ after the midpoint becomes stable.

We now evaluate this criterion using a core interaction motivated by the long-distance structure of a Ginzburg--Landau vortex.
Consider the standard GL free-energy density $f_{\rm GL}=\alpha'|\nabla\Delta|^2+m'|\Delta|^2+g'|\Delta|^4/2$ with $m'<0$, so that $\Delta_0^2=-m'/g'$ and $\xi^2=\alpha'/|m'|$.
For a singly quantized vortex, $\Delta(\mathbf r)=\Delta_0 f(r)e^{i\varphi}$, and the GL equation becomes
\begin{equation}
    f''+\frac{f'}{r}-\frac{f}{r^2}+\frac{1}{\xi^2}f(1-f^2)=0.
\label{eq:sm_GL_radial}
\end{equation}
At $r\gg\xi$, taking $f(r)=1-c\xi^2/r^2+\cdots$ and keeping the leading $1/r^2$ terms gives $c=1/2$.
Hence $f(r)=1-\xi^2/(2r^2)+\cdots$.
Since the local GL energetics depend on $|\Delta|^2$, we characterize the core by the normalized condensate-density depletion $h(r)\equiv1-f^2(r)$, which behaves as $h(r)\approx\xi^2/r^2$.
We therefore use the simple interpolation
\begin{equation}
    h(r)=\frac{\xi^2}{r^2+\xi^2}.
\label{eq:sm_GL_core_profile}
\end{equation}
We model the separation dependence of the core energy by the normalized overlap $F(z)=C(d)/C(0)$, where $C(d)=\int d^2r\,h(\mathbf r-\mathbf d/2)h(\mathbf r+\mathbf d/2)$.
Using Eq.~\eqref{eq:sm_GL_core_profile} and a Feynman parameter, the overlap can be evaluated as
\begin{equation}
    F(z)=\int_0^1\frac{ds}{1+s(1-s)z^2}=\frac{4}{z\sqrt{z^2+4}}\operatorname{arcsinh}\frac{z}{2},\qquad
    z=\frac{d}{\xi}.
\label{eq:sm_GL_overlap}
\end{equation}
This form has the algebraic long-distance decay inherited from the GL vortex profile.

For the midpoint curvature, we define $u_{mn}=(2m+n-1)/2$, $v_{mn}=\sqrt3 n/2$, and $r_{mn}=(u_{mn}^2+v_{mn}^2)^{1/2}$.
The core contribution is then
\begin{equation}
    \mathcal F_R''\left(\frac12\right)=\sum_{m,n\in\mathbb Z}
    \left[R^2F''(Rr_{mn})\frac{u_{mn}^2}{r_{mn}^2}+RF'(Rr_{mn})\frac{v_{mn}^2}{r_{mn}^3}\right].
\label{eq:sm_GL_lattice_curvature}
\end{equation}
The sum is convergent and can be evaluated using sixfold-symmetric triangular-lattice shells.
For concreteness, we take the locking scale $R_*=a_*/\xi=16$.
Numerically, Eq.~\eqref{eq:sm_GL_lattice_curvature} gives $\mathcal F_{16}''(1/2)\approx3.45747$, and Eq.~\eqref{eq:sm_general_Rstar} therefore yields
\begin{equation}
    q=\frac{\mathcal F_{16}''(1/2)}{-2G_\triangle''(1/2)}\approx0.18148.
\label{eq:sm_GL_q_example}
\end{equation}
To relate this value to the critical separation $d_c$ introduced in the main text, we use the asymptotic estimate $\Delta E(d)=\frac{\pi}{2}(\kappa-\rho)\ln(d/\xi)-E_c$.
The condition $\Delta E(d_c)=0$, together with the definition of $q$, gives
\begin{equation}
    q=\frac{1}{2\ln(d_c/\xi)},\qquad
    \frac{d_c}{\xi}=\exp\left(\frac{1}{2q}\right).
\label{eq:sm_q_dc_relation}
\end{equation}
For $q\approx0.18148$, this gives $d_c/\xi\approx15.72$, consistent with the values obtained near the PDW--DSM transition in the main-text calculation.
At the locking point, the half-vortex separation is $d_0/\xi=R_*/2=8$, which remains below $d_c/\xi$.
Thus the midpoint-locked configuration lies within the regime where the split half-vortex state is energetically favorable.

For $R>R_*$, the midpoint is unstable and the equilibrium displacement satisfies $d_0/a<1/2$, approaching the isolated-pair result in the dilute limit.
For $R<R_*$, the midpoint becomes locally stable in the regime of interest and the relative displacement locks to $d_0/a=1/2$.

For the Ginzburg--Landau core profile, this locking scale is parametrically distinct from the splitting scale $d_c$.
Since $F(z)\sim4\ln z/z^2$ at large $z$, the core contribution to the midpoint curvature behaves as
$\mathcal F_R''(1/2)\sim A\ln R/R^2$, with $A>0$.
The locking condition therefore gives $q\sim\ln R_*/R_*^2$, so that $R_*$ grows only algebraically, up to logarithmic corrections, as $q\to0^+$.
By contrast, we have $d_c/\xi=\exp[1/(2q)]$, and hence $\frac{R_*}{2d_c/\xi}\rightarrow0$ for $q\rightarrow0^+$.
Since the separation at locking is $d_*=R_*\xi/2$, the midpoint locks well inside the fractionalized regime.

\section{Majorana band structure of the half-vortex lattice}

\subsection{Effective Majorana hopping model}
\label{secS:majorana_model}

We next consider the Majorana band generated by hybridization of the half-vortex zero modes. 
We measure all distances in units of the triangular-lattice spacing $a$, so that the relative displacement is $\boldsymbol\tau=d\mathbf a_1$ with $0<d<1$.
The two half-vortex flavors form sublattices $A$ and $B$ at
\begin{equation}
    \mathbf r_{A,\mathbf R}=\mathbf R,\qquad
    \mathbf r_{B,\mathbf R}=\mathbf R+d\mathbf a_1,
\end{equation}
where $\mathbf a_1=(1,0)$ and $\mathbf a_2=(1/2,\sqrt3/2)$.

We retain the same-flavor nearest-neighbor hopping $t_s$ and the three inequivalent opposite-flavor hoppings $t_0,t_1,t_2$.
Their magnitudes are taken to decay exponentially with the Majorana separation,
\begin{equation}
    t_s=t_s^{(0)}e^{-1/\xi_M},\qquad
    t_0=t_c^{(0)}e^{-d/\xi_M},\qquad
    t_1=t_c^{(0)}e^{-(1-d)/\xi_M},\qquad
    t_2=t_c^{(0)}e^{-\sqrt{1-d+d^2}/\xi_M}.
\label{eq:sm_majorana_hoppings}
\end{equation}
The corresponding opposite-flavor bond vectors are $d\mathbf a_1$, $-(1-d)\mathbf a_1$, $d\mathbf a_1-\mathbf a_2$, and $d\mathbf a_1+\mathbf a_2-\mathbf a_1$, with the last two having the same length $\sqrt{1-d+d^2}$.

The effective Majorana Hamiltonian is
\begin{equation}
\begin{split}
    H_M={}&it_s\sum_{\alpha, \mathbf R}\eta_\alpha\Big[
    \gamma_{\alpha,\mathbf R}\gamma_{\alpha,\mathbf R+\mathbf a_1}
    +\gamma_{\alpha,\mathbf R}\gamma_{\alpha,\mathbf R-\mathbf a_2}
    +\gamma_{\alpha,\mathbf R}\gamma_{\alpha,\mathbf R+\mathbf a_2-\mathbf a_1}\Big]\\
    &+i\sum_{\mathbf R}\gamma_{A,\mathbf R}\Big[
    t_0\gamma_{B,\mathbf R}
    +t_1\gamma_{B,\mathbf R-\mathbf a_1}
    +t_2\gamma_{B,\mathbf R-\mathbf a_2}
    +t_2\gamma_{B,\mathbf R+\mathbf a_2-\mathbf a_1}
    \Big],
\label{eq:sm_majorana_realspace}
\end{split}
\end{equation}
where $\alpha=A,B$ for two flavors of Majorana modes, and $\eta_\alpha=\pm1$ specifies the orientation of the same-flavor hopping pattern.
We choose a translationally invariant gauge in which all opposite-flavor bonds are directed from $A$ to $B$. 
For the same-flavor hoppings, the positive directions on the $A$ sublattice are chosen as $\mathbf a_1$, $-\mathbf a_2$, and $\mathbf a_2-\mathbf a_1$, with $\eta_A=1$.
We denote by $\eta_B=\eta=\pm1$ whether the corresponding directed pattern on the $B$ sublattice is the same or reversed.

Introducing $k_1=\mathbf k\cdot\mathbf a_1$ and $k_2=\mathbf k\cdot\mathbf a_2$, we use the cell-coordinate Fourier transform
\begin{equation}
    \gamma_{\alpha,\mathbf R}=\frac{1}{\sqrt N}\sum_{\mathbf k}\gamma_{\alpha,\mathbf k}e^{i\mathbf k\cdot\mathbf R},
    \qquad
    \gamma_{\alpha,\mathbf k}^\dagger=\gamma_{\alpha,-\mathbf k}.
\end{equation}
Defining $\Psi_{\mathbf k}^T=(\gamma_{A,\mathbf k},\gamma_{B,\mathbf k})$, the Hamiltonian can be written as $H_M=\sum_{\mathbf k\in{\rm BZ}/2} \Psi_{\mathbf k}^\dagger\mathcal H_M(\mathbf k)\Psi_{\mathbf k}$, where ${\rm BZ}/2$ contains one representative of each pair $\mathbf k$ and $-\mathbf k$.
Then, $\mathcal H_M(\mathbf k)$ takes the form
\begin{equation}
    \mathcal H_M(\mathbf k)=
        \begin{pmatrix}
        g(\mathbf k) & if(\mathbf k)\\
        -if^*(\mathbf k) & \eta g(\mathbf k)
        \end{pmatrix},
\label{eq:sm_majorana_bloch}
\end{equation}
where
\begin{equation}
    g(\mathbf k)=2t_s\left[-\sin k_1+\sin k_2-\sin(k_2-k_1)\right],\qquad 
    f(\mathbf k)=t_0+t_1e^{-ik_1}+t_2e^{-ik_2}+t_2e^{i(k_2-k_1)}.
\label{eq:sm_gk_fk}
\end{equation}
The two bands are obtained analytically as
\begin{equation}
    E_\pm(\mathbf k)=\frac{1+\eta}{2}g(\mathbf k)\pm \sqrt{\left[\frac{1-\eta}{2}g(\mathbf k)\right]^2+|f(\mathbf k)|^2}.
\label{eq:sm_majorana_bands}
\end{equation}
Using $k_1=k_x$ and $k_2=k_x/2+\sqrt3k_y/2$, we have
\begin{equation}
    g(\mathbf k)=2t_s\left[-\sin k_x+2\sin\frac{k_x}{2}\cos\frac{\sqrt3k_y}{2}\right],\qquad
    |f(\mathbf k)|^2=\left[(t_0+t_1)\cos\frac{k_x}{2}+2t_2\cos\frac{\sqrt3k_y}{2}\right]^2+(t_0-t_1)^2\sin^2\frac{k_x}{2}.
\end{equation}
Thus the band structure is completely determined by the four overlap amplitudes in Eq.~\eqref{eq:sm_majorana_hoppings} and the chosen hopping-sign pattern (and hence the Majorana flux sector).

For $\eta=+1$, corresponding to the same directed same-flavor pattern on the two sublattices, $E_\pm(\mathbf k)=g(\mathbf k)\pm|f(\mathbf k)|$, whereas for $\eta=-1$, $E_\pm(\mathbf k)=\pm\sqrt{g(\mathbf k)^2+|f(\mathbf k)|^2}$.
The Majorana particle-hole constraint relates $E(\mathbf k)$ and $-E(-\mathbf k)$, so the spectrum need not be symmetric about zero at fixed momentum for $\eta=+1$.
Having obtained the effective Bloch Hamiltonian, we now examine how the relative displacement $d$ and hopping-sign sector $\eta$ control its symmetry and gap structure.

\subsection{Symmetry and band structure}
\label{secS:majorana_nodal}

The Majorana Bloch Hamiltonian satisfies
\begin{equation}
    \mathcal H_M(-\mathbf k)=-\mathcal H_M^T(\mathbf k),
\end{equation}
since $g(-\mathbf k)=-g(\mathbf k)$ and $f(-\mathbf k)=f^*(\mathbf k)$.
Accordingly, particle-hole symmetry relates $E(\mathbf k)$ to $-E(-\mathbf k)$, although the spectrum need not be symmetric about zero at fixed $\mathbf k$.
The energies are also invariant under $k_y\rightarrow-k_y$.
In addition, $d\rightarrow1-d$ exchanges $t_0$ and $t_1$ while leaving $|f(\mathbf k)|$ unchanged, so the spectrum is symmetric about the displacement $d=1/2$.

The special role of $d=1/2$ can be seen by defining $u=k_x/2$ and $v=\sqrt3k_y/2$, for which
\begin{equation}
    g(\mathbf k)=4t_s\sin u\,(\cos v-\cos u),\qquad
    f(\mathbf k)=e^{-iu}\left[(t_0+t_1)\cos u+2t_2\cos v
    +i(t_0-t_1)\sin u\right].
\label{eq:sm_f_factorized}
\end{equation}
At the symmetric displacement $d=1/2$, one has $t_0=t_1\equiv t_c$, and hence $f(\mathbf k)=2e^{-iu}\left(t_c\cos u+t_2\cos v\right)$.
Therefore $f(\mathbf k)=0$ defines a one-dimensional locus in momentum space.
For $d\neq1/2$, by contrast, $t_0\neq t_1$ and $f=0$ requires both $\sin u=0$ and $(t_0+t_1)\cos u+2t_2\cos v=0$.
For the exponentially decaying hoppings in Eq.~\eqref{eq:sm_majorana_hoppings}, $\sqrt{1-d+d^2}$ is larger than both $d$ and $1-d$, so $t_2<\min(t_0,t_1)$ and therefore $t_0+t_1>2t_2$.
These conditions cannot be satisfied simultaneously, implying that $f(\mathbf k)\neq0$ and $d\neq\frac12$.

For $\eta=-1$, the spectrum is $E_\pm=\pm\sqrt{g^2+|f|^2}$, so zero energy requires $g=f=0$ simultaneously.
At $d=1/2$, these conditions are satisfied at the two symmetry-related points
\begin{equation}
    \mathbf M_\pm=\left(\pi,\pm\frac{\pi}{\sqrt3}\right),
\end{equation}
and reciprocal-lattice-equivalent points.
Expanding about $\mathbf M_\pm$ gives a linear dispersion, so these are isolated Majorana Dirac nodes.
For $d\neq1/2$, $f(\mathbf k)\neq0$ immediately implies a fully gapped spectrum.
Thus, in the $\eta=-1$ sector, displacing the two half-vortex sublattices away from the symmetric configuration generates a mass for the Majorana Dirac nodes.

For $\eta=+1$, the spectrum instead takes the form $E_\pm=g\pm|f|$.
At $d=1/2$, the condition $f(\mathbf k)=0$ gives a one-dimensional direct band-touching locus, along which $E_+=E_-=g(\mathbf k)$.
This band-touching line is not generally at zero energy.
Zero-energy states satisfy instead
\begin{equation}
    g(\mathbf k)^2=|f(\mathbf k)|^2,
\label{eq:sm_fermi_line}
\end{equation}
which is one real condition in two-dimensional momentum space and therefore generically produces one-dimensional Majorana Fermi lines.
For $d\neq1/2$, the direct band-touching line is removed because $f(\mathbf k)$ is everywhere nonzero, but zero-energy Fermi lines can remain.
The $\eta=+1$ spectrum is fully gapped only when $|g(\mathbf k)|<|f(\mathbf k)|$, throughout the Brillouin zone.
For the parameters used in Fig.~\ref{figS:Majorana_band}, the $d=1/4$, $\eta=+1$ case remains gapless through such zero-energy contours.

These results are illustrated in Fig.~\ref{figS:Majorana_band}.
For $\eta=-1$, the symmetric configuration $d=1/2$ has isolated Majorana Dirac nodes, while the asymmetric configuration $d=1/4$ is fully gapped.
For $\eta=+1$, the symmetric configuration has an additional one-dimensional band-touching locus associated with $t_0=t_1$, together with zero-energy Majorana Fermi lines, and shifting to $d=1/4$ removes the direct band degeneracy but does not necessarily remove the zero-energy Fermi lines.
Such sensitivity of the Majorana spectrum to the hopping-sign structure is familiar from Majorana vortex-lattice models~\cite{Biswas2013Majorana,Murray2015Majorana,Liu2015Electronic}.

For the present PDW half-vortex lattice, the two flavors $(\delta\phi_{\rm sc},\delta\phi_K)=(\pi,\pm\pi)$ carry the same superconducting vorticity and the same half flux quantum.
If the symmetry relating the two PDW components does not reverse the effective Majorana hopping flux, the natural minimal choice is therefore $\eta=+1$, corresponding to the same directed same-flavor hopping pattern on the two sublattices.
We therefore consider both $\eta=\pm1$ in the effective Majorana model, while the physical hopping-sign sector should ultimately be fixed by projecting the microscopic vortex-lattice BdG states onto the Majorana subspace.
A projection of the microscopic periodic BdG problem onto the low-energy Majorana subspace can determine the physical flux sector unambiguously.

\begin{figure}[t]
    \centering
    \includegraphics[width=\textwidth]{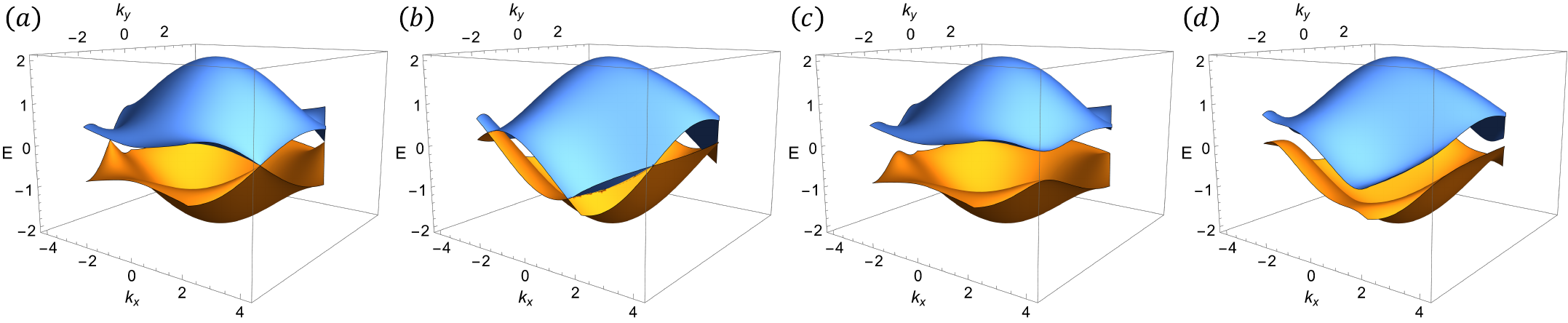}
    \caption{\textbf{Majorana band structure of the half-vortex lattice.}
    Three-dimensional dispersions for (a) $(\eta,d)=(-1,1/2)$, (b) $(1,1/2)$, (c) $(-1,1/4)$, and (d) $(1,1/4)$, with $t_c^{(0)}=1$, $t_s^{(0)}=0.3$, and $\xi_M=1$.
    For $\eta=-1$, the symmetric configuration $d=1/2$ hosts isolated Majorana Dirac nodes, while shifting to $d=1/4$ opens a full gap.
    For $\eta=+1$, $d=1/2$ exhibits an extended direct band-touching locus and zero-energy Majorana Fermi lines; the displacement $d=1/4$ lifts the direct band degeneracy, while zero-energy contours remain for the parameters shown.}
\label{figS:Majorana_band}
\end{figure}

\end{widetext}

\end{document}